\documentclass{article}

\usepackage[a4paper]{geometry}
\usepackage{graphicx}
\usepackage{natbib}
\usepackage{booktabs}
\title{Bayesian Inference with the $l_1$-ball Prior:\\
        Solving Combinatorial Problems with Exact Zeros}
        
\author{Maoran Xu\\
Department of Statistical Science, Duke University, Durham, USA
\\
Leo L. Duan\\
Department of Statistics, University of Florida, Gainesville, USA}
\usepackage{amsthm}
\usepackage{bookmark}
\usepackage{caption}
\usepackage{subcaption}
\usepackage{siunitx}
\usepackage{float}
\usepackage{enumerate}% http://ctan.org/pkg/enumerate 
\usepackage{soul}
\usepackage{amsmath, amssymb} 
\usepackage{mathrsfs}
\usepackage{mhequ}
\usepackage{xcolor}

\usepackage{subfiles}

\newcommand{\intB}{\textbf{int} \, \mathbb{B}_{r}}
\newcommand{\bdB}{\textbf{bd} \, \mathbb{B}_{r}}

\newcommand{\pr}{\text{pr}}

\renewcommand{\t}[1]{\text{#1}}

\newcounter{example}[section]

\newcommand{\T}{\textrm{T}}
\newcommand{\be}{\begin{equation*}
        \begin{aligned}}

\newcommand{\ee}{ \end{aligned}
        \end{equation*}
}

\newcommand{\bel}{\begin{equation}
        \begin{aligned}}

\newcommand{\eel}{ \end{aligned}
        \end{equation}
}

\DeclareMathOperator*{\argmin}{argmin}
\newtheorem{corollary}{Corollary}
\newtheorem{theorem}{Theorem}

\newtheorem{remark}{Remark}
\usepackage[framemethod=tikz]{mdframed}
\begin{document}
\maketitle
\begin{abstract}
The  $l_1$-regularization is very popular in high dimensional statistics --- it changes a combinatorial problem of choosing which subset of the parameter is zero, into a simple continuous optimization. Using a continuous prior concentrated near zero, the Bayesian counterparts are successful in quantifying the uncertainty in the variable selection problems; nevertheless, the lack of exact zeros makes it difficult for broader problems such as change-point detection and rank selection. Inspired by the duality of the $l_1$-regularization as a constraint onto an $l_1$-ball, we propose a new prior by projecting a continuous distribution onto the $l_1$-ball. This creates a positive probability on the ball boundary, which contains both continuous elements and exact zeros. Unlike the spike-and-slab prior, this $l_1$-ball projection is continuous and differentiable almost surely, making the posterior estimation  amenable to the Hamiltonian Monte Carlo algorithm. We examine the properties, such as the volume change due to the projection, the connection to the combinatorial prior, the minimax concentration rate in the linear problem. We demonstrate the usefulness of exact zeros that simplify the combinatorial problems, such as the change-point detection in time series, the dimension selection of mixture models, and the low-rank-plus-sparse change detection in medical images.
\end{abstract}
      
 \section{Introduction}
        The $l_1$-regularization has been a milestone in high dimensional statistics. Since its introduction in the lasso regression for solving the variable selection problem \citep{tibshirani1996regression}, it has inspired a rich class of algorithms and models --- an incomplete list of representative works cover areas of regression \citep{efron2004least,zou2005regularization, yuan2006model}, multivariate data analysis \citep{chen2001atomic,zou2006sparse}, graph estimation \citep{shojaie2010penalized,zhang2014sparse,fan2017high},  among others. For comprehensive reviews, see \cite{meinshausen2006high}, and more recently \cite{buhlmann2011statistics}. 
        
        One of the most appealing properties of $l_1$-regularization is that it induces exact zeros in the optimal solution, hence bypassing the need to decide which subset of the parameter should be zero. This is due to the well-known dual form of the $l_1$-norm penalty, as equivalent to constraining the parameter on an $l_1$-ball centered at the origin. The ``spikiness'' of the $l_1$-ball in high dimension makes it possible for a sparse recovery of the signals [See \cite{vershynin2018high} for a formal exposition].
        
        In recent years, it has been demonstrated that the sparse property can be exploited beyond the simple tasks of variable selection. In particular, some complicated combinatorial problems can be solved (or relaxed) via an {\em ``over-parameterize and then sparsify''} strategy, using the $l_1$-regularization. To give a few concrete examples, in the change-point detection of time series data, the fused lasso \citep{tibshirani2005sparsity} over-parameterizes each time point with an individual mean, then induces sparsity in  the temporal increments/decrements; effectively, this leads to a step function that captures any abrupt temporal changes.
For clustering problems, the sum-of-norms clustering \citep{lindsten2011clustering, tan2015statistical}  assigns a location parameter to every data point,  then sparsifies the pairwise distance matrix; this induces only a few unique locations as the cluster centers.
        In the low-rank matrix smoothing/imputation, one uses an unconstrained matrix as the smoothed mean, then adds the nuclear norm regularization
        \citep{grave2011trace} as equivalent to sparsifying the singular values; this effectively achieves a rank selection on the matrix. These are just a few examples; nevertheless, it is not hard to see the equivalent combinatorial problems would be quite difficult to handle directly.

Most of the above models have been developed with an optimization focus; in parallel, the Bayesian literature is expanding rapidly to address the uncertainty quantification problems, in particular: (i) how likely a parameter element is zero or non-zero? (ii) how much correlation there is between the non-zero elements?
These questions are important for the downstream statistical inference, such as building credible intervals and testing hypotheses.
Among the early work, the Bayesian lasso exponentiates the negative $l_1$-norm in a double exponential prior \citep{park2008bayesian}; however, it was discovered that except for the posterior mode, the posterior of the Bayesian lasso has very little concentration near zero, while the thin tails cause an under-estimation of the non-zero signal. To address these issues, a rich class of continuous shrinkage priors has been proposed, with a large concentration in a neighborhood near zero and a heavy tail to accommodate large signals. Examples include the horseshoe \citep{carvalho2010horseshoe}, generalized double Pareto \citep{armagan2013generalized}, Dirichlet-Laplace \citep{bhattacharya2015dirichlet}, Beta-prime \citep{bai2018beta}, spike-and-slab lasso \citep{rovckova2018spike}; among others. Due to the use of continuous priors, the posterior computation can be carried out efficiently using the Markov chain Monte Carlo methods \citep{bhattacharya2016fast}; this is advantageous compared to the classic spike-and-slab prior, which involves a combinatorial prior that selects a subset of parameters to be non-zero \citep{ mitchell1988bayesian}.

In these Bayesian models, although the posterior is not exactly zero, the close-to-zero estimates are adequate for the common variable selection problems.
However, for the above {over-parameterize-then-sparsify strategy}, it faces challenges to build the Bayesian equivalency using the continuous shrinkage priors. First, the success of {such a strategy} would require some {\em transform} of the parameter to be zero (for example, the sum of changes over a long period of time in the change-point detection case). To that end, the continuous shrinkage prior placed on the individual elements, collectively, does not have a sufficiently large probability for the transform to be near zero. Second, there are problems in assigning a prior directly on the transform. Most importantly, the transform is often in a constrained space (for example, the distance matrix for points in $\mathbb{R}^p$ being low-rank); hence the prescribed prior density is in fact incomplete, missing an intractable normalizing that could have an impact on the amount of shrinkage. 

To address this issue, as well as to encourage developing novel combinatorial models, we propose a new strategy: starting from a continuous random variable with unconstrained support, we project it onto the $l_1$-ball. This induces a distribution allowing the random variable to contain exact zeros. Since the projection is a continuous and differentiable transformation almost surely, we can use the popular off-the-shelf sampling algorithms such as the Hamiltonian Monte Carlo for the posterior computation. We are certainly not the first to consider a projection/transformation-approach for Bayesian modeling. For example, \cite{gunn2005transformation} used it for handling monotone and unimodal constraints, \cite{lin2014bayesian} for inducing monotone Gaussian process, \cite{jauch2020monte} for satisfying orthonormal matrix constraint, and most recently,  \cite{sen2018constrained} for a theoretic study on the projection of the posterior. Although we are inspired by those methods, our focus is not to satisfy the parameter constraints, but to use the {\em boundary} of a constrained set to induce a low-dimensional measure  ---  in this case, any point outside the $l_1$-ball will be projected onto the boundary, we obtain a positive probability for the random variable (or its transform) to contain both continuous elements and exact zeros. To our best knowledge, this idea is new.

We will illustrate the details of the projection, and its properties such as the volume change due to the projection, the connection to the combinatorial prior, and the minimax concentration rate in the linear problem. In the applications,
we will demonstrate the usefulness of {exact sparsity} in a few Bayesian models of combinatorial problems, such as piece-wise constant smoothing, dimension selection in the finite mixture model, and the low-rank matrix smoothing with an application in medical image analysis.

\section{The $l_1$-ball Prior}
    In this section, we propose a new prior construction of producing exact sparsity in a probabilistic framework. We start from several motivating examples, and then introduce the prior while addressing three questions : (i)  how to use an $l_1$-ball projection to change a continuous random variable into a mixed-type random variable, which contains both  continuous elements and zeros? (ii) how to calculate the associated probability after the projection? (iii) how to assign prior on the radius of an $l_1$-ball?    
    
    \subsection{Motivating Combinatorial Problems}
    To motivate a new class of priors on the $l_1$-ball, we first list a few interesting combinatorial problems that can be significantly simplified using our approach. We will use $\theta \in \mathbb{R}^p$  to denote the parameter of interest.
    
      (a) Sparse contrast models. Often in time series and image data applications, we want to estimate some underlying group structure, with each group defined as a continuous or uninterrupted temporal period or spatial region. Those models can often be re-parameterized as having sparse contrast $D\theta$, with a contrast matrix $D\in\mathbb{R}^{d\times p}$ (each row of $D$ adding to zero). For example, in image smoothing/boundary detection, one could use sparse $(\theta_i-\theta_j)$'s between all neighboring pixels  to induce a piece-wise constant structure in the mean. This idea has led to the success of fused lasso approaches using $l_1$-norm on $D\theta$  for obtaining point estimates in optimization \citep{tibshirani2011solution}. On the other hand, because there are more contrasts than pixels in an image ($d\approx 2p$ in a two-dimensional image), the sparse set $\{ x\in\mathbb{R}^d: x= D\theta\}$ is the column space of $D$ with intrinsic dimension at most $p$, hence is challenging to directly assign a sparse prior via conventional Bayesian approaches.
                    
(b) Reduced dimension models. It is common in Bayesian models to consider $\theta$ as a high-dimensional parameter residing near a low-dimensional space so that it can achieve approximate dimension reduction. However, there are cases when it is preferable to induce an exactly low dimension instead of approximation.
 For example, in clustering data analysis
using mixture models, often we want to estimate the number of clusters; and it has been shown that having a low-dimensional mixture weight  can lead to a consistent estimate on the number of clusters \citep{miller2018mixture}, whereas continuous shrinkage priors (such as Dirichlet process prior) can result in an overestimation. As another example, to parameterize a block-diagonal matrix $\theta$ (subject to row and column permutation), often used for community detection in network data analysis, one could control its Laplacian matrix $ L_\theta$ to be exactly low-rank \cite{anderson1985eigenvalues}.

(c) Structured/dependent sparsity models. There has already been rich literature on using sparsity models for variable selection in regression; nevertheless, there is a growing interest in inducing correlation/dependent structure within the parameter \citep{hoff2017lasso,griffin2019structured}.  For example, there may be prior knowledge that some elements in $\theta$ are more likely to be simultaneously zero.
     
Although there are some methods developed specifically for each scenario listed above, we will show that our new prior provides a  simple solution that enables an arguably more straightforward prior specification and tractable posterior estimation. Our proposal can be summarized as the following prior generating process:
        \bel\label{eq:generative}
        \beta &\sim \pi_\beta, \quad r\sim\pi_r,\\
        \theta & =  P_{\mathbb{B}_{h,r}}(\beta),
        \eel
        where $\beta\in \mathbb{R}^p$ is a continuous random variable from distribution $\pi_\beta$ (we slightly abuse notations by letting $\pi_\beta$  denote both distribution and associated density function), and $r>0$ is a scalar-valued random variable that we refer to as the radius and is from distribution $\pi_r$, $\mathbb{B}_{h,r} = \{ x:\|h(x)\|_1\le r\}$ is an $l_1$-ball associated with a function $h:\mathbb{R}^p\to\mathbb{R}^d$, and $ P_{\mathbb{B}_{h,r}}(\cdot)$ denotes a projection equal to  $ \argmin_x\|x-\beta\|_2^2$ with $x\in \mathbb{B}_{h,r}$. Later, we will show how to build models for cases (a) and (b) with some suitable choices of $h$, and how to induce dependence structure in $\theta$ for case (c) by adopting a correlated distribution for $\pi_\beta$.
        
                It is not hard to see that \eqref{eq:generative} gives a joint prior distribution $\pi_{\theta,r}(\theta, r) =  \pi_r(r)\pi_{\theta \mid r}(\theta)$, we will describe $\pi_{\theta \mid r}$ in Sections 2.2 and 2.3,  and $\pi_r$ in Section 2.4.
For now, for better clarity, we first focus on the identity function $h(x)=x$, show how an $l_1$-ball projection leads to sparsity, and characterize the probabilities associated with \eqref{eq:generative}.

        \begin{remark} To clarify, our approach is equivalent to reparameterizing a sparse $\theta$ [or $\theta$ with sparse $h(\theta)$] using continuous $\beta$. In a diagram, our modeling framework is
$$  \theta \sim \pi(\theta\mid y)=\frac{\mathcal L(y;\theta)\pi_\theta(\theta)}{\int \mathcal L(y;\theta)\pi_\theta(\theta) \textup{d}\theta} , \qquad \text{with }\theta=P_{\mathbb B}(\beta).$$
As $\beta$ effectively enters the likelihood $\mathcal L(y;\theta)$ and posterior $\pi(\theta\mid y)$, this is a fully Bayesian model that gives uncertainty quantification and enables inferences on sparsity. Further, this reparameterization does not depend on the form of likelihood, hence our method does not require the posterior to be log-concave.

To compare, there have been several post-processing approaches based on first sampling $\theta$ from a continuous posterior, then producing sparse $\theta^*$ via some transform \citep{bondell2012consistent,hahn2015decoupling,li2017variable}. In a diagram, they can be understood as two-stage estimators:
$$(i)\; \theta \sim \pi(\theta\mid y)= \frac{\mathcal L(y;\theta)\pi_\theta(\theta)}{\int \mathcal  L(y;\theta)\pi_\theta(\theta) \textup{d}\theta}. \qquad (ii)\; \theta^*=T(\theta),$$
for some post-processing mapping $T$. Note that the sparse $\theta^*$ does not influence the likelihood, and corresponds to zero posterior probability (due to $\theta$ being continuous posterior). As a result, these approaches cannot be used for inference tasks such as estimating the $(1-\tilde \alpha)$-credit interval on $\|\theta\|_0$ (number of non-zeros) and $(1-\tilde \alpha)$-prediction interval for $x^{* \rm T}\theta$ for a new $x^*$. We provide a detailed comparison in the supplementary materials. 
        \end{remark}

\subsection{Creating Sparse Prior via an $l_1$-ball Projection}
        
To ease notation, we now use $\mathbb{B}_{r}=\{x\in \mathbb{R}^p: \|x\|_1\le r \}$ as a shorthand for  the vector-norm $l_1$-ball $\mathbb{B}_{h,r}$ with $h(x)=x$.
         We denote the interior set by $\textbf{int} \, \mathbb{B}_r=\{x\in \mathbb{R}^p: \|x\|_1< r \}$, and boundary set by $\textbf{bd} \, \mathbb{B}_r=\{x:\|x\|_1= r \}$.       For any point $\beta\in \mathbb{R}^p $, we can project it
        onto the $l_1$-ball, by solving the following  problem,
        \be
        & \theta = P_{\mathbb{B}_r}(\beta)= \underset{\|x\|_1\le r}{\arg\min} \|\beta -x\|_2^2.
        \ee
        The loss function on the right hand side is strictly convex  ---  that is, for every $\beta$, there is only one optimal solution $\theta= P_{\mathbb{B}_r}(\beta)$  (the mapping is measurable).  For the sake of completeness, we present a simple algorithm [modifying from \cite{duchi2008efficient}]: if $\|\beta\|_1\le r$, let $\theta=\beta$;
 If $\|\beta\|_1>r$,
        \bel\label{eq:l1_proj}
        & \text{sort $\beta$ so that }   |\beta_{(1)}| \ge \ldots \ge |\beta_{(p)}|,\\
        & c:=  \max \bigg \{j: |\beta_{(j)}|>\frac{\mu_j}{j}, \; \mu_j=(\sum_{i=1}^j|\beta_{(i)}|-r)_{+}\bigg \},\\
        &\theta_i := \text{sign}(\beta_i) \text{}(|\beta_i| - \frac{\mu_c}{c})_{+},
        \eel
        where $(x)_+=\max\{x,0\}$.         
 
 We now examine the induced probability distribution of $\theta$. 
        Suppose  $\beta\in\mathbb{R}^p$ is a continuous random variable, in a probability space $(\mathbb{R}^p, \mathcal{B}(\mathbb{R}^p),
        \nu)$, with $\nu$ its measure absolutely continuous with respect to the  Lebesgue measure in $\mathbb{R}^p$, and $\pi_{\beta}$
        the associated density. We can compute  the probability
        for $\theta$ in any set $\mathcal{A}$ in $ \mathbb{B}_r$:
        \bel\label{eq:measure}
        \pr(\theta\in\mathcal{A}) = \int_{\mathbb{R}^p} \mathbb{I} [{P_{\mathbb{B}_r}(x)\in \mathcal
                A}]  \pi_\beta(x) \textup{d}x,
        \eel
        where $\mathbb{I}(E)=1$ if
        the event $E$ is true, otherwise takes value $0$.       
        Combining \eqref{eq:l1_proj} and \eqref{eq:measure}, we see two interesting results when we project from the outside $\beta: \|\beta\|_1>r$ :
                \begin{enumerate}
                \item It yields $\theta$ with $\|\theta\|_1=r$, hence we have $\theta$ in the boundary set.  Since all the points outside the ball will be projected to  $\theta\in\bdB$, the boundary set  has a positive probability.
                \item If the projection has $c<p$, there will be $(p-c)$ elements with $\theta_i=0$.
        \end{enumerate}
%        In addition, in the computation section, we will show %that:
%                        \begin{enumerate}\addtocounter{enumi}{2}
%                \item   The projection $P_{\mathbb{B}_r}$ is %a continuous and differentiable function of $\beta$ almost surely %with respect to $\nu$, which allows the use of the Hamiltonian %Monte Carlo for estimation.      
%        \end{enumerate}

       %  These results suggest a simple  way to develop an  exactly sparse prior: assigning a continuous prior $\pi_\beta$ on $\beta$ and set $\theta=P_{\mathbb{B}_r}(\beta)$. 

        \begin{figure}[H]
                \begin{subfigure}{.45\textwidth}
                        \includegraphics[width=1\linewidth]{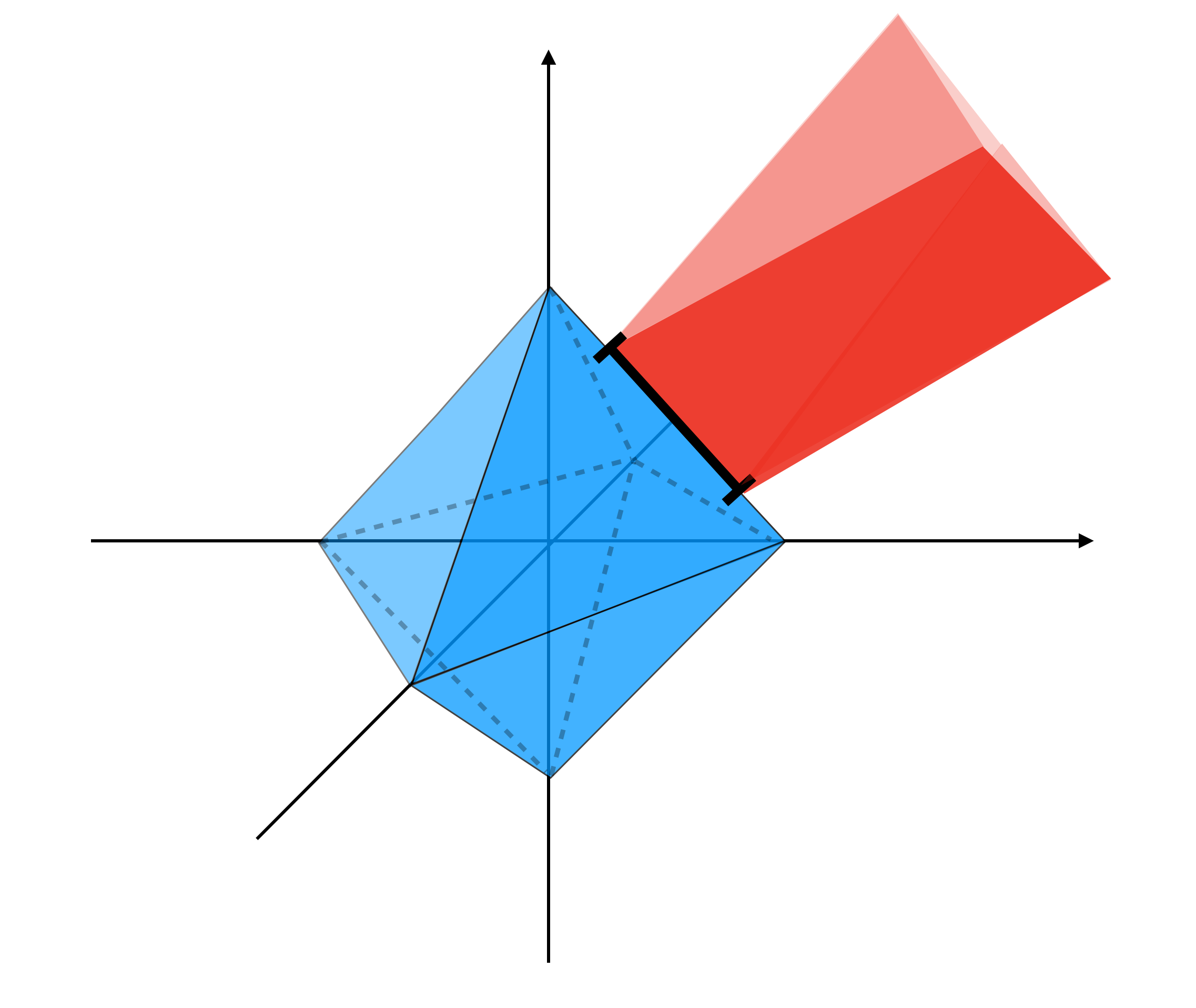}
                        \caption{All the points $\beta$ in the red area are projected to $\theta$ in the line segment $\{(x,0,r-x):0.2<x<0.8\}$, which contains exact zero $\theta_2=0$.}
                \end{subfigure}
                \quad
                \begin{subfigure}{.45\textwidth}
                        \includegraphics[width=1\linewidth]{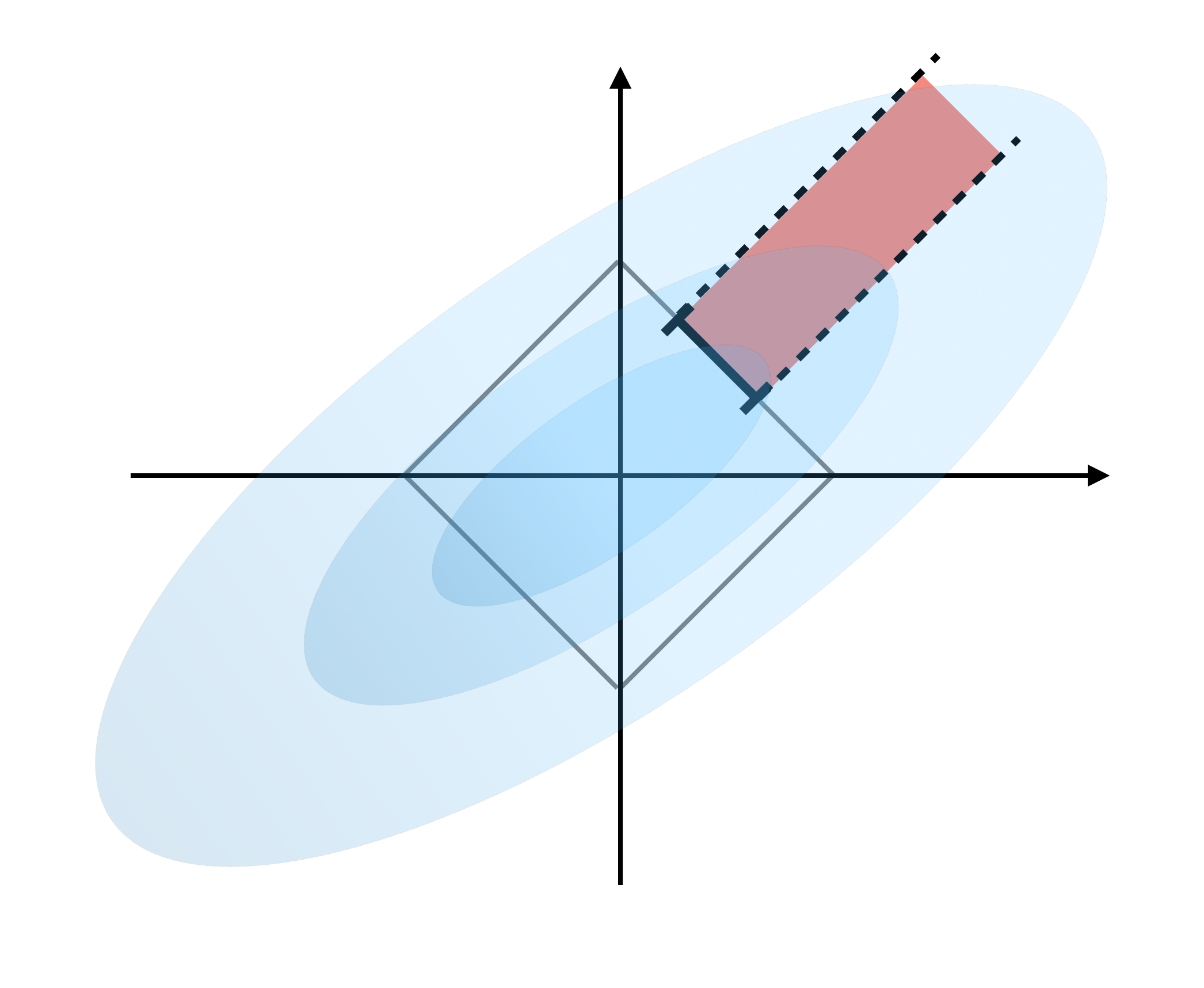}
                        \caption{Sectional view in 2-dimensions: the probability of  $\theta$ in the line segment is equal to the measure of
                                $\beta$  over the area in the red, which is positive.}
                \end{subfigure}
                \centering
                \caption{Projecting a continuous  Gaussian $\beta\in\mathbb{R}^3$ to $\theta$ onto an $l_1$-ball $\mathbb{B}_r$ (the right panel shows a density contour in 2-dimensions):
                        the boundary set of $\theta$ containing exactly zero has a  positive probability. For example, the probability of $\theta$ in the line segment $\{(x,0,r-x):0.2<x<0.8\}$ 
                        is equal to the probability of $\beta$ over the red region.
                        \label{fig:proj_illu}}
        \end{figure}

With  those two properties, we see that $\theta$ will be sparse with a prior probability greater than zero, at a given $r$.  Further, we now consider $r$ as another random variable. 

Note that the projection \eqref{eq:l1_proj} is equivalent to finding a threshold $\tilde \mu$ to make the smallest few elements  $|\beta_{(i)}| \le \tilde \mu$  zero, and having the rest shrink by $\tilde\mu$ and  $\sum_{i=1}^c [ |\beta_{(i)}| - \tilde \mu ] = r$. Therefore, using the joint prior  of $(\beta_1,\ldots,\beta_p, r)$,
we can find a lower bound probability of obtaining at least $k$ zeros after  projection:
\begin{align*}
        \text{pr}( \theta \text{ has at least $k$ zeros} ) &  \ge    \text{pr}[ \sum_{i=1}^{p-k} | \beta_{(i)} |  \ge r + (p-k) | \beta_{(p-k+1)} | ], \\
        & =     \text{pr} \big\{  r \le \sum_{i=1}^{p-k} \left [ | \beta_{(i)} | - | \beta_{(p-k+1)} | \right ] \big\}
\end{align*}
where the first line on the right corresponds to a sufficient condition for inducing exactly $k$ zeros: to have $\sum_{i=1}^{p-k} [| \beta_{(i)} | - \tilde \mu ] =r$, we need $\tilde \mu \ge | \beta_{(p-k+1)}|$, this makes all $(\beta_{(j)}- \tilde\mu)_+=0$ for $j=p-k+1,\ldots, p$. Since $| \beta_{(i)} | - | \beta_{(p-k+1)} | >0$ almost surely, with a suitable  prior $\pi_r$ for $r$, the above probability is strictly positive for any choice of $k=1,\ldots,(p-1)$.

  % For example, when $r$ is independent from $\beta$ and has a support $(0,\infty)$. In the theory section, we further show in some cases, we can compute the exact probability for $\theta$ to have $k$ zeros, when marginalizing over the distribution of $r$.

        To illustrate the geometric intuition, we show the projection in $\mathbb{R}^3$ (Figure~\ref{fig:proj_illu}) --- projecting $\beta$ from a multivariate Gaussian to $\theta\in \mathbb{B}_r$ gives us positive probability $\pr[\theta=(x_1,0,x_3)]>0$,  which equals to the Gaussian measure in the wedge region outside $\mathbb{B}_r$. Note that this is quite different from the conventional setting where $\theta$ is assigned a continuous prior in $\mathbb{R}^p$, for which fixing any $\theta_i=0$ would cause the probability to collapse to zero.

From now on, we refer to the  prior $\pi_\theta$ induced by a projection $P_{\mathbb{B}_r}$ to the vector-norm $l_1$-ball defined via $\|\theta\|_1$ as an ``$l_1$-ball prior''. For  the more general case using a projection  $P_{\mathbb{B}_{h,r}}$
to the $l_1$-ball defined via $\|h(\theta)\|_1$, we refer to it as a ``generalized $l_1$-ball prior'' and will defer its discussion to  Section \ref{sec:genl1ball}.
        
        \subsection{Closed-form Kernel for the Vector-norm $l_1$-ball Prior}
        
                   We now show that the  prior of $\theta$ induced by $P_{\mathbb{B}_r}$ has a closed-form kernel (the combination of probability mass and density functions), and we denote it by $\pi_{\theta\mid r}:\mathbb{B}_r\to [0,\infty)$. For now, we treat $r$ as given and will discuss its prior $\pi_r$ in the next subsection.

              % When $\theta\in \bdB$, the elements of $\theta$ contain both zeros and continuous random variables. Therefore, we associate $\theta$ with a prior {\em kernel} $\pi_\theta(\theta)$, , integrable to $1$ over $\mathbb{B}_r$ (we slightly abuse the notation to use $\pi_.$ to denote both the kernel and distribution).

 To introduce some notations, we let $[p]=\{1,\ldots,p\}$ be the full element indices, and $C =\{ i\in [p]: \theta_i\neq 0\}$ a subset for those non-zero elements with $c:=|C|$. And we use subscript to denote the non-zero sub-vector $\theta_C=(\theta_i)_{i\in C}$.
        
        We now divide the $l_1$-ball projection into two steps: (i) one-to-one transform of $\beta$ into a set of latent variables; (ii) integrating over those falling below zero, as equivalent to the zero-thresholding in \eqref{eq:l1_proj}.
        
   Step (i) produces the following latent variables:
        \be
        & t_i :=  |\beta_i| - \frac{\mu}{c}, \qquad s_i := \text{sign}(\beta_i)\qquad \t{ for } i\in[p],\\
        & \mu := \mu_c, \quad c=  \max\{j: |\beta_{(j)}|>\frac{\mu_j}{j}, \; \mu_j=(\sum_{i=1}^j|\beta_{(i)}|-r)_{+}\}.
        \ee
We denote the above transform by $f(\beta)=(t,s,\mu)$, which will be shown in the theory section is one-to-one, hence we can denote the inverse function as $f^{-1}$ and use the change-of-variable method to compute the probability kernel for  $ \pi_{t,s,\mu \mid r}(t,s,\mu)$.
        \begin{theorem}  [volume preserving transformation] With $(t,s,\mu)=f(\beta)$ defined as the above, for
                any proper density $\pi_\beta$, the absolute determinant of the Jacobian, denoted by $|J_f|$, is one.
              Therefore,
%                \begin{mdframed}[hidealllines=true,backgroundcolor=blue!20]
                \be
            \pi_\beta(\beta)=    \pi_{t,s,\mu \mid r}[f(\beta)]. %\qquad \pi_\theta(t,s,\mu)=\pi_{\beta}[f^{-1}(t,s,\mu)]
                \ee
%                \end{mdframed}
        \end{theorem}
        \begin{remark}
                The constant $|J_f|=1$ shows that $f$ is a 
                volume-preserving transform, hence the induced kernel is invariant to the number of non-zeros $c$. 
                This is especially useful for the posterior computation, as the posterior kernel is continuous even when the number of zeros changes from $c$ to $c'$.
        \end{remark}
        %  Since $(t,s,\mu)$ are the augmented form for $\theta$, we denote the
        % associated prior kernel as $\pi_{t,s,\mu}(t,s,\mu)$ and refer it as the augmented prior. Note that .
        
  Step (ii) produces a sparse $\theta$ via the signed  zero-thresholding $\theta_i = s_i(t_i)_{+}$. Equivalently, we can view $\pi_\theta(\theta)$ as the marginal form for $\pi_{t,s,\mu}(t,s,\mu)$, summed over those $(s_i, t_i): t_i<0$ and $\mu>0$:
        \be
        \pi_{\theta \mid r}(\theta)=&\pi_{\theta \mid r}( \theta_i = s_i t_i \text{ for } i  \in C;
        \theta_i = 0 \text{ for } i \not \in C
        ) \\
        &= \sum_{s_i=\pm 1 \text{ for } i\not \in C}\int_{0}^{\infty} \int_{(-\mu/c,0)^{p-c}}  \pi_{t,s,\mu \mid r}(t,s,\mu) \textup{d} t_{[p]\setminus C}
        \textup{d} \mu.
        \ee
        
        To show the details of the above results, we use a working example: consider an independent double exponential  $\beta_i\sim \text{DE}(0,\lambda_i)$,
        $
        \pi(\beta_{i}) = {1}/{(2\lambda_{i})} \exp( -{|\beta_i|}/{\lambda_{i}})$, with
        $\lambda_{i}>0$.
%        We choose this form for its ease of integration in the theoretic analysis. One can freely choose other continuous   $\pi_\beta(\beta)$, such as multivariate Gaussian. 
                Transforming $\beta$ to $(t,s,\mu)$, we obtain
                the prior kernel:
        \bel
        \label{eq:workingmodel}
        \pi_{t,s,\mu  \mid r}(t,s,\mu) =  \prod_{i=1}^p\frac{1}{2\lambda_i} \exp( -\frac{\mu/c}{\lambda_{i}})\exp(
        -\frac{t_i}{\lambda_{i}}),
        \eel
        subject to constraints $ t_i> -\mu/c$ and $\sum_{i=1}^p (t_i)_+=r$. In this special case, we can take a step further and integrate out $t_i: i \in [p]\setminus C$:
        \bel\label{eq:adaptive_ss}
        & \pr(\theta_i=0)=  1- \exp( -\frac{\mu/c}{\lambda_{i}}), \quad \pr(|\theta_i|>0)=  \exp( -\frac{\mu/c}{\lambda_{i}}),\\
        &\pi_\theta(\theta_i \mid |\theta_i|>0)=\frac{1}{2\lambda_{i}}
        \exp( -\frac{|\theta_i|}{\lambda_{i}}).
        \eel
                    %   \begin{remark}
                %However, a major difference in using the individual $\lambda_i$ and a common threshold is that --- now the ``spike'' probability  changes with  the scale of the ``slab'': as $\lambda_{i} \to 0$, $\pr(\theta_i=0)\to 1$. Therefore,  this gives adaptiveness that increases the chance of shrinking small noise to zero, while decreasing the chance of removing large signals. We will show a clear difference in the data experiments.
      %  \end{remark}

                \begin{remark}
        To clarify, in this article, we choose to present  the double exponential $\pi_\beta$ for the ease of integration, which is useful for a tractable theoretic analysis later. In practice,        
            we can choose any continuous $\pi_\beta$, such as the multivariate Gaussian. The volume preserving property and convenient computation will hold in general.
              \end{remark}

        In general, the above marginal kernel may not be available in closed-form
for other choice of $\pi_\beta$ or more general $\mathbb{B}_{h,r}$, however,
                 we can use the data augmentation  \citep{tanner1987calculation}
for the posterior estimation. To elaborate,  let $\mathcal L(y;\theta, \eta)$ be the likelihood,  $y$ the data, $\eta$ some other parameter, we can sample the posterior $\beta, r$ and $\eta$ via:
%         \be
%         \pi(t,s,\mu,\eta \mid y)& \propto  \pi_\eta[\eta\mid
% f^{-1}(t,s,\mu)] \pi_{t,s,\mu}(t,s,\mu) \mathcal L[y;\theta, \eta: 
%         \theta_i= s_i(t_i)_+ \t{ for } i\in [p] ].
% %        & = \pi_\eta(\eta) \pi_\beta(\beta) L[y; P_{\mathbb{B}_r}(\beta), \eta]
%         \ee
%         Equivalently, changing the variable $(t,s,\mu)$ into $\beta$, we have a very simple posterior of $\beta$:
                \bel\label{eq:posterior}
        \pi(\beta,\eta,r \mid y) \propto   \pi_{\eta\mid\beta}(\eta) \pi_{r}(r) \pi_\beta(\beta) \mathcal L[y; P_{\mathbb{B}_{h,r}}(\beta), \eta].
        \eel
   This means, we can first obtain the posterior samples of $(\eta,r,\beta)$, compute $\theta=  P_{\mathbb{B}_{h,r}}(\beta)$ for each sample of $\beta$, then
  discard the other information.

        %  with augmented $\mu$ and $(t_i,s_i)$ for $i\not\in C$, the joint kernel can be evaluated directly using the density of $\beta$ before the $l_1$-ball projection. For example, when assigning multivariate Gaussian to $\pi_{t,s,\mu}(t,s,\mu)$, the augmented prior kernel is equal to,
        % \be
        % \pi_0[t(\beta),s(\beta),\mu(\beta)] = (2\pi)^{-p/2}|\Sigma|^{-1/2} \exp \bigg[ - \frac{1}{2}\beta^{\rm T} \Sigma^{-1} \beta \bigg],
        % \ee
        % where we put $(\beta)$ as $t,s,\mu$ are the result of augmented $l_1$-ball transform of $\beta$; similarly, as $\theta_i= s_i \max(t_i,0)$, we will use $\theta(\beta)$ notation as well. 
        % In the theory section, we will carefully quantify the behavior of $\pi_\theta(\theta)$ by marginalizing over $s_i,t_i$ and $\mu$.

        \subsection{Generalized $l_1$-ball Prior}
        \label{sec:genl1ball}
        
        We now discuss the general cases that use $l_1$-ball projection to create priors such that some transform of $\theta$ is sparse. Specifically, let $h:\mathbb{R}^p\to \mathbb{R}^d$, we use the following projection:
            \bel\label{eq:gen_l1_ball}
                \theta=P_{\mathbb{B}_{h,r}}(\beta)=\argmin_{z\in \mathcal{Z}:\|h(z)\|_1\le r}\|z-\beta\|_2^2.
                \eel
                For regularities, we require $\|h(z)\|_1$ to be
convex and $\mathcal{Z}$ to be a convex set in $\mathbb{R}^p$. When these conditions are satisfied, the level set $\{z:\|h(z)\|_1\le r\}$ is a convex set, making the Euclidean projection unique hence a measurable transform.  
 This includes a large class of useful functions, such as $h(z)=Dz$ with $D\in \mathbb R^{d\times p}$ as in the sparse contrast models, and $\|h(z)\|_1= \sum_k \sqrt{\sum_j z^2_{k(j)}}$ as in the grouped shrinkage, $\t{tr}[(ZZ^{\rm T})^{1/2}]$ as the nuclear norm to control the number of non-zero eigenvalues for square matrix $Z$. Further, we can consider $\mathcal Z$ as a low-dimensional constrained space such as the one for positive definite matrix, for example, for modeling a sparse covariance/precision matrix.

         % more than variable selection and shrinkage, but can be used in many other $l_1$-tricks. We can extend this prior to more general cases, for example the generalized $l_1$-regulation regression, which is firstly proposed by \cite{tibshirani2011solution}. Instead of pure sparsity, we expect certain structure in the parameter $\beta\in\mathbb R^p$. Depending on the scenario, we choose a penalty matrix $D\in \mathbb R^{m\times p}$ such that sparsity in $D\beta$ corresponds to some other desired behavior. For example, in standard sparse regression problems, $D$ is chosen as the identity matrix $I_p$; in the sparse fused lasso model, $D$ is . When $D$ is an invertible matrix, or with full row rank with $m\le p$, then we could reparameterize $\beta$ with $\theta = D\beta$, and solve a typical sparse regression problem by enforcing $\theta$ to be sparse. If exact zeros are essential, then any spike-and-slab methods could be applied.

The projection may not have a closed-form solution, however, it can be efficiently calculated using the splitting technique:
            \be
                \theta=\argmin_{z:h(z)=s,\|s\|_1\le r}\|z-\beta\|_2^2 +\eta^{\rm T}[h(z)-s]+ \frac{1}{2\rho}\|h(z)-s\|^2_2,
                \ee
where $\rho>0$ and $\eta\in\mathbb{R}^d$ is Lagrangian multiplier (the values of $\rho$ and $\eta$ do not impact the convergence). Using  $\kappa=\rho\eta$, the optimal solution can be computed using the alternating direction method of multipliers (ADMM) algorithm \citep{boyd2011distributed}, that iterates in:
\bel\label{eq:admm}
z & \leftarrow \argmin_{z\in\mathcal{Z}}[ \|z-\beta\|_2^2 + \frac{1}{2\rho}\|h(z)-s+\kappa\|^2_2], \\
s & \leftarrow  P_{\mathbb{B}_r}[ h(z)+\kappa], \\
\kappa  & \leftarrow  \kappa+ h(z)-s,
\eel
until it converges, and then set $\theta$ to be equal to $z$. Note that this algorithm contains a projection step to the vector-norm
$l_1$-ball, as in \eqref{eq:l1_proj}.

       It is important to point out that, even though $P_{\mathbb{B}_{h,r}}(\beta)$ may not have a closed-form, $P_{\mathbb{B}_{h,r}}(\beta)$ is  a continuous function of $\beta$, and differentiable almost surely with respect to the distribution of $\beta$, as explained in the next section.
         
\subsection{Prior Specification on the Radius} 

We now discuss how to choose $\pi_r$ for the radius $r$. To induce a principled choice in simple linear models and to allow a straightforward prior calibration in complex models, we propose to use an exponential prior:
$$\pi_r(r) = \frac{1}{\alpha}e^{-r/\alpha},$$
with $\alpha>0$ a calibration parameter chosen based on either some theory-guided conditions, or some prior assumption on the dimensionality of $\theta$.

As we will show in the theory section, for the vector-norm $l_1$-ball prior, with an exponential $\pi_r$  and 
 $\beta_i\sim \text{DE}(0,\lambda)$, we can obtain a closed-form expression on the cardinality of $\theta$:
$\text{pr}(|C|=j\mid \lambda, \alpha) = {\lambda/\alpha}{(1+\lambda/\alpha)^{-j}}$
for $j=1,\ldots,(p-1)$. The tractable form enables us to choose $\alpha$ via the asymptotic theory of signal recovery in linear models.

\begin{figure}[H] 
                \begin{subfigure}{.45\textwidth}
                        \includegraphics[width=.8\linewidth]{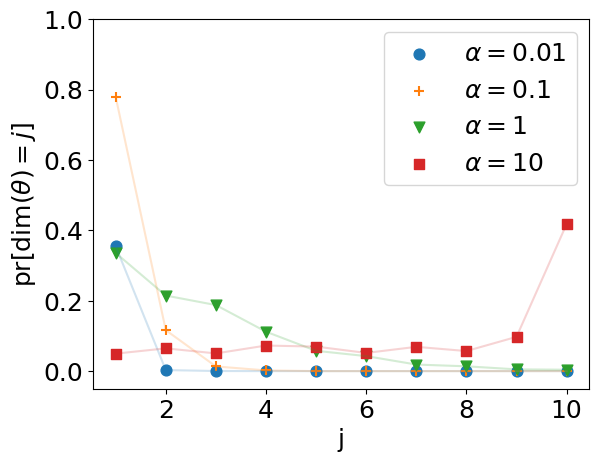}
                        \caption{Plotting $\text{pr}[\text{dim}(\theta)=j\mid \alpha]$ against $j$, with $\theta$ a $10$-dimensional vector and $\alpha$ ranging from $\{0.01,0.1,1,10\}$.}
                \end{subfigure}\quad
               \begin{subfigure}{.45\textwidth}
                        \includegraphics[width=.8\linewidth]{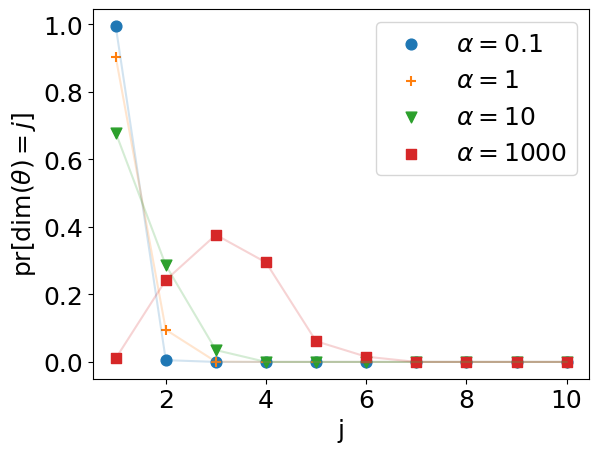}
                        \caption{Plotting $\text{pr}[\text{dim}(\theta)=j\mid \alpha]$ against $j$, with $\theta$ a $10\times10$ matrix and $\alpha$ ranging from $\{0.1,1,10,1000\}$.}
                \end{subfigure}                \centering
                \caption{Prior distribution of dim$(\theta)$ with varying $\alpha$. Panel (a) shows the cardinality of a vector with vector-norm $l_1$-ball prior, and panel (b) shows the rank of a symmetric matrix with nuclear-norm $l_1$-ball prior. \label{fig:pi_radius}}
        \end{figure}

For general cases involving other forms of $\pi_\beta$ or $l_1$-balls defined via $\|h(\theta)\|_1$, one can easily use numerical simulations to plot the induced prior distribution for the dimension of $\theta$, varying according to the value of $\alpha$. This allows one to calibrate $\alpha$ according to their prior belief.
 To illustrate this approach, in Figure \ref{fig:pi_radius} we show the prior distribution of the dimension of $\theta$, for the vector-norm $l_1$-ball prior based on normal $\beta_i \stackrel{\text{iid}}\sim N(0,1)$ for a $10$-dimensional vector $\beta$; and the one of the dimension of matrix $\theta$ (rank) for the nuclear-norm $l_1$-ball prior based on normal $\beta_{i,j} \stackrel{\text{iid}}\sim N(0,1)$, $i\le j$ for a $10\times 10$ symmetric matrix $\beta$.

%\begin{remark} Beyond the exponential, we can choose other distributions supported on $(0,\infty)$ to be the radius prior. For example, choose a half-Cauchy for a heavier tail in $r$, or even make $\pi_r$ dependent on $\beta$, for more direct control of sparsity. Practically, using an exponential prior with a large calibration scale $\sigma_r$ suffices in most cases. We postpone the discussion about other choices of radius priors in the supplementary materials section B.
%\end{remark}

\section{Continuous Hamiltonian Monte Carlo for Posterior Computation}
The Hamiltonian Monte Carlo is a powerful method for sampling the posterior distribution. It uses the Hamiltonian dynamics to propose a new state of parameter and accept it using the Metropolis-Hastings criterion. Due to the energy-preserving property, the Hamiltonian dynamics is capable of producing a proposal that is far away from the current state, while often enjoying a high acceptance rate.  \cite{neal2011mcmc} provides an introduction on this algorithm.
        
One limitation, however, is that the common algorithm in the Hamiltonian Monte Carlo (such as the one implemented in Stan) only works on the continuous random variable under a continuous posterior density. Although some data augmentation methods are proposed for the binary discrete random variable, such as \cite{pakman2013auxiliary}, they create discontinuity in the augmented posterior. To briefly explain, when a binary variable changes from $0$ to $1$, the likelihood will have a sudden jump that breaks the energy-preserving property of Hamiltonian dynamics. To handle this issue, one needs a modified algorithm such as the discontinuous Hamiltonian Monte Carlo \citep{nishimura2020discontinuous}.

Interestingly, the $l_1$-ball priors (including the generalized $l_1$-ball priors) are free from this issue, as long as $\pi_\beta$ is continuous in $\beta$ and the likelihood is continuous in $\theta$. Intuitively, for the $l_1$-ball prior, as $\beta_i$ reduces in magnitude, $\theta_i$ gradually changes toward $0$ and stays at $0$ after passing below the threshold --- therefore, the projection function is continuous. We now formalize this intuition using the property of the proximal mapping, then briefly explain the Hamiltonian Monte Carlo algorithm.

\subsection{Almost Sure Smoothness of the Posterior}
We first introduce the concept of the proximal mapping (for a more
comprehensive introduction, see \cite{beck2017first}), defined as
\be
\t{prox}_g(x) = \text{argmin}_{z}\left\{g(z)+\|x-z\|_2^2/2\right\},
\ee
with $g$ a  lower-semicontinuous and convex function.  Now we choose $g= \mathcal X_{\mathbb{B}_{h,r}}$, the characteristic function of a set (an $l_1$-ball), which takes value 0 in $\mathbb{B}_{h,r}$ and takes $\infty $ otherweise. Since $\mathbb B_{h,r}$ is a convex set, this means that the projection $P_{\mathbb B_{h,r}}(\beta)$ is a proximal mapping.

As a useful property, the proximal mapping is Lipschitz continuous with the Lipschitz constant $1$ (\cite{beck2017first}, Theorem\ 6.42); in our case,
\be
\|P_{\mathbb{B}_{h,r}}(\beta_1)-P_{\mathbb{B}_{h,r}}(\beta_2)\|_2\le \|\beta_1-\beta_2\|_2.
\ee
Further, by the Rademacher's Theorem (\cite{federer2014geometric}, Thm.\ 3.1.6), any Lipschitz continuous function is differentiable almost everywhere with respect to the measure on its input  --- in our case, the chosen $\pi_\beta$ before the projection.

As a result, when the likelihood function is continuous in $\theta$, it is continuous in $\beta$ using variable transformation $\theta=P_{\mathbb{B}_{h,r}}(\beta)$, hence the posterior is a continuous function of $\beta$ as well. Further, if the likelihood is smooth with respect to $\theta$ almost surely, then the posterior is smooth almost surely as well. Therefore we can simply run the continuous Hamiltonian Monte Carlo (HMC). We provide a brief review of the HMC algorithm and describe further details of implementing this algorithm in the supplementary materials.

\subsection{Point Estimate and Credible Region for the Low Dimensional Parameter}
Uncertainty quantification often relies on the calculation of the credible region: for a certain function of the random variable $g(\theta)$ (such as $\theta$ itself, fitted value $X\theta$, etc.), we want a region $R_{g}$, such that
\be
 \t{pr} [g(\theta) \in R_{g} \mid y]=1-\tilde \alpha,
\ee
with some given $\tilde \alpha\in (0,1)$. 

A common way to approximate $R_{g}$ is to take the posterior samples of $g(\theta)$, and take point-wise quantiles in the elements of the $g(\theta)$ output, while adjusting for multiplicity. However, this is sub-optimal for a sparse $\theta$ and/or an intrinsically low-dimensional $g(\theta)$ (such as the piece-wise linear $X\theta$), for two reasons: (i) the multiplicity adjustment is often too conservative, making the credible region too large (that is, the associated probability is in fact much larger than $1-\tilde \alpha$); (ii) the combination of the point-wise credible intervals is often no longer low-dimensional.

To bypass these issues, we use the solution from \cite{breth1978bayesian} based on the \emph{top $(1-\tilde \alpha)$ posterior density region}:
\be
R_g = \{ g(\theta): \pi_{\theta} (\theta \mid y) \ge \kappa_{\tilde \alpha}\},
\ee 
where  $\kappa_{\tilde \alpha}$ is a threshold that makes the region having a probability $1-\tilde \alpha$. In practice, we can approximate $\kappa_{\tilde \alpha}$ by simply calculating the posterior kernels for  all the samples, then taking the $\tilde \alpha$ quantile. 

Similarly, for a point estimate, since $\theta$ may reside on a low-dimensional space $\mathcal M$, the sample mean of $\theta$ or $g(\theta)$ is not ideal as it may end up being high-dimensional. Therefore, we use the  
Fr\'echet mean:
\be
\overline{g(\theta)} = \underset{ g(z):z\in \mathcal M}{\arg\min}\mathbb{E}_{\theta\sim \pi(\theta\mid y)}\| g(z) - g(\theta)\|^2.
\ee
As we often do not know $\mathcal M$, we can approximate the above using the posterior samples $\{\theta^j\}_{j=1\ldots m}$, which give the estimator $\arg\min_{g(z):z\in  \{\theta^j\}}\sum_{\theta\in\{\theta^j\}}\| g(z) - g(\theta)\|^2$. We will illustrate these in numerical examples.

        \section{Theoretical Study on $l_1$-ball Prior}
        \label{sec:theory}
        We now focus on a more theoretical study on the $l_1$-ball prior. For ease of analysis, we focus on the vector-norm $l_1$-ball in this section.

First, we show that the augmented transform $f$ in the $l_1$-projection is indeed invertible. Recall that $f:\mathbb{R}^p\to \{ (t,s,\mu)\in\mathbb{R}^p \times
        \{1,-1\}^ p \times \mathbb{R}_+: \sum_{i=1}^p(t_i)_+=r,\;t_i\ge -\mu/|C| \}$.
        \begin{theorem}
                Consider another transform $\beta=g(t,s,\mu)$ with  $C=\{i:t_i>0\}$, and let
                \bel\label{eq:back_proj}
                \beta_i = s_i( t_i + \frac{\mu}{|C|}),
                \eel
                where  its domain satisfies the following: $s_i\in \{-1,1\};$  if $\sum_{i\in C} t_i =r$, then for $i\not\in
                C$, $-\mu/|C|\le t_i \le 0$ and $\mu \ge 0$;
                if all $t_i \ge 0$ and $\sum t_i <r$, then  $\mu = 0$. If $|C| = 0$,  all
                $\beta_i=0$. Then $g$ is the inverse mapping of $f$, that is:
                \be
                f[g(t,s,\mu)]=(t,s,\mu), \quad g[f(\beta)]=\beta.
                \ee
        \end{theorem}
        \begin{remark}
               Broadly speaking, this gives an ``projection-based data augmentation'' scheme for any sparse $\theta$: we can augment  $(t_i, s_i)$'s for  those $\theta_i=0$ and a $\mu \ge 0$, and  then apply $\beta=g(t,s,\mu)$. The produced $\beta$ is a continuous embedding for $\theta$.
        \end{remark}
        
        Next, we establish a link between some special form of the $l_1$-ball prior to the combinatorial prior that chooses a subset of $\theta$ to be zero.
When  $\beta_{i}\stackrel{\text{iid}}\sim\text{DE}(0,\lambda)$, we can further integrate and obtain two simple marginal forms.
       \begin{theorem}\label{thm: marginal}
                If $\pi_\beta(\beta) = \prod_i (2\lambda)^{-1}\exp(-|\beta_i|/\lambda)$
                with $\lambda>0$, then for $\theta\in \intB$,
                $\pi_{\theta \mid r, \lambda}(\theta) = \prod_i (2\lambda)^{-1}\exp(-|\theta_i|/\lambda)\mathbb{I}(\|\theta\|_1 <r) $, 
                and for $\theta\in \bdB$,
                \bel\label{eq:de}
                \pi_{\theta \mid r, \lambda}(\theta) =   \frac{  (2\lambda)^{-|C|} }{\left(\begin{array}{c}
                                p\\|C|
                        \end{array}\right)}\lambda \exp \left (-\frac{r}{\lambda}\right)\mathbb{I}(\|\theta\|_1
                =r),
                \eel
                where $C=\{i \in [p]:\theta_i\neq 0\}$.
        \end{theorem}
        Further marginalizing over $\theta$ on
        $\{\theta:\|\theta\|_1=r, \sum_{i=1}^p \mathbb{I}(\theta_i\neq 0)=j\}$,  we can obtain a discrete prior distribution on $|C|$.
        \begin{corollary}\label{cor: model}
                If $\pi_\beta(\beta) = \prod_i (2\lambda)^{-1}\exp(-|\beta_i|/\lambda)$
                with $\lambda>0$, then the marginal prior $\pi(|C|;r)$ follows a truncated Poisson distribution, with
                \bel\label{eq:model_selection}
                \pr(|C|=j  \mid r) =\frac{(r /\lambda)^{j-1}}{(j-1)!} \exp \left (-\frac{r}{\lambda}\right),
                \eel
                for $j=1,\ldots,(p-1)$; and $\pr(|C|=p \mid\ r)= 1- \sum_{j=1}^{p-1}{(r /\lambda)^{j-1}}/{(j-1)!} \exp  (-{r}/{\lambda}).$
        \end{corollary}
        In the above, we can see how the radius impacts the level of sparsity:
\be
\mathbb{E}(|C|-1) &=\sum_{j=1}^{p-1}(j-1)\frac{(r /\lambda)^{j-1}}{(j-1)!} \exp \left (-\frac{r}{\lambda}\right)+(p-1) \sum_{j=p}^{\infty}\frac{(r /\lambda)^{j-1}}{(j-1)!} \exp \left (-\frac{r}{\lambda}\right)\\
& \le r/\lambda,
\ee
where the inequality is due to $(p-1)\le (j-1)$ for $j\ge p$ and the expectation of an untruncated Possion $(r/\lambda)$ is $r/\lambda$. Therefore, 
 a smaller $r$ favors a smaller $|C|$ and more $\theta_i$'s to be zero.
        
Lastly, we study the posterior convergence rate using the above prior. Since the rate is highly dependent on the form of the likelihood, we choose to narrow our focus on the well-studied linear regression model, and demonstrate an equivalently optimal rate as the existing approaches. For a comprehensive review on this topic, see \cite{castillo2012needles}.

         We   follow the standard theoretic analysis and assume $y_i$ and $\theta_i$ are re-scaled by $1/\sigma$, so that 
        $y_i\sim \text{N}(x_i^{\rm T}\theta, 1)$, while assuming there is an oracle $\theta^0\in\mathbb{R}^p$,
        with the true cardinality $c_0\neq 0$. 
        In practice, since we do not know $\sigma^2$ we can assign a prior on $\sigma^2\sim \text{Inverse-Gamma}(\gamma_{\sigma^2,1},\gamma_{\sigma^2,2})$, and additionally let
        $\theta$ scale with $\sigma$.       
        To provide a straightforward result, we use $\pi_{r}(r) =\alpha^{-1} \exp(- r/ \alpha)$ with $\alpha$ a parameter to determine. Multiplying it to \eqref{eq:model_selection} and integrating over $r$, 
        and we obtain the marginal model selection probability under the  $\pi_{r}(r)$:
        \bel \label{eq:c_marginal_cardinalty}
        \pr(|C|=j ; \lambda, \alpha) =\frac{\lambda /\alpha}{ (1+\lambda/\alpha)^{j}},
        \eel
        for $j=1,\ldots,(p-1)$; and $\pr(|C|=p; \lambda ,\alpha)= 1- \sum_{j=1}^{p-1} {\lambda/ \alpha}{ (1+\lambda/\alpha)^{-j}}= (1+\lambda/\alpha)^{-(p-1)}.$ And
        \bel\label{eq:c_marginal_density}
        \pi_{\theta}(\theta) =   \frac{  (2\lambda)^{-|C|} }{\left(\begin{array}{c}
                        p\\|C|
                \end{array}\right)}\lambda/\alpha \exp \left [-  (1/\lambda+1/\alpha)^{-1}{\|\theta\|_1} \right],
        \eel
        for $|C|=1,\ldots,(p-1)$.
        \begin{remark}
                The equations \eqref{eq:c_marginal_cardinalty} and \eqref{eq:c_marginal_density} show that, even after marginalizing over the radius $r$, the prior still has a positive probability for $\theta$ to contain $k=p-|C|$ zeros.
        \end{remark}

         We now present the convergence result.
        
 \begin{theorem}
        
        \label{thm: main}
        If the data are generated from $y_i=X_{i,\cdot}\theta^0+\epsilon_i, \epsilon_i \stackrel{\text{iid}}{\sim} \text{N}(0, 1)$, with
        $(\lambda,\alpha)$  chosen as $\lambda = b_{1}p^{b_{2}} /\|X\|_{2,\infty}  , \alpha = p^{b_3}/\|X\|_{2,\infty},   b_1>0, b_{2}>b_{3}$,  $b_{3}\le 1,$ and  $\|X\|_{2,\infty}:=\max_j \sqrt{\sum_i X_{i,j}^2}$, with sufficiently large $M$,
        then as $n,p\to \infty$:
        \begin{itemize}
                \item  (Cardinality) For estimating the true cardinality $c_0$, 
                \be
                \sup _{\theta^{0}} \mathbb{E}_{\theta^{0}} \pi\left(\theta:\left|C_{\theta}\right|>c_{0}\left[1+\frac{M}{b_{2}-b_{3}}\left(1+\frac{16}{\phi\left(C_{0}\right)^{2}} \frac{\lambda^*}{2 \|X\|_{2,\infty}\sqrt{\log p}}\right) \right ]  \bigg | y\right) \rightarrow 0,
                \ee
                where $\lambda^*=\|X\|_{2,\infty} ({b_{1}p^{b_{2}}
                        +p^{b_{3}}})/({b_{1}p^{b_{2}} p^{b_{3}}})$.
                
                \item  ($l_2$-recovery) The recovery of true $\theta^0$ has
                \be
                % \begin{array}{c}
                        & \sup _{\theta^{0}} \mathbb{E}_{\theta^{0}} \pi\left(\theta:\left\|\theta-\theta^{0}\right\|_{2}>\frac{M}{{\psi}\left(C_{0}\right)
                                ^{2}} \frac{\sqrt{c_{0} \log p}}{\|X\|_{2,\infty} \phi\left(C_{0}\right)} \bigg| y\right) \rightarrow 0, 
                        % \end{array}
                \ee
                % for prediction, the error has
                % \be
                % & \sup _{\theta^{0}} \mathbb{E}_{\theta^{0}} \Pi\left(\theta:\left\|X\left(\theta-\theta^{0}\right)\right\|_{2}>\frac{M}{m_{2}}
                % \frac{\sqrt{c_{0} \log p}}{\phi\left(C_{0}\right)} \bigg| Y\right) \rightarrow
                % 0,
                % \ee
                % where $m_1$ and $m_2$ are some constants in $(0,1]$ (details provided in the appendix).
                \item  ($l_1$-recovery) The recovery of true $\theta^0$ has 
                \be
                \sup _{\theta^{0}} \mathbb{E}_{\theta^{0}} \pi\left(\theta:\left\|\theta-\theta^{0}\right\|_{1}>\frac{M}{\bar{\psi}\left(C_{0}\right)^{2}} \frac{c_0 \sqrt{\log p}}{\|X\|_{2,\infty} \phi\left(C_{0}\right)^{2}} \bigg| y\right) \rightarrow 0.
                \ee 
                \item ($l_\infty$-recovery)  
                For every $\eta>0,$ any $d_{0}<\eta^{2}\left[ 1+2 /(b_{2}-b_{3})\right]^{-1} / 8,$ and $c_{n}$ such that \\
                $ c_{n}({b_{1}p^{b_{2}}
                        +p^{b_{3}}}) \sqrt{\log p} /({b_{1}p^{b_{2}} p^{b_{3}}}) \rightarrow 0$, for the set $\mathcal{C}^*=\{C_0:\phi\left(C_{0}\right) \geq\eta, {\psi}\left(C_{0}\right)
                \geq\eta, c_0\le c_n, c_0\le d_0 \t{mc}(X)^{-1}\},$ then the recovery of true $\theta^0$ has
                $$
                \sup _{\theta^{0} : C_0\in \mathcal{C}^* } \mathbb{E}_{\theta^{0}} \pi\left(\theta:\left\|\theta-\theta^{0}\right\|_{\infty}>M \frac{\sqrt{\log p}}{\|X\|_{2,\infty}} \bigg| y\right) \rightarrow 0.
                $$
        \end{itemize}
        In the above, $\t{mc}(X) = \max_{i\neq j} \frac{|X_{.,i}^{\rm T}X_{.,j}|}{\|X_{.,i}\|_2\|X_{.,j}\|_2}$ is the mutual coherence, and $\phi(C), \bar\psi(C), \psi(C)$ are the compatibility numbers for matrix $X$ that we give the definitions in the supplementary materials.  
    \end{theorem}

                Taking one step further, we now characterize the uncertainty via examining
                the
                 asymptotic posterior distribution in linear regression.                 For a given model $C\subset[p]$, we let $X_C$ be the $n\times |C|$ subset matrix consisting of the columns $X_{\cdot, i}$ with $i\in C$, and $\hat\theta_{C}$ be the least square estimator in the restricted model
                $\hat\theta_{C}\in\argmin_{\theta_{C}\in\mathbb R^{|C|}}\|Y-X_C\theta_C\|_2$. We obtain the following Bernstein von-Mises theorem, which shows the posterior converging to a Gaussian distribution concentrated on the true model when $n\to \infty$.

                \begin{theorem}%(Bernstein von-Mises)
                        \label{thm: bvm}
                        Let $(\lambda,\alpha)$ be chosen as $\lambda = b_1p^{b_2}/\|X\|_{2,\infty},\alpha= p^{b_3}/\|X\|_{2,\infty}$ with $b_1>0,b_2^{-1} = o_p(1),b_3\le 1$, and we assume $\theta^0$ is in the parameter space $\Theta^0 = \big\{ \theta^0: 
                        |\theta^0_i|\ge \frac{M}{\bar\psi\left(C_{0}\right)^{2}}\frac{c_{0} \sqrt{\log p}}{\|X\|_{2, \infty} \phi\left(C_{0}\right)^2} \forall i\in C_0, \t{with sufficiently large } M;
                       c_0\sqrt{\log p}/\|X\|_{2,\infty}\to 0; \phi(C_0)\ge a_0;  \\ \bar\psi(C_0)\ge a_0;{\tilde \sigma_{\min}(X^\T_{C_0}X_{C_0})}/{\|X\|_{2,\infty}}\ge a_0
                        \big\}$ with $\tilde \sigma_{\min}(\cdot)$ denotes the smallest eigenvalue of a matrix. Then for any $a_0>0$, as $n,p\to\infty:$
                        
                        \[\sup_{\theta^0 \in \Theta^0} \|\pi(\theta \large\mid Y) - N(\hat\theta_{C_0}, (X_{C_0}^\T X_{C_0})^{-1})\otimes\delta_{[p] \setminus C_0}\|_{TV} \to 0,\]
                        where $\delta_{[p] \setminus C_0}$ denotes the Dirac measure at a zero vector for those $\theta_i: i \not \in C_0$. 
                \end{theorem} 
    \begin{remark}
        To justify the assumptions about $ \Theta^0$, the first condition is that the absolute value of each non-zero entry in $\theta^0$ is larger than a threshold; the second condition and the positive lower bound on $\phi$ and $\bar\psi$ make this threshold go to zero when $p\to \infty$; the last condition on the first eigenvalue ensures the positive definiteness of $X_{C_0}^\T X_{C_0}$.
    \end{remark}

\section{Comparison with Some Existing Methods}
\subsection{Comparison with the Spike-and-Slab Priors}
The spike-and-slab priors are well known for solving Bayesian variable selection problems, and they are also capable of inducing exact zeros in $\theta$ with a positive probability. Therefore, we provide a detailed comparison between the $l_1$-ball and the spike-and-slab priors.

As there are multiple variants under the name of spike-and-slab, we focus on the ones in the following form  \citep{lempers1971posterior,mitchell1988bayesian}:
\bel\label{eq:spike_and_slab}
& (\theta_i \mid  \tau_i) \stackrel{\text{indep}}\sim   (1 - w) \delta_0(.) +  w \delta_{\tau_i}(.), \\
& \tau_i \stackrel{\text{iid}}\sim \pi_\tau ,
\eel
for $i=1,\ldots,p$,
where $\delta_x(.)$ denotes a point mass at point $x$, and $\pi_\tau$ denotes a continuous distribution centered at zero such as Gaussian $N(0,\sigma^2_\tau)$, $w \in (0,1)$ is a probability, typically assigned with a beta prior. Therefore, this prior is a two-component mixture of point mass at $0$ (``spike'') and a continuous distribution $\pi_\tau$ (``slab'').  The later versions \citep{george1995stochastic,ishwaran2005spike} improve the computational performance by replacing $0$ with another continuous distribution concentrated near zero; however, they lose the positive probability at zero. Therefore, for a direct comparison with the $l_1$-ball prior, we will focus on the Lempers-Mitchell-Beauchamp version here.

On the one hand, we show that the classic spike-and-slab prior can be in fact viewed as a special case of the $l_1$-ball prior, under three restrictions: (i) isotropic $\Pi_\beta$,  (ii) projection to a vector $l_1$-norm ball, (iii) quantile-based threshold.  We now construct an $l_1$-ball prior that has the same marginal form as \eqref{eq:spike_and_slab}. Consider a distribution for $\beta$ isotropic in each coordinate/element and dependent on $w$:
\bel\label{eq:spike_slab_augmented}
& \beta_i \stackrel{\text{iid}} \sim \pi_{\tilde \tau}, \\
& \pi_{\tilde\tau} (x; w)  = \left\{ 
\begin{array}{l l}
(1-w){\pi_0( x) }, & \text{ if } |x|\le \tilde\mu \\
{w}\pi_\tau [ \text{sign}(x) (|x| - \tilde \mu)_+], &\text{ if } |x|>\tilde\mu 
\end{array}
\right.
,
\eel
where $\pi_0( x)$ is an augmented density that accounts for the probability of producing $\theta_i=0$, and we can use any  proper density that integrates to 1 and has support over  $[- \tilde \mu, \tilde \mu ]$,  with $\tilde\mu>0$  determined by the choice of $\pi_0(x)$.

 Using the \emph{soft-thresholding} representation $\theta_i = \text{sign}(\beta_i)(|\beta_{i}|-\tilde \mu)_+$, it is not hard to see that we have $\theta_i=0 $ with probability $1-w$, and  $\theta_i \sim \pi_{\tau}$ if $\theta_i\neq 0$, independently for $i=1,\ldots, p$. Therefore, this distribution is marginally equivalent to  \eqref{eq:spike_and_slab}.
 
 %To illustrate this via a simple example, we consider the spike-and-slab prior with the double exponential slab $\pi(\theta_i \mid \theta_i \neq 0)\propto 
 %       \exp( -{|\theta_i|}/{\lambda})$, as recently used by \cite{castillo2012needles}. We can augment a truncated double exponential $\pi_0(x) \propto   \exp( -{|x|}/{\lambda})  \mathbb{I}(|x|\le \tilde \mu)$ and $\tilde \mu =  -\log(w) \lambda$, to make a simple $\pi_{\tilde\tau} (x \mid w) \propto \exp( -{|x|}/{\lambda}) $. Therefore, this spike-and-slab prior is equivalent to projecting $\beta_i \stackrel{\text{iid}}{\sim} \text{DE}(\lambda)$ to an $l_1$-ball, with the radius dependent on $w$ and $\beta$.
       
     Based on this connection, we can use continuous Hamiltonian Monte Carlo
       for the posterior computation under a spike-and-slab prior. In the supplementary materials, we show that empirically, this leads to much faster mixing performance in the Markov chains, compared to the conventional combinatorial search based on the update of binary inclusion variables.

On the other hand, by relaxing those restrictions, the $l_1$-ball prior and the generalized $l_1$-ball prior can induce a much more flexible model than the spike-and-slab  --- in particular, those zero elements $\theta_i=0$ (the ``spikes'') no longer need to be independent, but can satisfy some complicated dependence and/or combinatorial constraints as we motivated in the beginning.

First, we can easily induce dependence among the zeros in $\theta$ by replacing the isotropic $\pi_\beta$  with a correlated one. For example, using $\beta\sim {N}(0, \Sigma_{\beta})$, we can see that, after the projection,
\be
\text{pr}( \theta_i = \theta_{i'}=0) = \text{pr}( |\beta_i| \le \tilde \mu, |\beta_i' |\le  \tilde \mu),
\ee
which is the probability for finding a correlated bivariate Gaussian random variable inside a box $[-\tilde \mu,\tilde \mu]^2$ --- if their correlation is close to $1$, then when $\theta_{i'}$ is zero, $\theta_i$ is very likely to be zero as well. This could be very useful if one wants to impose prior assumption on where the zeros could simultaneously appear. 
%For example, in the linear regression one could use the design matrix $X$ to form an empirical prior for $\beta \sim \text{N}[ 0, g (X^{\rm T}X + I \alpha)^{-1}]$ with $g>0, \alpha>0$ some hyper-parameters, as mimicking the one in \cite{zellner1986assessing} (that uses $\theta \sim \text{N}[ 0, g (X^{\rm T}X + I \alpha)^{-1}]$) but now enabling variable selection via the $l_1$-ball projection. Although potentially, one could use other approaches to extend the classic spike-and-slab for a similar latent correlation among the zeros, the one based on the $l_1$-ball projection (and thresholding) has the advantages of being straightforward and having a low modeling complexity. 
In the numerical experiments,
we will show an example of inducing dependence with a correlated $\pi_\beta$ in a brain connectivity study.

Second, by replacing the vector norm $\|\theta\|_1$ with more general $\|h(\theta)\|_1$ we can now induce sparsity on a constrained or low-dimensional space, such as the sparse contrast and reduced-dimension examples we presented early. Such a task would be very difficult to do via  \eqref{eq:spike_and_slab}, for two reasons
% To illustrate this, we now revisit the two examples described at the end of last section. In the fused-lasso type sparsity,  it is not hard to see that those $\theta_i$'s are subject to a set of combinatorial constraints ---  if $\theta_i -\theta_{i-1}=0$, then $\theta_i$ and $\theta_{i-1}$ must be simultaneously zero or non-zero. In fact, the space of this sparse vector $\{s\in\mathbb{R}^{2p-1}:s=D\theta\}$ is low-dimensional, with the dimensionality at most $p$.  In the other example of modeling the rank-constrained sparse matrix, the parameter  space $\{\theta: \text{rank}(\theta)\ge d\}$ is a highly constrained space, with an unknown rank  from $\{d,\ldots, p\}$. In both cases, it would be very difficult to use the classic spike-and-slab prior on these parameters ---
(i)  assigning an element-wise  spike-and-slab prior on a constrained parameter would create an intractable normalizing constant (that involves the parameter $w$ and the ones in $\pi_\tau$), making the prior specification/calibration very difficult; (ii)  the computation would become formidably challenging, due to the need to satisfy the constraints when updating the value of each $\theta_i$ [hence early methods tend to rely on approximation, such as \cite{banerjee2013bayesian} for estimating sparse covariance matrix]. To compare, the $l_1$-ball prior does not have these issues since the prior is fully defined on the unconstrained parameter $\beta$ without any intractable constant, and we can easily sample from the posterior distribution.

\subsection{Comparison with the Post-processing Algorithms}

There are several works in the literature on post-processing continuous posterior samples to obtain exact zeros \citep{bondell2012consistent,hahn2015decoupling,li2017variable}. In the supplementary materials, we provide a detailed comparison of both methodology and numerical experiments on point estimates.

        \section{Numerical Experiments}

        In this section, we illustrate three interesting applications related the motivating combinatorial problems as described in Section 2.1.
        
%         In all the experiments, we use $\pi_\beta$ as $\beta_i \sim \t{DE}(0,\lambda_i)$, with $\lambda_i\sim \t{Exp}(\alpha_\lambda)$ ($\alpha_\lambda=1$ unless stated otherwise); the prior on the radius/threshold for $\tilde\mu$ as described in section \ref{sec:threshold} with $a_w=p$, $b_w=1$, to reflect the prior belief that $\theta$ or its transform should be very sparse.

\subsection{Sparse Contrast Modeling: Piece-wise Constant Smoothing}
In the first example, we conduct a task of image denoising/segmentation. Consider each pixel measurement of an image as $y_{i,j}\in\mathbb R$ (for simplicity, we focus on one color channel) modeled by:
$$y_{i,j} = \mu + \theta_{i,j} + \varepsilon_{i,j}, \quad\varepsilon_{i,j}\stackrel{\text{iid}}{\sim} N(0,\sigma^2),$$
for all pixels indexed by horizontal $i=1,\ldots,p_1$ and vertical $j=1,\ldots,p_2$ and some scalar $\mu\in \mathbb{R}$. A common strategy is to impose segment-wise (piece-wise) constant $\theta_{i,j}$'s as a smoothing for the image, nevertheless, the issue is that the boundary of each segment is not predefined; hence finding the boundary of and the number of segments is a combinatorial problem. 

A nice solution, popularized by the fused lasso \citep{tibshirani2005sparsity}, is to induce sparsity in contrasts:
$$
\theta_{i,j} - \theta_{i+1,j}, \qquad \theta_{i,j}-\theta_{i,j+1}
$$
for $i=1,\ldots,p_1-1$ and $j=1,\ldots,p_2-1$; as well as sparsity in $\theta_{i,j}$'s. Applying $l_1$-norm on each and summing up, we can represent the regularization by $\|D\theta\|_1$, with matrix $D\in\mathbb{R}^{d \times (p_1p_2)}$ and $d=(p_1-1)p_2 + p_1 (p_2-1)+ p_1p_2$.

Besides point estimate, a common inference task is to assess whether a non-zero difference across the boundary is indeed significant, or just random variation. A popular frequentist solution is to develop a series of hypothesis tests [for recent work, see \cite{jewell2019testing} and references within]. On the other hand, using our $l_1$-ball prior based on $\mathbb{B}_{h,r}$ with $h(\theta) = D\theta$, we can obtain a simple Bayesian solution. Using the ADMM algorithm we described in \eqref{eq:admm}, we have the first step in closed-form:
$$
z \leftarrow [ (2\rho)^{-1}D^{\rm T}D + I]^{-1} [ \beta + (2\rho)^{-1}D^{\rm T}(s-\kappa)],
$$
hence the projection of $\beta$ to $\theta\in\mathbb{B}_{h,r}$ can be evaluated rapidly.

        \begin{figure}[H]
        \centering
%                \begin{subfigure}[t]{3.5cm}
%                        \centering                        \includegraphics[width=1.1\linewidth]{figure/fused_a}
%                        \caption{Original image.}
%                \end{subfigure}
                \begin{subfigure}[t]{4.5cm}
                        \centering
\includegraphics[width=1.1\linewidth]{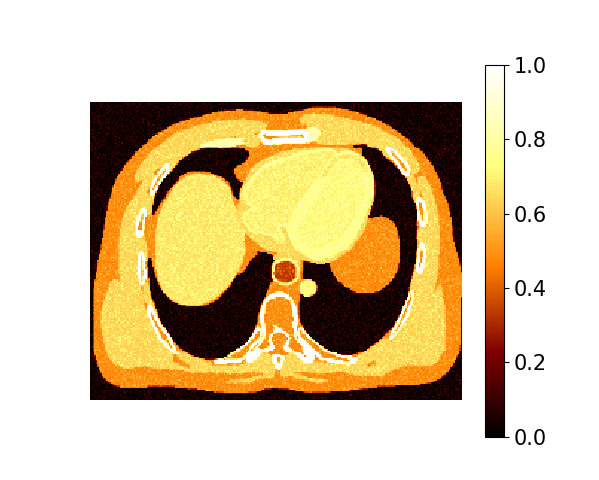}
                \caption{Noisy image of a chest scan.}
                \end{subfigure}         \;
                \begin{subfigure}[t]{4.5cm}
                        \centering
                \includegraphics[width=1.1\linewidth]{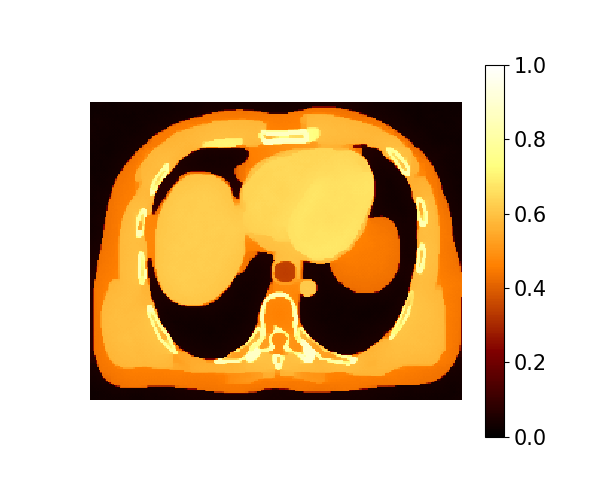}
                \caption{Posterior Fr\'echet mean corresponding to a piece-wise constant smoothing.}
                \end{subfigure}         \;
                \begin{subfigure}[t]{4.5cm}
                        \centering
               \includegraphics[width=1.1\linewidth]{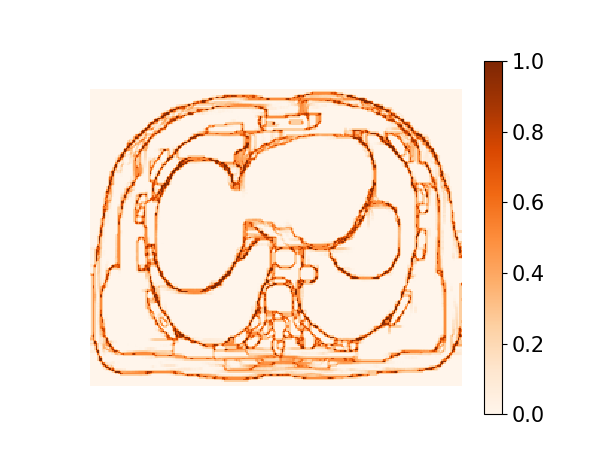}
                \caption{Pixel-wise probability for each pixel being not equal to at least one of the neighboring pixels.}
                \end{subfigure}  
                \caption{\label{fig: fused2d}An example of the sparse contrast modeling using the generalized $l_1$-ball prior. Our model finds a piece-wise constant smoothing (b) underneath a noisy image (a), while quantifying the uncertainty (c).}
        \end{figure}
        
We apply the above model with a generalized $l_1$-ball prior using $\beta_{i,j}\stackrel{\text{iid}}\sim \t{DE}(0,1)$, and {$\alpha=10.$}. The scan image is from \cite{gong2017low} and previously used by \cite{https://doi.org/10.1111/rssb.12407}, with the noisy raw data plotted in Figure \ref{fig: fused2d}(a). In the posterior samples, 
 the Fr\'echet mean indeed recovers a segment-wise constant structure, with each part clearly shown (b). Importantly, we quantify the posterior probability that a smoothed pixel is not equal to at least one of its neighbors, $\t{pr}[\theta_{i,j} \neq \theta_{(i+k,j+k')} \; \t{ for at least one }(k,k') =(0,1),(1,0), (-1,0), (0,-1) \mid y]$ for $i=2,\ldots,p-1$ and $j=2,\ldots,p-1$. Indeed, those large probabilities are located near the boundary; and interesting extension could be further explored, such as one for Bayesian  multiplicity control on image boundary detection.

\subsection{Reduced Dimension Modeling: Finding the Number of Mixture Components}
        To illustrate the usefulness of exact zeros outside the linear models, we consider the prior specification problem for the finite mixture model. We focus on the $K$-component Gaussian mixture likelihood:
        \be
        \pi(y_j; \theta, \mu, \sigma^2) = \sum_{k=1}^K  \theta_k \phi(y_j \mid \mu_k, \sigma^2_k),
        \ee
        for $j=1,\ldots,n$, where $\phi(.\mid a,b)$ denotes the normal density with the mean $a$ and variance $b$.

        Suppose the data are generated from a $K_0$-component model, but we do not know $K_0$ exactly; hence a common Bayesian modeling practice is to assign a prior on the $\theta_k$'s and shrink some of them close to zero. For example, it has been popular to use  the infinite mixture model, which considers $K$ as unbounded and assigns a stick-breaking construction on $\{\theta_k\}_{k=1}^{\infty}$, such as the one in the Dirichlet process. However, this was recently discovered to yield an inconsistent result \citep{miller2014inconsistency} for the number of components, as the posterior probability $\t{pr}(K=K_0)$ goes to zero as $n\to \infty$. Later, \cite{miller2018mixture} show that instead of putting an infinite mixture prior on $\theta_k$, if we treat $K$ as a finite number and put a prior $\Pi_K(K)$, this can yield a consistent estimation at $K_0$. They refer to it as the ``mixture of finite mixtures'' model.
The major drawback is that this involves a combinatorial search over different $K$'s.

        Using the $l_1$-ball prior, we can significantly simplify this problem. Assuming we know a $K_1$ large enough to have $K_1>K_0$, starting from $\beta_i\stackrel{\text{ind}}\sim  \t{DE}(0,\lambda_i)$, we project it to the  $l_1$-ball and apply the transformation:
        \be
        w =  P_{\mathbb{B}_{r}}(\beta),\qquad
        & \theta_k = \frac{|w_k|}{\sum_{i=1}^{K_1} |w_i| } \t{ for } k=1,\ldots, K_1,
        \ee
note that in the second step, we have $\sum_{i=1}^{K_1} |w_i|=r$ if $\sum_{i=1}^{K_1} |\beta_i|\ge r$. Compared to the parameter space in an infinite mixture model $\{ \theta: \theta_k>0 \;\forall k\in \mathbb{Z}_+,\sum_{k=1}^{\infty} \theta_k=1\}$, besides being finite dimensional, a key difference here is that the space of $\theta$ is the {\em closure} of the probability simplex (which includes the  case for some $\theta_k=0$),       
        \be
        \Delta^{K_1-1} =\{\theta :\theta_{k}\ge 0  \;\;\forall k,  \sum_{k=1}^{K_1} \theta_k=1\},
        \ee
 Therefore, using the projection, we assign a positive probability 
for each $K \in \{1,\ldots, K_1\}$; hence this gives a continuous version of the mixture of finite mixtures model.

When simulating the data, we use $K_0=3$ with  $(0.3,0.3,0.4)$ as the mixture weights; to have the components overlap, we use $\mu_1=0, \mu_2=4, \mu_3=6$, and all $\sigma^2_1 = \sigma^2_2=\sigma^2_3=1$. We generate $n=1,000$ in the simulation.

        \begin{figure}[H]
                \centering
                \begin{subfigure}[t]{0.3\textwidth}
                        \centering
                        \includegraphics[width=0.9\textwidth, height = .6\textwidth]{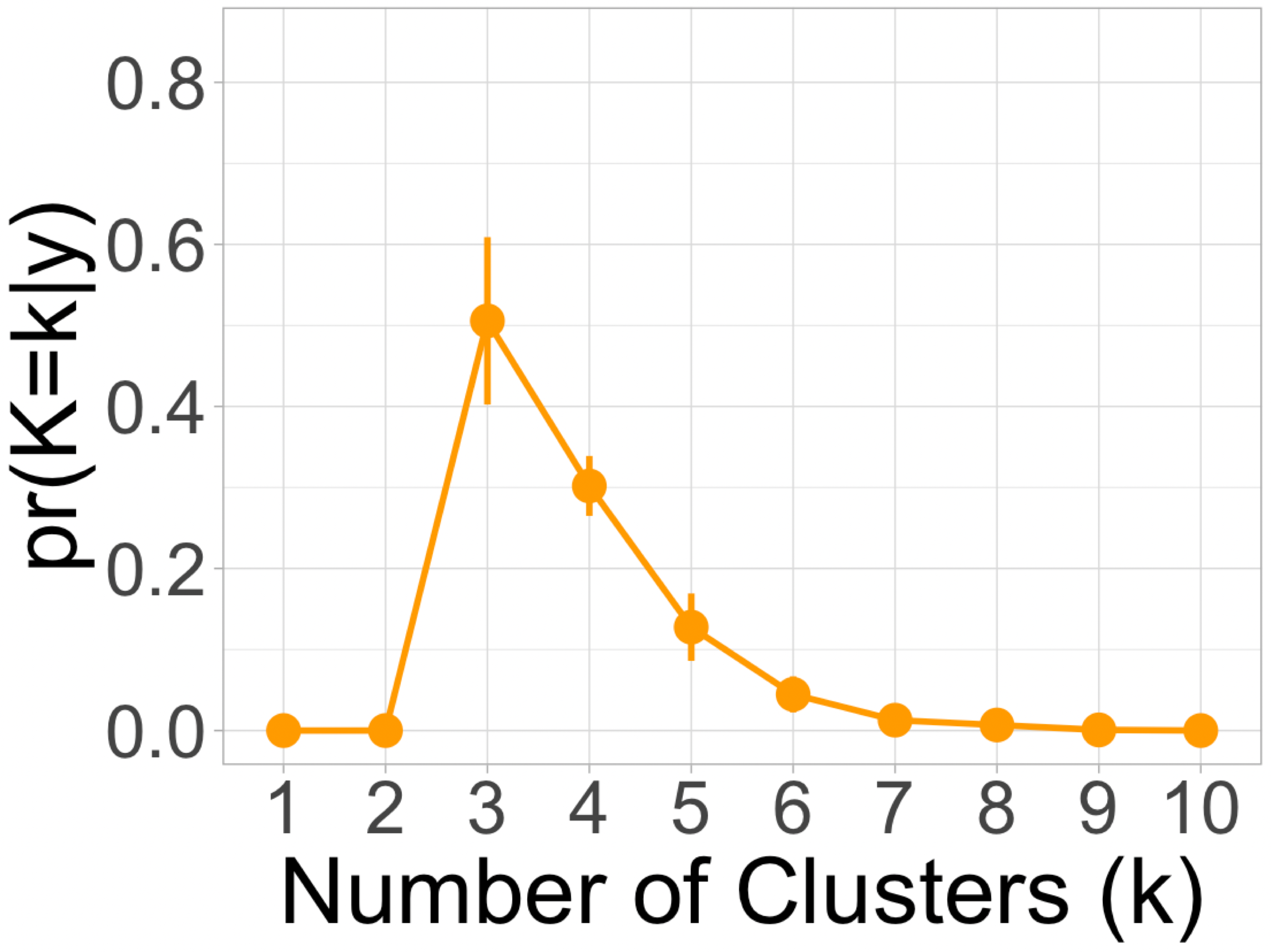}
                        \caption{The $l_1$-ball mixture of finite mixtures model.}
                \end{subfigure}
                 \hspace{.3cm}
                \begin{subfigure}[t]{0.3\textwidth}
                        \centering
                        \includegraphics[width=0.9\textwidth, height = .6\textwidth]{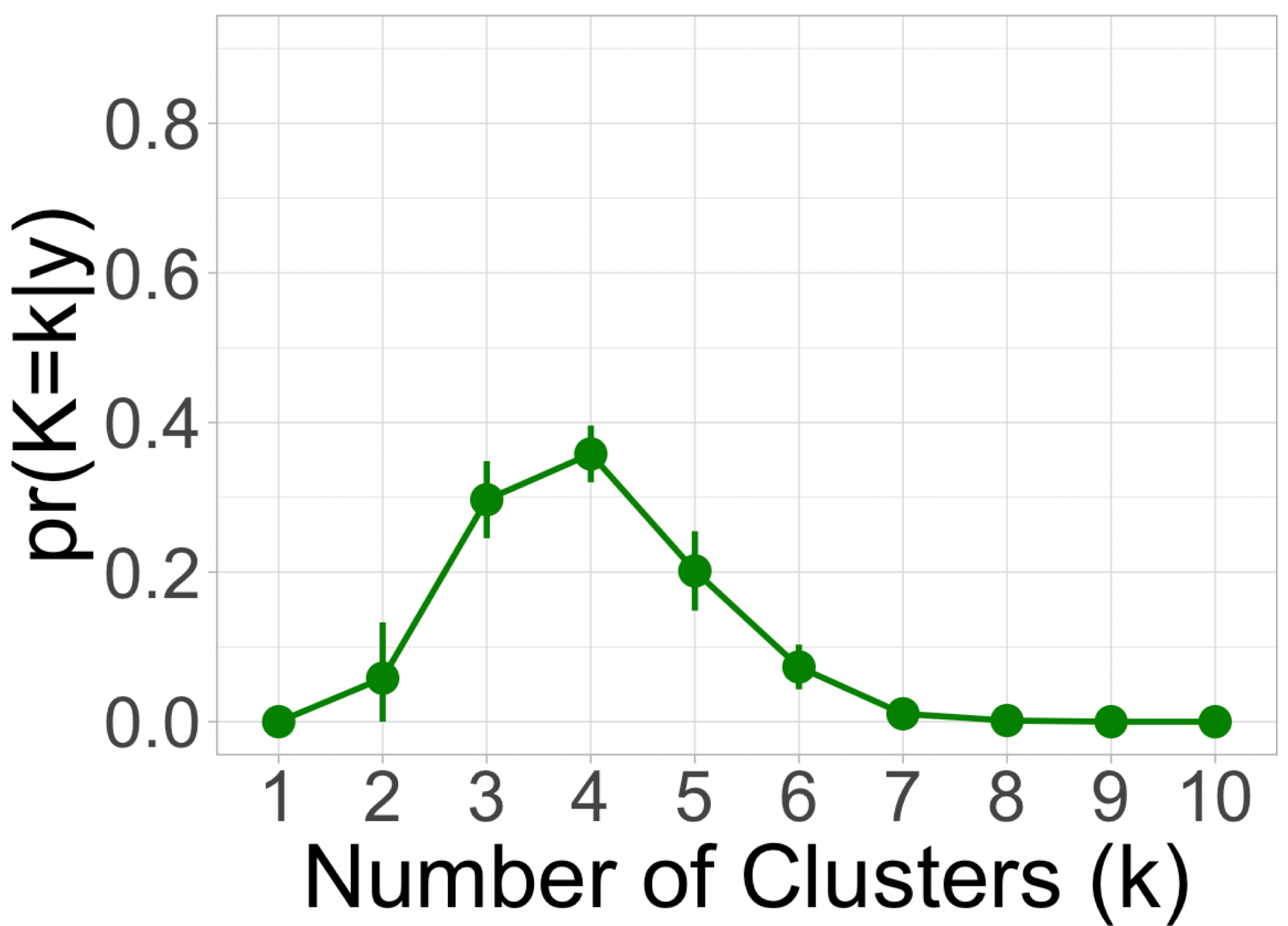}
                        \caption{The finite mixture model with finite Dirichlet prior.}
                \end{subfigure}
                 \hspace{.3cm}
                \begin{subfigure}[t]{0.3\textwidth}
                        \centering
                        \includegraphics[width=0.9\textwidth, height = .6\textwidth]{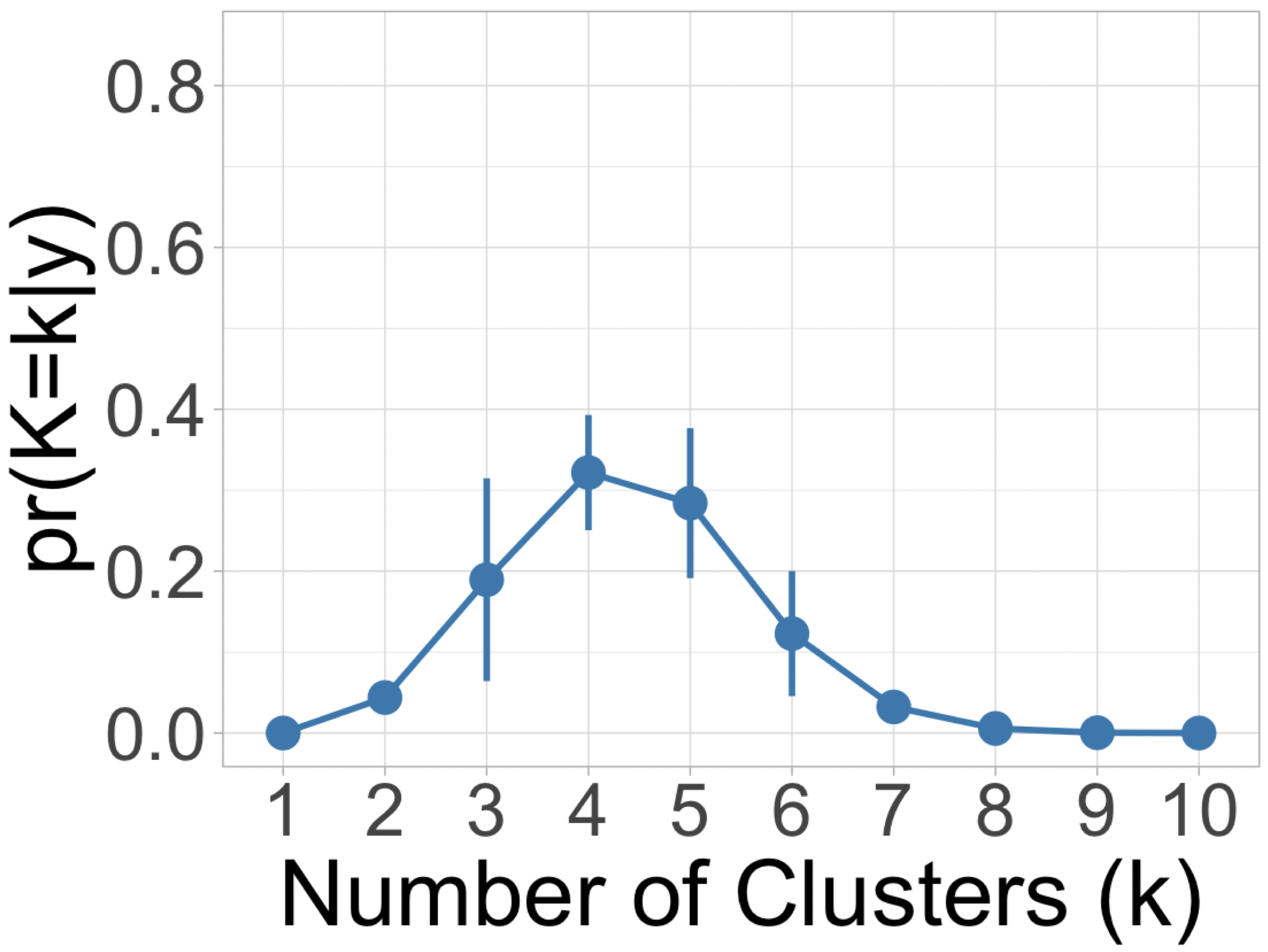}
                        \caption{The Dirichlet process mixture model with $\tilde \alpha=1$.}
                \end{subfigure}
                                \begin{subfigure}[t]{0.4\textwidth}
                        \centering
                        \includegraphics[width=.9\textwidth, height = .4\textwidth]{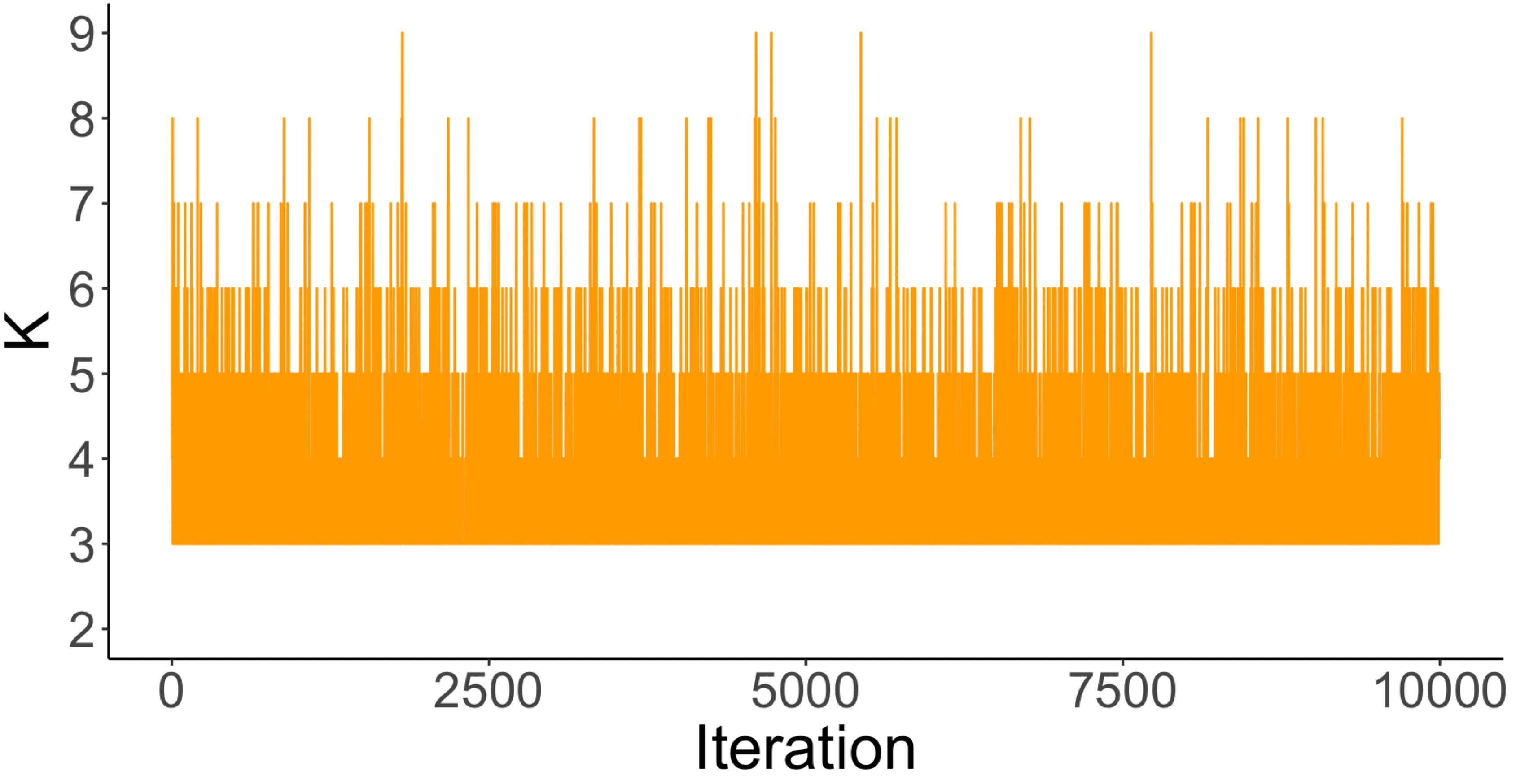}
                        \caption{Traceplot on the number of clusters of the $l_1$-ball mixture of finite mixtures model using the Hamiltonian Monte Carlo.}
                \end{subfigure}
                \quad
                                                \begin{subfigure}[t]{0.4\textwidth}
                        \centering
                        \includegraphics[width=.9\textwidth, height = .4\textwidth]{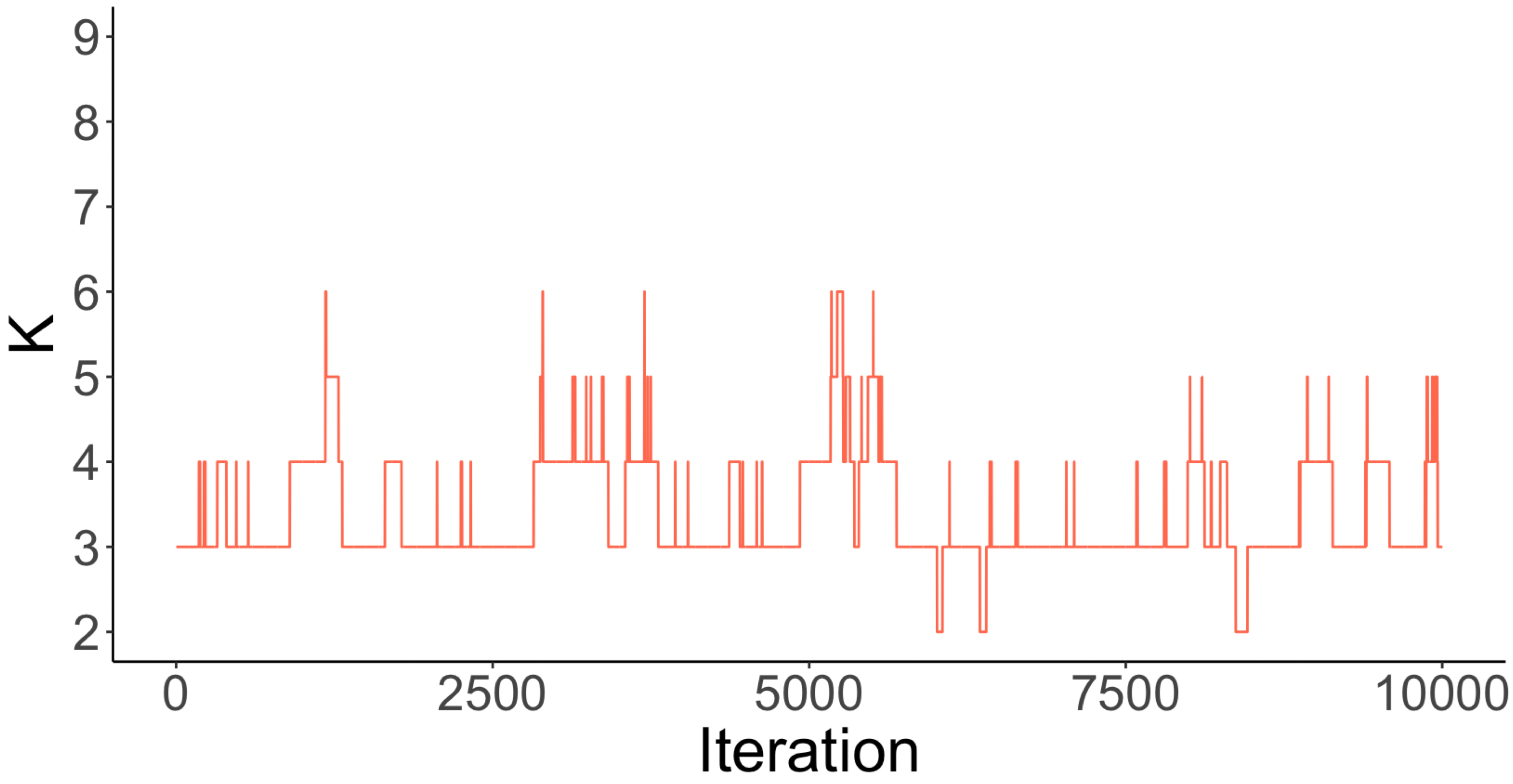}
                        \caption{Traceplot on the number of clusters of the mixture of finite mixtures model with a prior on $K$, using the combinatorial  split-merge algorithm \citep{jain2007splitting}.}
                \end{subfigure}
                \caption{Comparing the performances of applying $l_1$-ball prior to the finite mixture weight and the other mixture models. The experiment is repeated five times, and the mean of the number of clusters with its standard error is presented. In each experiment, we run the Markov chain for 20,000 iterations and discard the first 10,000 as burn-ins. }
                \label{fig: mixture}
        \end{figure}
        
                For the $l_1$-ball prior, we use $K_1=10$, Inverse-Gamma$(1,1)$ prior for $\sigma^2_k$ and $\t{N}(0,10^2)$ prior for $\mu_k$. To compare, we also use (i) the Dirichlet process mixture model with the concentration parameter $\tilde\alpha=1$; (ii) the finite mixture model with the same dimension $K_1=10$, but with a finite Dirichlet distribution prior $(\theta_1,\ldots,\theta_{K_1})\sim \text{Dir}(\tilde\alpha)$ with $\tilde\alpha=0.001$ to favor sparsity in $\theta$. We use the same prior for $\mu_k$ and $\sigma^2_k$; for the radius prior in the $l_1$-ball prior, we use {$\alpha=3$}.

Figure~\ref{fig: mixture} shows the posterior distribution of $K$ in all three models. Clearly, the one uses the $l_1$-ball has the largest probability assigned to $K_0=3$. Both the Dirichlet process mixture model and the finite mixture model with a Dirichlet prior put the largest probability at $K=4$.
{The good result is because we effectively assign a discrete prior on the number of mixture components, hence the consistency theory of \cite{miller2022consistency} directly applies.}
 In addition, we compare the computing performance using the $l_1$-ball prior against the combinatorial search using the split-merge algorithm \citep{jain2007splitting}. The $l_1$-ball mixture of finite mixtures enjoys faster mixing and less autocorrelation.

 \subsection{Nonlinear Modeling: Discontinuous Gaussian Process Regression}
Gaussian process regression is a useful non-parametric method to model nonlinear functions. For outcome $y_i\in \mathbb R$ and predictors $x_i\in\mathbb R^p$, we model the outcome 
 $$y_i = f(x_i)+\epsilon_i, \quad \epsilon_i\stackrel{\text{iid}}{\sim}N(0,\sigma^2_e), \quad f(x) \sim GP[0,K_\theta(\cdot, \cdot)] $$
for $i=1,\ldots,n$, where $GP$ represents a Gaussian process, such that any finite dimensional realization follows a multivariate normal with mean $0$ vector and covariance determined by the covariance function $\t{Cov}[f(x_i),f(x_j)]= K_\theta(x_i,x_j)$. An often seen property of those popular covariance functions, such as squared exponential and Mat\'{e}rn, is that the correlation $\t{Corr}[f(x_i),f(x_j)]\to 1$ as $\|x_i-x_j\|_2\to 0$; as the result, it imposes almost everywhere continuity: for every $\delta_1>0$, there is a $\delta_2$ such that for any $(x_i,x_j):\|x_i-x_j\|_2\le \delta_2 \Rightarrow |f(x_i)-f(x_j)|\le \delta_1$. This  continuity may not be desirable, if we want to model $f$ as a function containing one or several points of discontinuity. To address this shortcoming, \cite{gramacy2008bayesian} proposed to use multiple independent Gaussian processes, each supported on one of the partitioned regions; to obtain the partition, they used a predictor-partition tree that requires a combinatorial search in the computation.
 
 We now develop a simple alternative that uses one Gaussian process, based on a slight modification of a popular squared exponential covariance function, and an application of a generalized $l_1$-ball prior. With a predictor-dependent graph $\mathcal G=(\mathcal V, \mathcal E_x)$  with $\mathcal V$ containing $n$ nodes, and edges formed by radial neighbors $\mathcal E_x = \{(i,j): \|x_i-x_j\|_2\le d\}$ with some pre-set $d$ (we choose the smallest $d$ so that each node has at least one connected neighbor on $\mathcal G$), we use $\theta= \{\sigma_K^2,\lambda_1,\lambda_2, \eta\}$ and specify
\be
&  K_\theta(x_i,x_j) = \sigma_K^2 \exp[- \frac{\|x_i-x_j\|_2^2}{\lambda_1}] \exp[-\frac{|\eta_i-\eta_j|^2}{\lambda_2}],\\
&  \eta = P_{ \{ z: \sum_{(i,j)\in  \mathcal E_x} |z_i-z_j| \le r \} }(\beta), \qquad \beta_i \stackrel{\text{iid}}\sim \text{DE}(0,1),
 \ee
 where $\sigma_K^2 >0$, $\lambda_1>0$, $\lambda_2>0$ are the parameters for a squared exponential covariance function. For the radius prior in the $l_1$-ball prior, we use $\alpha=0.1$. We put an $\text{Inverse-Gamma}(2,0.1)$ prior on each of those parameters. At the same time, we introduce $\eta_i\in\mathbb{R}$ as a \emph{latent jittering coordinate}, that arises from a generalized $l_1$-ball prior related to the graph-fused lasso \citep{tibshirani2011solution}. As the result of the projection, most of $(i,j)\in\mathcal E_x$ would have $\eta_i=\eta_j$; however, a few $(i,j)$ would have $\eta_i \neq \eta_j$. On any edge with $\eta_i\neq \eta_j$, we have $\text{Corr}[f(x_i), f(x_j)]\ll 1$ even if $x_i\approx x_j$, which allows discontinuity to occur. 
 
\begin{figure}[H] 
                      \begin{subfigure}[t]{.3\textwidth}
                        \includegraphics[width=1\linewidth, height =4.0cm]{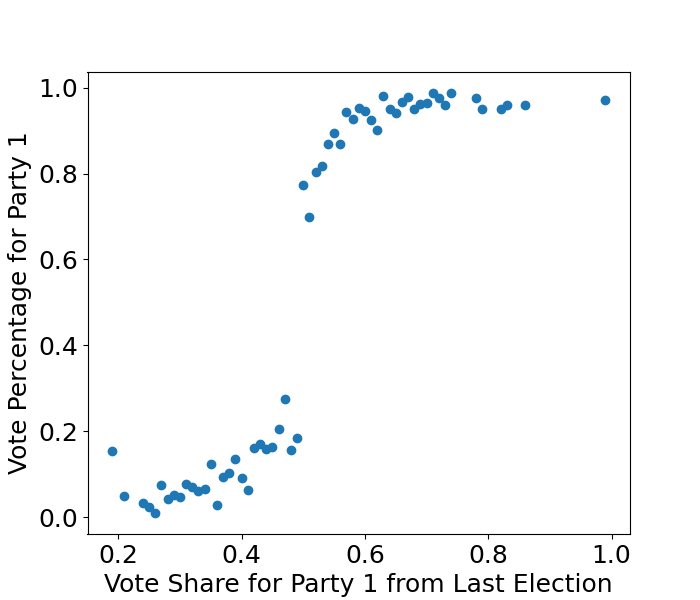}
                        \caption{Scatter plot comparing the vote percentage for Party 1 and the vote share in the last election.}
                \end{subfigure}
                \;
                \begin{subfigure}[t]{.3\textwidth}
                        \includegraphics[width=1\linewidth, height =4.0cm]{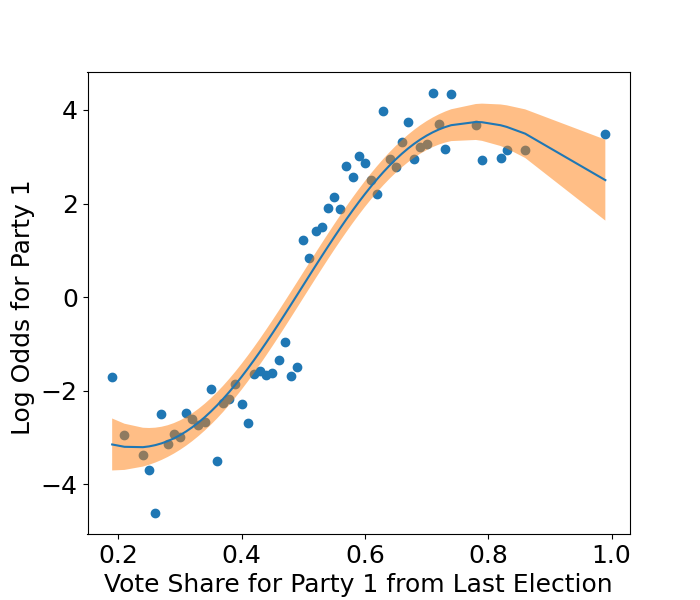}
                        \caption{Fitting a continuous Gaussian process with squared exponential covariance.}
                \end{subfigure}
                \;
                \begin{subfigure}[t]{.3\textwidth}
                        \includegraphics[width=1\linewidth, height =4.0cm]{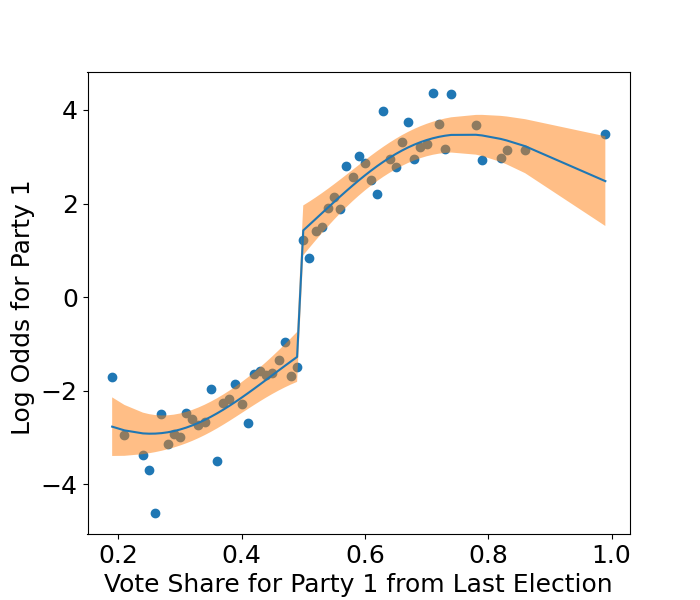}
                        \caption{Fitting a discontinuous Gaussian process with latent jittering coordinates, which have a generalized $l_1$-ball prior.}
                \end{subfigure}
                \centering
                \caption{Discontinuous Gaussian process regression on the election data (Panel a). With a latent jittering coordinate $\eta_i$ regularized by a generalized $l_1$-ball prior, we can change a Gaussian process with squared exponential covariance function (Panel b) to have points of discontinuity (Panel c), giving an improved fit to the data.\label{fig: GPR}
                The fitted curves and 95\% point-wise credible bands are shown.
                }
        \end{figure}

We use the above model in a political science data application. The data were collected from $n=59$ voting districts. The goal is to find a relationship between the vote percentage for  Party 1  and several predictors, including the  vote share of Party 1 in the last election, minorities percentage in the voters, and urban percentage \citep{lee2004voters}.  Since we observe the vote percentages directly, we use their log-odds transforms as $y_i$'s and model them by Gaussian processes. The scatter plot in Figure \ref{fig: GPR}(a) shows that there is a sudden jump in the current-year vote percentage when the vote share from the last election passes near 50\%, which is likely due to the increased polarization of political preferences since the last election.

For a clear illustration, in the main text, we fit regression models using the vote share for Party 1 in the last election as the only predictor (the single predictor model allows the fitted curves to be smooth in $x_i$ except for points of discontinuity). To compare, we also fit the regression models using the continuous Gaussian Process with a squared exponential covariance function. As shown in Figure \ref{fig: GPR}(b) and (c), the discontinuous Gaussian process successfully discovers two distinct values in $\eta_i$'s and gives a better fit to the data, compared to the continuous one.  The root-mean-square deviation (RMSD) is 0.478 for the discontinuous model, and 0.589 for the continuous one. When using all three predictors, the discontinuous Gaussian process finds three distinct values in $\eta_i$'s, and we provide the results and comparison in the supplementary materials.

\subsection{Additional Numerical Results}
In the supplementary materials, we provide some additional results related to (i) benchmarking the $l_1$-ball prior method in linear regression setting; (ii) simulation in change point detection with comparison to continuous shrinkage prior; (iii) numerical comparison with post-processing methods on the accuracy of linear model selection; (iv) assessing the mixing performance of running Hamiltonian Monte Carlo via the $l_1$-ball parameterization for a spike-and-slab prior, with comparison to the Gibbs sampling algorithm; (v) application of the $l_1$-ball prior to induce structured sparsity; (vi) simulation on rank recovery with nuclear-norm $l_1$-ball prior.

   \section{Data Application: Sparse Change Detection in the Medical Images}
For the application, we use the $l_1$-ball prior on the analysis of a medical imaging dataset.
The data are the abdominal dynamic contrast-enhanced magnetic resonance imaging   \citep{otazo2015low}, collected on a healthy human subject during normal breathing. It is in the form of a video, acquired via a whole-body scanner to record the aorta, portal vein and liver enhancement. There are $384\times 384$ pixels in each frame (corresponding to $0.94$ seconds) and $T=75$ frames in total.

An important scientific task is to detect the locations of the large changes, corresponding to important organ activities. However, there are a few challenges: (i) most parts of the image are not fixed but also dynamically changing (such as the overall brightness, although to a less degree compared to the sharp changes), hence we need to model a ``background'' time series; (ii) the video is noisy hence there are uncertainties on the detected changes. To handle this problem, we consider the low-rank plus sparse model:
\be
M_t = \sum_{k=1}^d \tilde \alpha_{t,k} \psi_k + S_t +  E_t,
\ee
for $t=1,\ldots, T$; where the $\psi_k\in\mathbb{R}^{384\times 384}$ corresponds to some latent component shared by all frames, and $\tilde \alpha_{t,k}$ is the loading dynamically changing over time; $S_t \in\mathbb{R}^{384\times 384}$ is a sparse matrix corresponding to the sharp changes that we wish to detect; $E_t$ corresponds to the noise and we model it as independent $N(0,\sigma^2_e)$ for each of its element.

A common problem for low-rank modeling is to determine the rank, in this case, the number of latent components $d$.  \cite{bhattacharya2011sparse} previously proposed to view $d$ as unbounded, while applying a continuous shrinkage prior on the scale of loading, closer towards zero as $k$ increases. We are inspired by this idea, nevertheless, we achieve an exact rank selection by using a generalized $l_1$-ball prior based on the nuclear norm. This has two advantages: we can treat the low-rank part using one matrix parameter $L$ replacing $\sum_k\tilde  \alpha_{t,k} \psi_k $, which avoids the potential identifiability issues when estimating $\tilde \alpha_{t,k}$ and $\psi_k$ separately; having an exactly low-rank part reduces the confoundingness between the near-low-rank and sparse signals.

Specifically, we reparameterize the $75$ matrices $\{ \sum_{k=1}^d \tilde \alpha_{t,k} \psi_k\}_{t=1}^{\rm T}$ via a single matrix of size $75 \times 384^2$:
\be
L  = \sum_{k=1}^d ( \tilde \alpha_{1,k},\ldots, \tilde \alpha_{T,k})  [\t{vec}(\psi_k)]^{\rm T}.
\ee
Without specifying $d$, we can treat $L$ as the output of projecting a dense matrix $\beta \in \mathbb{R}^{75 \times 384^2}$  to a generalized $l_1$-ball:
\be
L =  \argmin_{ Z\in\mathbb R^{75 \times 384^2}: \|Z\|_* \le r}\|Z-\beta\|_F^2
\ee
where $\|Z\|_*$ denotes the nuclear norm, as the sum of the singular values $\sum_{k=1}^{75} \rho_{k}(Z)$, with $Z = U_Z \t{diag}[\rho_k (Z)] V^{\rm T}_Z$.

 Since having exact $\rho_k(Z)=0$ for some $k$'s will lead to an effective rank reduction to $d=| \{\rho_k(Z): \rho_k(Z)>0\}|$, the nuclear norm regularization is very popular in the optimization literature \citep{hu2012fast}. This projection has a closed-form solution: if $\sum_k \rho_k(\beta) \ge r$, $L= U_\beta \t{diag}[ (\rho_k(\beta) - \tilde\mu)_+ ] V^{\rm T}_\beta$, for some $\tilde \mu>0$ such that $\sum_k [\rho_k(\beta) - \tilde\mu]_+ = r$; $L=\beta$ if $\sum_k \rho_k < r$.  For the radius prior in the $l_1$-ball prior, we use {$\alpha=100$}.

Assigning  $\beta_{i,j}\sim \t{DE}(0,\lambda_{i,j})$, we can quantify the uncertainty regarding the rank. Further, we assign an element-wise $l_1$-ball prior to each element of $S_t$, so that we can obtain sparse estimates on the sharp changes. In the supplementary materials, we show that the background time series is mostly based on the linear combination of $2$ latent components (panel a), each corresponding to a dense image (b-c). By visualizing the estimated backgrounds at three different time points, we can see some very subtle differences, such as the brightness between (d) and (f), which involves most of the pixels. Indeed, these small changes are what we wish to find and separate from the sparse part $S_t$.

\begin{figure}[H]
     \begin{subfigure}[b]{.3\textwidth}
\centering
        \includegraphics[height=3.3cm, width=4.2cm]{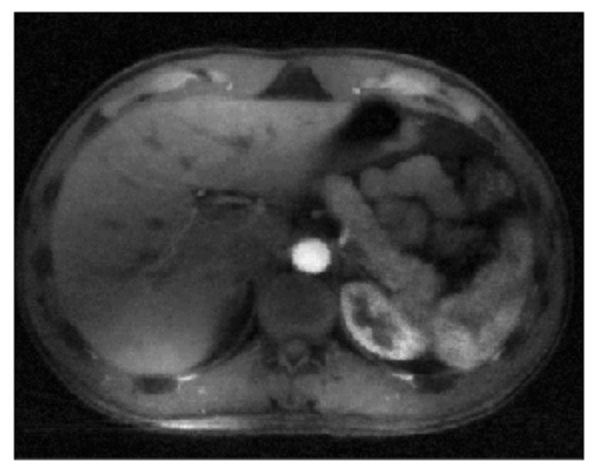}
        \caption{Raw image at $t=15$.}
\end{subfigure}
\quad
        \begin{subfigure}[b]{.3\textwidth}
\centering
        \includegraphics[height=3.3cm, width=4.2cm]{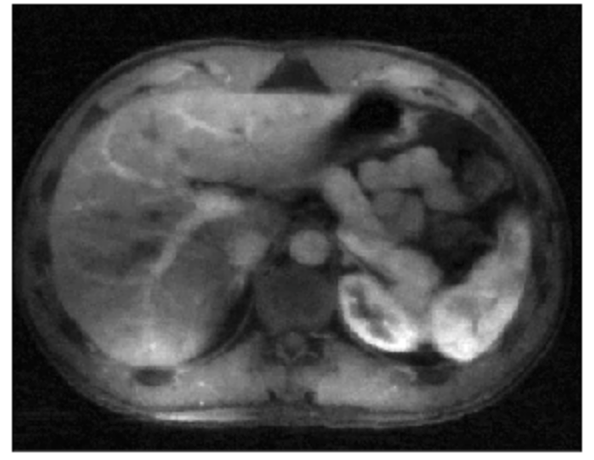}
        \caption{Raw image at $t=35$.}
\end{subfigure}
\quad
        \begin{subfigure}[b]{.3\textwidth}
\centering
        \includegraphics[height=3.3cm, width=4.2cm]{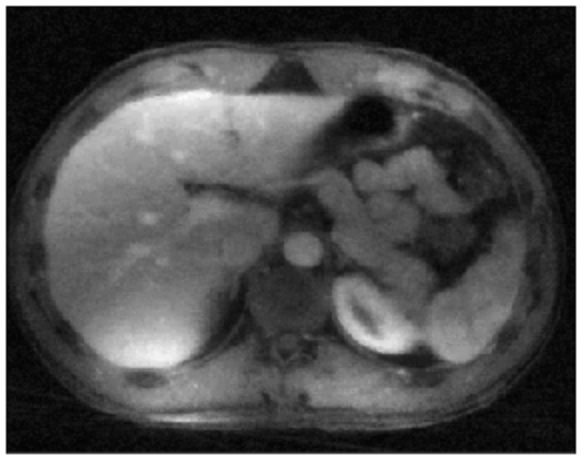}
        \caption{Raw image at $t=75$.}
\end{subfigure}
     \begin{subfigure}[b]{.3\textwidth}
\centering
        \includegraphics[height=3.3cm, width=4.2cm]{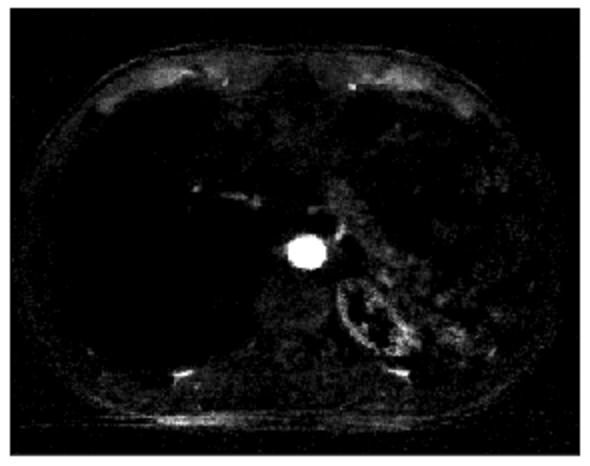}
        \caption{Estimated sparse change at $t=15$.}
\end{subfigure}
\quad
        \begin{subfigure}[b]{.3\textwidth}
\centering
        \includegraphics[height=3.3cm, width=4.2cm]{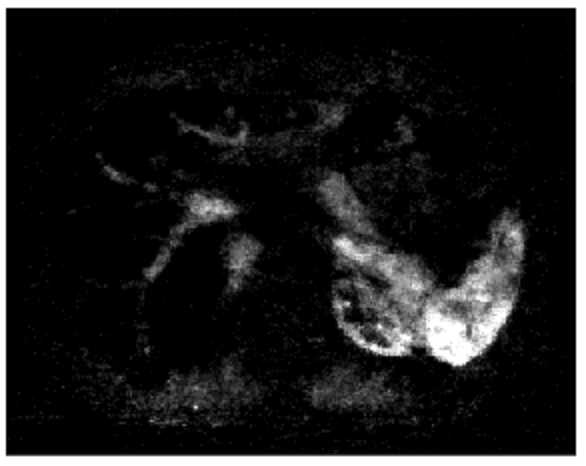}
        \caption{Estimated sparse change  at $t=35$.}
\end{subfigure}
\quad
        \begin{subfigure}[b]{.3\textwidth}
\centering
        \includegraphics[height=3.3cm, width=4.2cm]{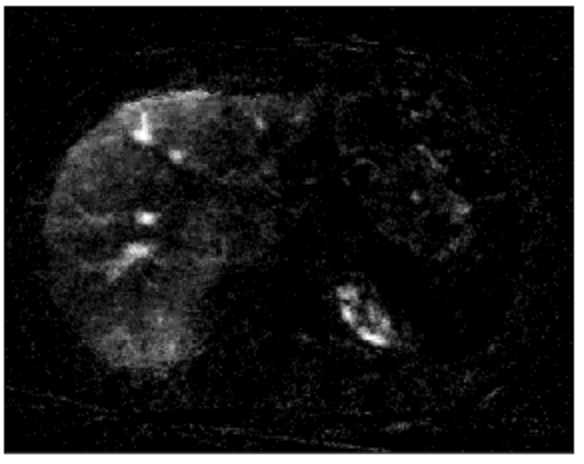}
        \caption{Estimated sparse change  at $t=75$.}
\end{subfigure}
     \begin{subfigure}[b]{.3\textwidth}
\centering
        \includegraphics[height=3.3cm, width=4.2cm]{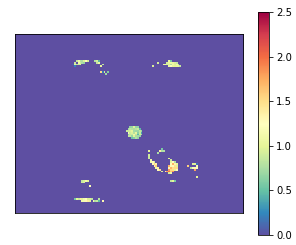}
        \caption{The pixel-wise variance at $t=15$.}
\end{subfigure}
\quad
        \begin{subfigure}[b]{.3\textwidth}
\centering
        \includegraphics[height=3.3cm, width=4.2cm]{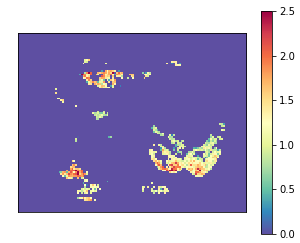}
        \caption{The pixel-wise variance at $t=35$.}
\end{subfigure}
\quad
        \begin{subfigure}[b]{.3\textwidth}
\centering
        \includegraphics[height=3.3cm, width=4.2cm]{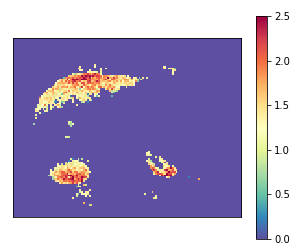}
        \caption{The pixel-wise variance at $t=75$.}
\end{subfigure}
\caption{\label{fig:video_S} The sparse change component $S_t$ and its pixel-wise variances are visualized. The three frames correspond to three time points, when the aorta, liver and portal vein are in their enhancement phase respectively. The posterior of the sparse parameter shows the locations of these changes, as well as the uncertainties.}
\end{figure}

Figure~\ref{fig:video_S} shows the locations of the sharp changes we estimate from the sparse $S_t$.
The results are very interpretable as they correspond to when the aorta, portal vein, and liver are in their enhancement phases, respectively. The vessels and organs are distinct from the abdominal background. Further, we compute the pixel-wise variance for these sparse estimates, as a measurement of the uncertainty. It can be seen that the aorta and portal vein (panels g and h)  have a relatively low uncertainty on the changes; whereas the liver (panel i) has a higher uncertainty, hence some caution should be applied when making a conclusion based on the last frame. In the supplementary materials, we experiment with a simpler alternative that replaces the low-rank $\sum_{k=1}^d \tilde \alpha_{t,k} \psi_{k}$ with a time-invariant background $\psi_0$. This leads to much less sparse estimates that are more difficult to interpret.

%%%%%%%
        
        \section{Discussion}
          In this article, we propose a new prior defined on the boundary of the $l_1$-ball. This allows us to build many interesting applications that implicitly involve a combinatorial selection, such as in the change points, the number of mixture components and the rank of a matrix. We show that the $l_1$-ball projection is continuous hence giving a convenient ``continuous embedding'' for these combinatorial problems, establishing a connection between the rich optimization and Bayesian literature.
          
There are several interesting extensions worth pursuing. First, when projecting a continuous distribution into the boundary of a certain geometric set, it often leads to a degenerate distribution that is useful but difficult to parameterize directly. This convenient property is not limited to the $l_1$-ball. For example, in the high-dimensional optimal transport problem, when estimating the contingency probability table given two marginal probability vectors, the solution table is likely to be very sparse \citep{cuturi2013sinkhorn}. This is due to the projection to the high-dimensional polytope will likely end up in the vertex corresponding to a high sparsity. Therefore, our projection idea can be generalized to such useful geometric sets. Second, our data augmentation strategy can be viewed as ``augmentation-by-optimization'', as opposed to the conventional ``augmentation-by-integration''. \cite{polson2016mixtures} have previously explored the connection and difference between maximizing and integrating over a latent variable, under the scope of comparing a frequentist/optimization-based model and its Bayesian counterpart; here, we demonstrate that the barrier can be in fact removed, and the Bayesian models can leverage a maximization over a certain latent variable, creating a new class of useful priors. 
%The use of optimization techniques for the Bayesian approach has recently become popular, especially with a computational focus, such as the proximal Markov chain Monte Carlo \citep{pereyra2016proximal}; on the other hand, our method shows that one could use such a technique for the purpose of model/prior building. 
Lastly, for the theory, we chose to focus on linear regression because of the tractability of vector-norm $l_1$-ball projection, and we can analytically integrate out several parameters to obtain a simple combinatorial prior; for generalized $l_1$-ball projection, a direct analysis would be difficult hence it is interesting to explore different strategies.

\section*{Supplementary Materials}
\appendix
\section{Proofs}

\begin{proof}
Since permutation of indices does not affect $|J|$, without loss of generality, we assume $\sum^{c}_{i=1} |\theta_{i}|=r$ and $|\theta_{i}|>0$ for $i=1,\ldots,c$.

Now $f^{-1}$ is a mapping from $(\theta_{1},\ldots,\theta_{c-1},t_{c+1},\ldots,t_{p},\mu)$ to $(\beta_{1},\ldots,\beta_{c-1},\beta_{c+1},\ldots,\beta_{p},\beta_{c})$, where  $\beta_{c}= s_{c}(r-\sum^{c-1}_{i=1} |\theta_{i}|+{\mu}/{c})$. The Jacobian matrix $J$ is
%\begin{table}[H]\centering
\begin{center}
  \begin{tabular}{l | c c c c c c c }
 ${\partial}$   & $\beta_{1}$ &  $\cdots$ & $\beta_{c-1}$ &  $\beta_{c+1}$  &  $\cdots$ & $\beta_{p}$ & $\beta_{c}$ \\
   \hline
  $/\partial\theta_{1}$ &  $1$  &  $\cdots$ & 0 &0 & $\cdots$  & 0 &  $-s_{1}s_{c}$ \\
 $\vdots$ &  $\vdots$ &  $\ddots$ & $\vdots$ & $\vdots$ & $\vdots$ & $\vdots$ & $\vdots$  \\
 $/\partial \theta_{c-1}$ & 0  & $\cdots$ & $1$ &0 &$\vdots$  & 0 & $-s_{c-1}s_{c}$ \ \\
 $/\partial  t_{c+1}$ & 0  & & 0 & $s_{c+1}$ &0  & 0 & 0 \\
 $\vdots$ &  $\vdots$ &  $\vdots$ & $\vdots$ & $\vdots$ & $\ddots$ & $\vdots$ & $\vdots$\\
 $/\partial t_{p}$ & 0  & $\cdots$ & 0 &0 &$\cdots$  & $s_{p}$ & 0 \\
  $/\partial \mu$ &  $s_{1}/c$  & $\cdots$ &$s_{c-1}/c$ &$s_{c+1}/c$ &$\cdots$  & $s_{p}/c$ & $s_{c}/c$ 
 \end{tabular}
 \end{center}

%\end{table}
Split the matrix into four blocks, with $A=J_{1:(p-1),1:(p-1)}$, $B=J_{p,1:(p-1)}$, $C=J_{1:(p-1),p}$ and $D=s_c/c$. We know
\be
|J|& = |D-BA^{-1}C||A|\\
&= |s_{c}/c +\sum_{i=1}^{c-1} s_{i}^2s_{c}/c | \times 1\\
&=|s_c |\\
&=1.
\ee
\end{proof}

\subsection*{Proof of Theorem 2}
\begin{proof}
        With $\beta_i$'s re-ordered $ |\beta_{(1)}| \ge \ldots \ge |\beta_{(p)}|$,
        we will prove $|\beta_{(j)}|> (\sum_{i=1}^j|\beta_{(i)}|-r)\}/j$ for all $j\le |C|$ and $|\beta_{(j)}|< (\sum_{i=1}^j|\beta_{(i)}|-r)\}/j$ for $j>|C|$. This is equivalent to comparing $(j-1)|\beta_{(j)}|- (\sum_{i=1}^{j-1}|\beta_{(i)}|-r)$ against $0$.
        
        When $j\le |C|$,
        \be
        &       (j-1)|\beta_{(j)}|- (\sum_{i=1}^{j-1}|\beta_{(i)}|-r)\\
        &       = (j-1) (|\theta_{(j)}|+\frac{\mu}{|C|})- \{\sum_{i=1}^{j-1} (|\theta_{(i)}|+\frac{\mu}{|C|})-r\}\\
        & = (j-1) |\theta_{(j)}|- (\sum_{i=1}^{j-1} |\theta_{(i)}|-r)\\
        & >0,
        \ee
        since $\sum_{i=1}^{j-1} |\theta_{(i)}|<r$ for $j\le |C|$.
        
        When $j>|C|+1$,
        \be
        &       (j-1)|\beta_{(j)}|- (\sum_{i=1}^{j-1}|\beta_{(i)}|-r)\\
        &       = (j-1) (t_{(j)}+\frac{\mu}{|C|})- \left\{\sum_{i=1}^{|C|} (|\theta_{(i)}|+\frac{\mu}{|C|})+ \sum_{i=|C|+1}^{j-1} (t_{(i)}+\frac{\mu}{|C|})-r\right\}\\
        &= (j-1) t_{(j)}-\left\{\sum_{i=1}^{|C|} |\theta_{(i)}|+ \sum_{i=|C|+1}^{j-1} t_{(i)}-r\right\}\\
        &       \stackrel{(a)}{=} (j-1) t_{(j)}-  \sum_{i=|C|+1}^{j-1} t_{(i)}\\
        &       \stackrel{(b)}{<} (j-1-|C|) t_{(j)}-  \sum_{i=|C|+1}^{j-1} t_{(i)}\\
        &       = \sum_{i=|C|+1}^{j-1} (t_{(j)}-   t_{(i)})\\
        & \stackrel{(c)}\le 0,
        \ee
        where $(a)$ is due to $\sum_{i=1}^{|C|} |\theta_{(i)}|=r$, $(b)$ due to $t_{(j)}<0$ and $(c)$ due to $|\beta_{(j)}|-\mu/|C|\le |\beta_{(i)}| -\mu/|C| $ for $j>i$.
        
        When $j=|C|+1$, $(j-1)|\beta_{(j)}|- (\sum_{i=1}^{j-1}|\beta_{(i)}|-r) = |C|t_{(j)}<0$.
        
        Therefore, we have $c=|C|$, and it can be verified that
        \be
        \mu_c= \sum_{i=1}^{|C|} (|\theta_{(i)}|+\frac{\mu}{|C|})-r=\mu.
        \ee
\end{proof}

\subsection*{Proof of Theorem 3}

\begin{proof}
        For ease of notation, we denote $\vec t_{\bar C} :=  t_{\sigma_1},\ldots, t_{\sigma_{p-|C|}}$
        $$\begin{aligned}
        \pi_\theta(\theta)=
& \sum_{ \substack{ s_{\sigma_1},\ldots, s_{\sigma_{p-|C|}}\\ \in \{-1,1\}^{p-|C|}}}\int_0^{\infty}\int_{(-\mu/|C|,0)^{p-|C|}} \pi_\beta\{ g(t, s, \mu)\} \textup d \vec t_{\bar C} \textup d \mu\\
        &= 2^{p-|C|}(2\lambda)^{-p}
\int_0^{\infty}\int_{(-\mu/|C|,0)^{p-|C|}}\prod_{i\in C} \left\{e^{-\frac{|\theta_i|+\mu/|C|}{\lambda}}\right\}
        \prod_{i\in \bar C} \left\{e^{-\frac{t_i+\mu/|C|}{\lambda}}\right\} \textup d \vec t_{\bar C} \textup d \mu \\
        & = 2^{p-|C|}(2\lambda)^{-p}
        \prod_{i\in C}\left\{\exp{\left (-\frac{|\theta_i|}{\lambda}\right)}\right\}
        \int_0^{\infty} e^{-\frac{p\mu}{\lambda|C|}}\int_{(-\mu/|C|,0)^{p-|C|}} \prod_{i\in \bar C}e^{-\frac{t_i}{\lambda}}d \vec t_{\bar C} \textup d \mu\\
        & = 2^{p-|C|}(2\lambda)^{-p}
       \exp \left (-\frac{r}{\lambda}\right)
        \cdot \lambda^{p-|C|}\int^\infty_0 e^{-\frac{p\mu}{\lambda|C|}}\left(e^{\frac{\mu}{\lambda|C|}}-1\right)^{p-|C|}d\mu.
        \end{aligned}$$
Let $u= e^{-\frac{\mu}{\lambda|C|}}$, then $\textup d u=- (\lambda|C|)^{-1}e^{-\frac{\mu}{\lambda|C|}} \textup d \mu$, we have
        
        $$\int^\infty_0 e^{-\frac{p\mu}{\lambda|C|}}\left(e^{\frac{\mu}{\lambda|C|}}-1\right)^{p-|C|}d\mu = {\lambda |C|}\int_0^1 u^{|C|-1}(1-u)^{p-|C|}du =\lambda \frac{\Gamma(|C|+1)\Gamma(p-|C|+1)}{\Gamma(p+1)}.$$
        
        Combining the results,
        
        $$\pi_{\theta}(\theta) =   \frac{  (2\lambda)^{-|C|} }{\left(\begin{array}{c}
                p\\|C|
                \end{array}\right)}\lambda \exp \left (-\frac{r}{\lambda}\right)$$
\end{proof}

\subsection*{Proof of Corollary 1}
\begin{proof}
We first focus on when $\|\theta\|_1<r$, since under which, $|C|<p$ happens with probability zero, therefore,
\be
& \pr(|C|=p, \|\theta\|_1<r) = \int_{\mathbb{R}^p} \prod_i (2\lambda)^{-1}\exp(-|\theta_i|/\lambda)\mathbb{I}(\|\theta\|_1 <r)  \textup{d}\theta \\
& = \int_{\mathbb{R}^p_+} \prod_i (\lambda)^{-1}\exp(-x_i/\lambda)\mathbb{I}(\sum x_i <r)  \textup{d}x \\
& \stackrel{(a)}{=} \int_{0}^{r} \frac{1}{\Gamma(p)\lambda^p} y^{p-1}\exp(-y/\lambda)\textup{d}y \\
%& = \frac{1}{\Gamma(p)} \gamma(p, \frac{r}{\lambda}) \\
& \stackrel{(b)}{=} 1- \sum_{j=0}^{p-1} \frac{1}{j!} (\frac{r}{\lambda})^j \exp(-r/\lambda),
\ee
where (a) uses the fact that sum of $p$ iid Exp$(\lambda)$'s is a Gamma$(p,\lambda)$ with $\lambda$ the scale parameter, and (b) uses the CDF formula as $p$ is an integer.

When $\|\theta\|_1=r$, and $|C|=j$, denote the non-zero indices by $\{i_1,\ldots,i_j\}$, note that $x=( |\theta_{i_1}|/r,\ldots, |\theta_{i_j}|/r)$ is on a probability simplex with dimension $(j-1)$, $\Delta^{j-1}$, hence we can use Dirichlet distribution integral $\int_{\Delta^{j-1}}1 \textup{d}x = 1/\Gamma(j)$. We have
\be
&\pr(|C|=j, \|\theta\|_1 = r) =  \frac{  (2\lambda)^{-j} }{\left(\begin{array}{c}
                p\\j
                \end{array}\right)}\lambda \exp \left (-\frac{r}{\lambda}\right)
                2^j\left(\begin{array}{c}
                p\\j
                \end{array}\right) r^{j-1}/\Gamma(j) \\
              &  =  (\frac{  r }{\lambda})^{j-1} \exp \left (-\frac{r}{\lambda}\right)
                 /(j-1)!,
\ee
for $j=1,\ldots,p$.
Combining the above gives the result.
\end{proof}

\subsection*{Proof of Theorem 4}
\begin{proof}
        The compatibility numbers are
        \be
        & \phi(C)=\inf_{\theta}
        \left\{\frac{\|X \theta\|_{2}|C|^{1 / 2}}{\|X\|_{2,\infty}\left\|\theta_{C}\right\|_{1}}:\left\|\theta_{[p]\setminus
        C}\right\|_{1} \leq 7\left\|\theta_{C}\right\|_{1}, \theta_{C} \neq 0\right\},\\
        %& \bar{\phi}(c):=\inf_{\theta} \left\{\frac{\|X \theta\|_{2}\left|C_{\theta}\right|^{1 / 2}}{\|X\|_{2,\infty}\|\theta\|_{1}}: 0 \neq\left|C_{\theta}\right| \leq c\right\},\\
        & \psi(C)=\widetilde{\phi}\left[\left(2+\frac{3}{a+b}+\frac{33}{\phi(C)^{2}} \frac{\lambda^*}{2 \|X\|_{2,\infty}\sqrt{\log p}}\right)|C|\right],\\
        & \bar\psi(C)=\bar{\phi}\left[\left(2+\frac{3}{a+b}+\frac{33}{\phi(C)^{2}} \frac{\lambda^*}{2 \|X\|_{2,\infty}\sqrt{\log p}}\right)|C|\right],\\
        & \widetilde{\phi}(c):=\inf_{\theta} \left\{\frac{\|X \theta\|_{2}}{\|X\|_{2,\infty}\|\theta\|_{2}}: 0 \neq\left|C_{\theta}\right| \leq c\right\}\\
         & \bar{\phi}(c):=\inf_\theta \left\{\frac{\|X \theta\|_{2}\left|C_{\theta}\right|^{1 / 2}}{\|X\|_{2,\infty}|\theta\|_{1}}: 0 \neq\left|C_{\theta}\right| \leq c\right\}.
        \ee

        Our results are based on the early work of \cite{castillo2015bayesian},
        Theorems 1 and 2:
        For a constant $\lambda^*$ and a discrete distribution $g(c)$ supported on $\{0,\ldots, p\}$, when 
        \begin{enumerate}
        \item ${\|X\|_{2,\infty}}/{p} \leq \lambda^{*} \leq 4{\|X\|_{2,\infty}} \sqrt{\log p,}$
        \item There exist constants $a_{1}, a_{2}, a_{3}, a_{4}>0$
        with
        $(2.2) \quad a_{1} p^{-a_{3}}  \leq \frac{g(c)}{g(c-1)} \leq a_{2} p^{-a_{4}}$ for  $c=2, \ldots, p$.
        \end{enumerate}
        Then for a prior kernel of the form
        \be
        \pi_{\theta}(\theta ;  \lambda^{*} ,g) =g(|C|)   \frac{  1 }{\left(\begin{array}{c}
                        p\\|C|
                        \end{array}\right)} (\frac{\lambda^*}{2})^{|C|}\exp \left (-{\lambda^*\|\theta\|_1}
        \right),
        \ee
        with $g(|C|=j)=\pr(|C|=j),$ would enjoy the results in the theorem.
        We now check these two conditions and compute the associated constants.
        % \todo[inline]{Check if the other terms match with our result -> Yes, checks out}
        
        Using the chosen $\lambda$ and $\alpha$, we have
        \be
        \lambda^* =
         \frac{\lambda+\alpha}{ \lambda\alpha } =\|X\|_{2,\infty} \frac{b_{1}p^{b_{2}}
        +p^{b_{3}}}{b_{1}p^{b_{2}} p^{b_{3}}},
        \ee
        Since $b_3\le 1 $, we have $\lambda^* \ge \|X\|_{2,\infty} /p^{b_3} \ge \|X\|_{2,\infty} /p$. Since $b_2> b_3$, for $p$ large enough, 
        $b_{1}p^{b_{2}}> p^{b_{3}}$, hence $\|X\|_{2,\infty} ( b_{1}p^{b_{2}}
        +p^{b_{3}})/ (b_{1} p^{b_{2}} p^{b_{3}})\le 2/p^{b_{3}} \|X\|_{2,\infty} \le 4{\|X\|_{2,\infty}} \sqrt{\log p}$.

        On the other hand, When $c=1,\ldots,p-1$.
        \be
        \frac{g(c)}{g(c-1)}=  (1+\lambda/\alpha)^{-1} 
        % = (1+ \frac{p^{a+b}}{\|X\|_{2,\infty}})^{-1}
        = \frac{1}{1 + b_{1}p^{b_{2}-b_{3}}}.
        \ee
        Clearly, ${g(c)}/{g(c-1)} \le   1/(b_{1}p^{b_{2}-b_{3}}$), satisfying $a_2=1/b_1$ and $a_4=b_2-b_3$. For $p$ large enough $b_{1}p^{b_{2}-b_{3}}>1$, we have $ {g(c)}/{g(c-1)} \ge 1/(2b_{1}p^{b_{2}-b_{3}}) $, satisfying $a_1=1/(2b_1)$
        and $a_{3}=b_2-b_3$. When $c=p$, ${g(c)}/{g(c-1)}= \alpha/\lambda = 1/(b_{1}p^{b_{2}-b_{3}}$), hence also satisfying the above results. Therefore, we apply $a_4=b_{2}-b_{3}$ in the two theorems of \cite{castillo2015bayesian}, and arrive at our results.
        \end{proof}

\subsection*{Proof of Theorem 5}
\begin{proof} 
	Let $\pi_1$ denote the restricted normal distribution under the true model \\ $N(\hat\theta^0_{C_0}, (X_{C_0}^\T X_{C_0})^{-1})\otimes\delta_{[p]\backslash C_0}$, and
	\be  
	\pi_2(\theta) &=  (2\pi)^{(c_0-|C_\theta|)/2}\frac{|X_{C_0}^\T X_{C_0}|^{1/2}}{|X_{C_\theta}^\T X_{C_\theta}|^{1/2}}e^{-\frac{1}{2}\left(\|Y-X_{C_\theta}\theta_{C_\theta}\|^2_2 - \|Y-X_{C_0}\theta_{C_0}\|^2_2+\|X_{C_\theta}\hat\theta_{C_\theta}\|^2_2-\|X_{C_0}\hat\theta_{C_0}\|^2_2\right)},\\
	\pi_3(\theta)&=  \frac{\pi_\theta(\theta)}{\pi_\theta(\theta^0)}. 
	\ee
	%\frac{\otimes\delta_{C_0^c}(0)}{\otimes\delta_{C^c}(0)}
	Thus, the posterior $\pi(\theta \large\mid Y) \propto \pi_1\pi_2\pi_3$.
	Let $A$ be the set  
	\be
	\bigg\{\theta:&\left|C_{\theta}\right|\le c_{0}\left[1+\frac{M}{b_{2}-b_{3}}\left(1+\frac{16}{\phi\left(C_{0}\right)^{2}} \frac{\lambda^{*}}{2\|X\|_{2, \infty} \sqrt{\log p}}\right)\right], \\
	& \text{ and }\left\|\theta-\theta^{0}\right\|_{1}\le \frac{M}{\bar\psi\left(C_{0}\right)^{2}}\frac{c_{0} \sqrt{\log p}}{\|X\|_{2, \infty} \phi\left(C_{0}\right)^2}\bigg\}.\ee  By Theorem 4, $\pi (A\mid Y)\to 1$ under $\mathbb P_{\theta^0}$. 
	We now prove that $\pi_1(A\mid Y)\to 1$ under $\mathbb P_{\theta^0}$ as well. Let $Z \sim N(\hat\theta_{C_0}, (X_{C_0}^\T X_{C_0})^{-1})\otimes \delta_{[p]\backslash C_0}$, then $Z$ immediately satisfies $\left|C_{Z}\right|\le c_{0}\left[1+\frac{M}{b_{2}-b_{3}}\left(1+\frac{16}{\phi\left(C_{0}\right)^{2}} \frac{\lambda^{*}}{2\|X\|_{2, \infty} \sqrt{\log p}}\right)\right]$. Denote $B_n = \frac{M}{\bar\psi\left(C_{0}\right)^{2}}\frac{c_{0} \sqrt{\log p}}{\|X\|_{2, \infty} \phi\left(C_{0}\right)^2}$. We have
	\be \text{pr}\left(\|Z-\theta^0\|_1\le B_n \right) &=  \text{pr}\left(\|Z_{C_0}-\hat\theta_{C_0}+\hat\theta_{C_0}-\theta^0_{C_0}\|_1 \le B_n \right) \\
	&=  \text{pr}\left(\|(X^\T_{C_0}X_{C_0})^{-1} X^\T_{C_0}(\varepsilon+\eta)\|_1 \le B_n \right)\\
	& \ge 
	\text{pr}\left(\text{tr}[(X^\T_{C_0}X_{C_0})^{-1}]\|\varepsilon+\eta \|_2 \le B_n\right)\\
	&\ge \text{pr}(c_0[\tilde\sigma_{\min}(X^\T_{C_0}X_{C_0})]^{-1}\|\varepsilon+\eta \|_2 \le B_n)\\
	& \ge \text{pr}\left( \|\varepsilon+\eta \|_2 \le\frac{M}{\bar\psi\left(C_{0}\right)^{2}} \frac{\sqrt {\log p}}{ \phi\left(C_{0}\right)^2}\frac{\tilde\sigma_{\min}(X^\T_{C_0}X_{C_0}))}{\|X\|_{2,\infty}} \right)\\
	&\ge \text{pr}\left( \|\varepsilon+\eta \|_2 \le\frac{M}{\bar\psi\left(C_{0}\right)^{2}} \frac{\sqrt {\log p}}{ \phi\left(C_{0}\right)^2}a_0 \right) \to 1,
	\ee 
	where $\varepsilon$ and $\eta$ are two independent standard normal vectors in $\mathbb R^{n}$. The second equality holds because both $Z_{C_0}-\hat\theta_{C_0}$ and $\hat\theta_{C_0}-\theta^0_{C_0}$ follows $ N(0,(X^\T_{C_0}X_{C_0})^{-1})$, and the first inequality holds due to Cauchy-Schwarz inequality.
	
	Since the total variation distance between a probability
	measure $\pi (\cdot)$ and its renormalized restriction $\pi _A(\cdot) = \pi (\cdot) / \pi (A)$ is bounded above by $2\pi ^c(A)$, we can replace the two measures in the total variation distance by their renormalized restrictions to $A$. Therefore, it is sufficient to show 
	\be
	&\int_A | \pi_1\pi_2\pi_3 - N(\hat\theta^0_{C_0}, (X_{C_0}^\T X_{C_0})^{-1})\otimes\delta_{[p]\backslash C_0} | d\theta\to 0.
	\ee 
	We now show that  $ \max_{\theta\in A} |\pi_2(\theta)-1|\to 0$ and $ \max_{\theta\in A} |\pi_3(\theta)-1|\to 0$, where we need to firstly prove $\pi(C_\theta= C_0\mid Y) \to 1$ for $\theta\in A$. By Theorem 4, 
	we have
	\be \sum_{C_0\not\subseteq C}\pi (\theta: C_\theta = C \mid y)&	\le \pi(\|\theta-\theta^0\|_1 > \frac{M}{\bar\psi\left(C_{0}\right)^{2}}\frac{c_{0} \sqrt{\log p}}{\|X\|_{2, \infty} \phi\left(C_{0}\right)^2}\mid Y)\to 0.\ee
	
	Therefore $\pi (C_\theta\supseteq C_0\mid Y) \to 1$. In light of assertion 1 in Theorem 4 and the choice of $\lambda$, we have $\pi (|C_\theta|\le c_0\mid Y)\to 1$. Combining the last two conclusion, we have $\pi (C_\theta= C_0\mid Y) \to 1$. Then
	\be
	&|\|Y-X_{C_\theta}\theta_{C_\theta}\|^2_2 - \|Y-X_{C_0}\theta_{C_0}\|^2_2|  \le \|X_{C_\theta\backslash C_0}\theta_{C_\theta\backslash C_0}\|_2^2 + 2(Y-X_{C_0}\theta_{C_0})^\T X_{C_\theta\backslash C_0}\theta_{C_\theta\backslash C_0} \\
	&\le \|X_{C_\theta\backslash C_0}\theta_{C_\theta\backslash C_0}\|_2^2 +2\|\epsilon\|_2\|X_{C_\theta\backslash C_0}\theta_{C_\theta\backslash C_0}\|_2 \to 0 \text{ in probability}.
	\ee
	Since
	$\|X_{C_\theta}\hat\theta_{C_\theta}\|^2_2-\|X_{C_0}\hat\theta_{C_0}\|^2_2$ is the square length of the projection of $Y$ on a subspace of dimension $|C_\theta|-c_0$, this also converge to 0 in probability, hence $\max_{\theta\in A}|\pi_2(\theta)-1|\to 0$.

	Since $$\pi_3(\theta) = (2\lambda)^{c_0-|C_\theta|}\frac{\left(\begin{array}{c}p \\C_0\end{array}\right)}{\left(\begin{array}{c}p \\C_\theta\end{array}\right)}\exp\left[-(1/\lambda + 1/\alpha)(\|\theta\|_1-\|\theta^0\|_1)\right],$$
	and that $|\|\theta\|_1-\|\theta^0\|_1|\le \|\theta-\theta^0\|_1\le \frac{M}{\bar\psi\left(C_{0}\right)^{2}}\frac{c_{0} \sqrt{\log p}}{\|X\|_{2, \infty} \phi\left(C_{0}\right)^2}\to 0$, we have $ \max_{\theta\in A} |\pi_3(\theta)-1|\to 0$.

	Combining the above results, we have
	\be &\int_A | \pi_1\pi_2\pi_3 - N(\hat\theta^0_{C_0}, (X_{C_0}^\T X_{C_0})^{-1})\otimes\delta_{[p]\backslash C_0}  | d\theta\\
	&=\int_A|\pi_1\pi_2\pi_3+\pi_1\pi_2-\pi_1\pi_2-\pi_1|d\theta\\
	&\le \int_A |\pi_1\pi_2 - \pi_1| d\theta + \int_A |\pi_1\pi_2 (\pi_3 - 1)| d\theta\\
	&\le  \max_{\theta\in A} |\pi_2-1|\int_A |\pi_1|d\theta +  \max_{\theta\in A}|\pi_3 - 1|\int_A |\pi_1\pi_2| d\theta \to 0.
	\ee    
\end{proof}

\section{Review of the HMC Algorithm}
For completeness, we now briefly review the Hamiltonian Monte Carlo (HMC) algorithm. To sample the parameter $q=(\beta,\eta,r)$ from target distribution $q\sim\pi_{q\mid y}(\cdot)$, the HMC algorithm takes an auxiliary momentum variable $v$ with density $\pi_v(v)$,  and samples from the joint distribution $\pi(q,v)=\pi_{q\mid y}(q)\pi_v(v)$. The potential energy and kinetic energy are defined as $U(q) = -\log\pi_{q\mid y}(q)$ and $K(v) = -\log\pi_v(v)$, and the total Hamiltonian energy function is denoted by $H(q,v) = U(q)+K(v)$. Our choice of $\pi_v(v)$ is the multivariate Gaussian density  $N(0, I)$, with the kinetic energy $K(v) = v^{\rm T}v/2$. 
 
 At each state $(q,v)$, a new proposal is generated by simulating Hamiltonian dynamics, which satisfy Hamilton's equations:
  \bel
  \label{eq: diff_equation}
  \frac{\partial q}{\partial t} = \frac{\partial H(q,v)}{\partial v} = v,\qquad \frac{\partial v}{\partial t} = -\frac{\partial H(q,v)}{\partial q} = \frac{\partial\log\pi_{q\mid y}(q)}{\partial q}.
 \eel 
The exact solution for (\ref{eq: diff_equation}) is often intractable, but we can numerically approximate
the solution to the differential equations by algorithms such as the leapfrog scheme. The leapfrog is a reversible and volume-preserving integrator, which updates the evolution $(q^{(t)},v^{(t)})\to (q^{(t+\epsilon)},v^{(t+\epsilon)})$ via
\bel
v\leftarrow v+\frac{\epsilon}{2} \frac{\partial\log\pi_{q\mid y}(q)}{\partial q},\quad q\leftarrow  q+\epsilon v,\quad v\leftarrow v+\frac{\epsilon}{2} \frac{\partial\log\pi_{ q\mid y}( q)}{\partial q}.
\eel 
The proposal $( q^{*},v^{*})$ is generated by taking $L$ leapfrog steps from current state $( q^{(0)},v^{(0)})$, then accepted using the Metropolis-Hastings adjustment, with the acceptance probability:
 \be
 \min \{1, \exp[-H( q^{*},v^{*})+H( q^{(0)},v^{(0)})]\}.
 \ee
For the step size $\epsilon$ and the leapfrog steps $L$, we use the No-U-Turn Sampler \citep{hoffman2014no} to automatically adapt these working parameters. 

When an $l_1$-ball projection has a closed form, its gradient would have a closed form as well. For example, for the vector-norm $l_1$-ball projection with $\theta_i = \text{sign}(\beta_i)(|\beta_i|-\tilde \mu)_+$, the gradient is $\partial \theta_i /\partial \beta_i = (1-1/c)  \mathbb{I}(|\beta_i|> \tilde \mu)$, $\partial \theta_i /\partial \beta_j = (-1/c)  \mathbb{I}(|\beta_i|> \tilde \mu)  \mathbb{I}(|\beta_j|> \tilde \mu)$ for $i\neq j$. In practice, the gradient calculation is conducted via the auto-differentiation framework. Many other $l_1$-ball projections have closed forms, including those for rank selection or group sparsity; Chapter 6 of \cite{beck2017first} contains many useful examples. 

On the other hand, when the projection lacks a closed form and requires an iterative algorithm for its calculation, we need to numerically evaluate its gradient. When $\beta$ is in a low-dimensional space, we use finite difference approximation for the $j$-th entry: $\partial P_{\mathbb{B}_{h,r}}(\beta)/\partial \beta_j\approx  [P_{\mathbb{B}_{h,r}}(\beta+\varepsilon e_j)-P_{\mathbb{B}_{h,r}}(\beta)] /\varepsilon$, where $e_j$ is the $j$-th standard basis and $\varepsilon>0$.  When $\beta$ is in a high-dimensional space, to avoid the high cost of evaluating the projection for $(p+1)$ times, we use the simultaneous perturbation stochastic approximation  \citep{spall1992}, which reduces the times of projection evaluation to $(m+1)$:
$$  \frac{ \partial P_{\mathbb{B}_{h,r}}(\beta)}{\partial \beta_j}\approx\frac{1}{m}\sum_{k=1}^m   [P_{\mathbb B_{h,r}}(\beta+\varepsilon\Delta^{(k)})-P_{\mathbb{B}_{h,r}}(\beta)] /(\varepsilon\Delta_j^{(k)}),$$
where $\Delta^{(k)} = \{\Delta^{(k)}_1,\ldots,\Delta^{(k)}_p\}$ has each $\Delta^{(k)}_j\in\{-1, 1\}$ independently
generated from Rademacher distribution. It is worth clarifying that, even when an approximate gradient is used, the HMC algorithm satisfies the detailed balance condition thanks to the Metropolis-Hastings adjustment. Therefore, the accuracy of the approximate gradient would only impact the acceptance rate and not the invariant distribution of the Markov chains. Empirically, we find that $m=20$ and $\varepsilon=10^{-5}$ achieve good acceptance rate.

\section{Benchmark of Algorithms on Sampling from Spike-and-Slab Posterior}

	\begin{table}  
\parbox[t]{\linewidth}{\caption{\label{table: compare_algo}Comparing running time and effective sample size.} }\\
	\begin{subtable}[t]{0.48\textwidth}
		\centering
		\begin{tabular}{*{3}{c}}
		 \toprule
			{$(n, p, c_0)$} &{ $l_1$-ball-HMC } &  {SS-Gibbs} \\ \midrule
			{(200, 500, 25)}  & {216.92}      & {183.30}   \\ 
			{(500, 1000, 50)} &  {238.82} &{ 2621.69}                \\ \bottomrule
		\end{tabular}
		\subcaption{Running time (in seconds) for 1,000 iterations of Markov Chain. SS-Gibbs runs very slowly when $p$ is large.}
	\end{subtable} 
	\begin{subtable}[t]{0.48\textwidth}
		\centering
		\begin{tabular}{*{3}{c}} 
			\toprule
			{$(n, p, c_0)$} &{ $l_1$-ball-HMC } &  {SS-Gibbs} \\ \midrule
			{(200, 500, 25)}  & {11.30$\%$}     &  {2.45$\%$}   \\ 
			{(500, 1000, 50)} & { 3.68$\%$}     &  {0.8$\%$}                \\\bottomrule
		\end{tabular}
		\subcaption{Effective sample size of $|C|$ in the Markov chains.}
	\end{subtable}
\end{table}   
As the spike-and-slab prior can be written as a special case of $l_1$-ball prior, we compare the computational efficiency using the Hamiltonian Monte Carlo (henceforth named the $l_1$-ball-HMC) with the one using the Gibbs sampling algorithm [henceforth named the SS-Gibbs]. The latter is implemented as a Gibbs sampler that draws each variable inclusion indicator given the parameters and then draws the parameter given the indicators.

{To clarify, for a linear model under a normal likelihood and conjugate priors for coefficients $\theta$ and variance $\sigma^2$, one could integrate out the values of $\theta$ and $\sigma^2$, and obtain a marginal posterior on $b_i=\mathbb{I}(\theta_i \neq 0)$ for $i=1,\ldots,p$. This leads to the stochastic search variable selection algorithm \citep{george1995stochastic}, which enjoys excellent mixing performance.
On the other hand, since in this article we  compare with the general class of spike-and-slab priors that (i) may not necessarily lead to posterior conjugacy, and (ii) may be used in non-linear models, we use the Gibbs sampler that updates all parameters $(\theta, \sigma^2)$ without any marginalization.
}

 We use linear models $y \sim N(X\theta^0, \sigma^2I_n)$ with different $(n,p)$ and let the number of non-zeros $c_0$ to be $0.05p$. The entries in $X$ are iid standard normal. We let the non-zero entries  be 1 and set $\sigma = 0.1$. We compare running time per thousand iterations and examine the mixing performance (the ability of each Markov Chain to explore alternative high-probability models and the effective sample size) in each setting.  As shown in Table \ref{table: compare_algo} (a), SS-Gibbs requires much longer running time at $n=500$ and $p=1000$. The combinatorial search in the variable indicator causes heavy burden in computation  (2622 seconds for 1000 iterations). Meanwhile, the $l_1$-ball-HMC is less affected by the increase in dimension. 

\begin{figure}[H]
	\centering
		\begin{subfigure}[htbp]{7cm}
			\centering
			\includegraphics[height=5cm]{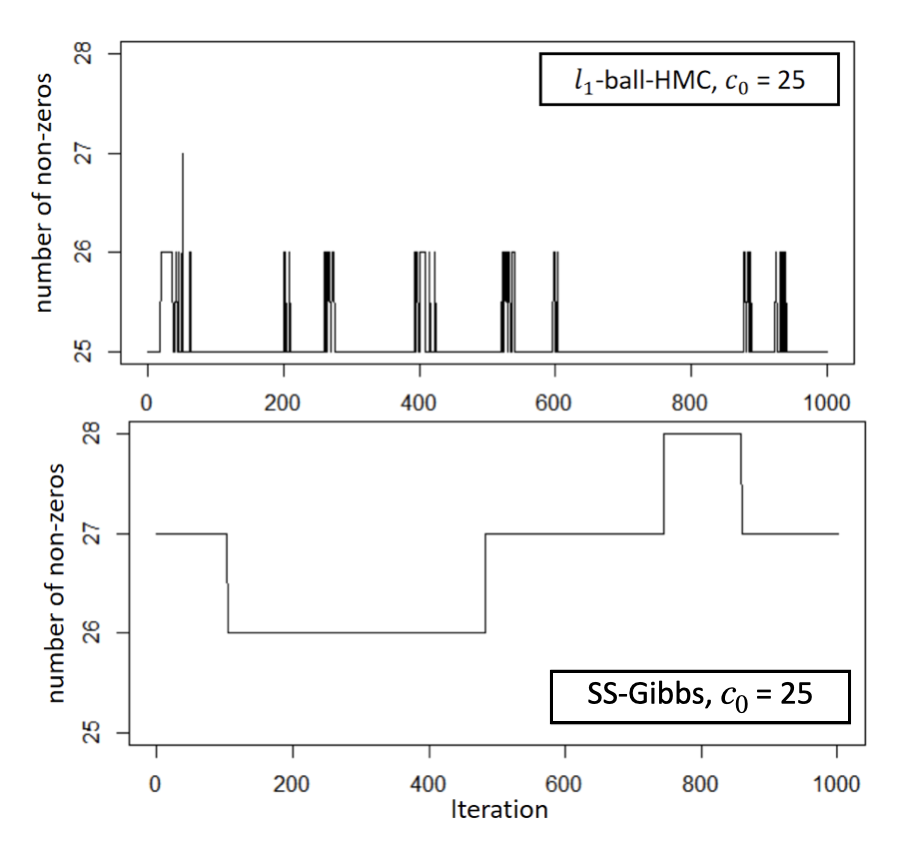}
			\caption{Traceplot of number of nonzeros ($|C|$) in the posterior when $c_0 = 25$.}
		\end{subfigure}		
		\begin{subfigure}[htbp]{7cm}
			\centering
			\includegraphics[height=5cm]{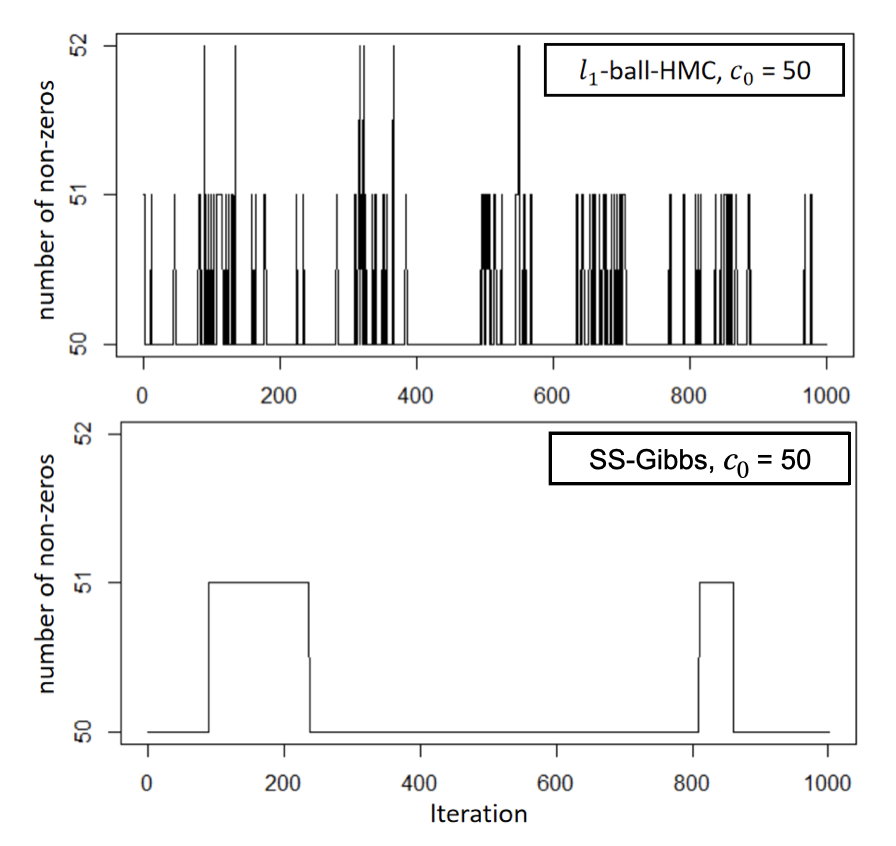}
			\caption{Traceplot of number of nonzeros ($|C|$) in the posterior when $c_0 = 50$.}
		\end{subfigure}	 
		\begin{subfigure}[htbp]{7cm}
			\centering
			\includegraphics[height=5cm]{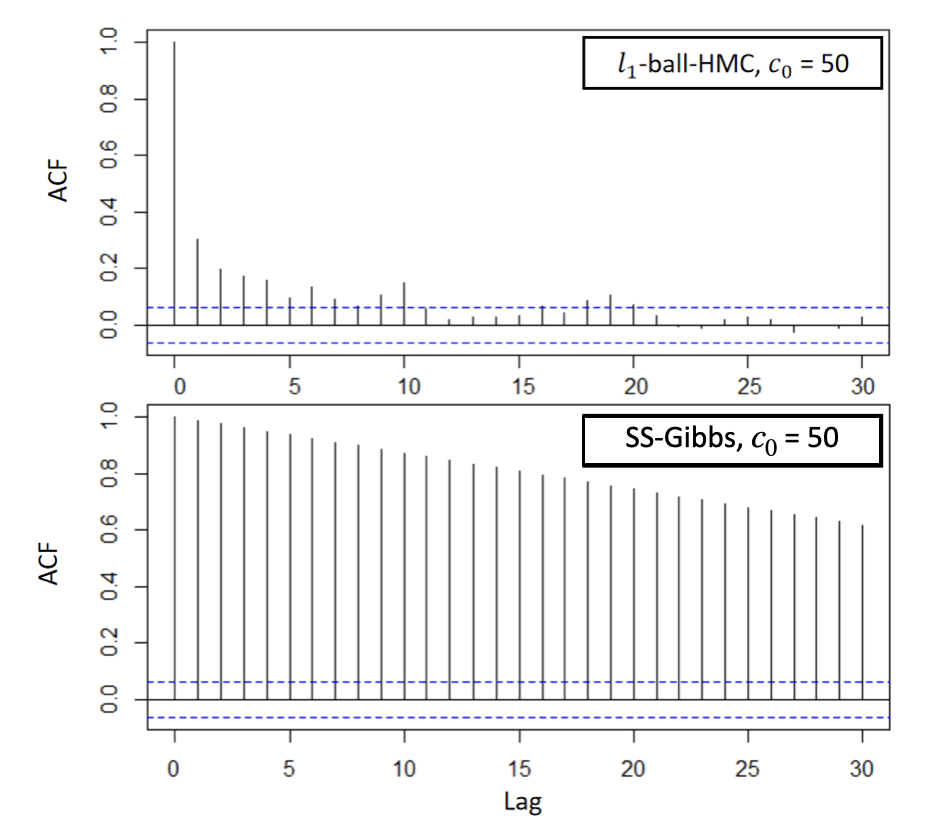}
			\caption{ACF plot of $|C|$ in the posterior when $c_0=25$.}
		\end{subfigure}
		\begin{subfigure}[htbp]{7cm}
			\centering
		\includegraphics[height=5cm]{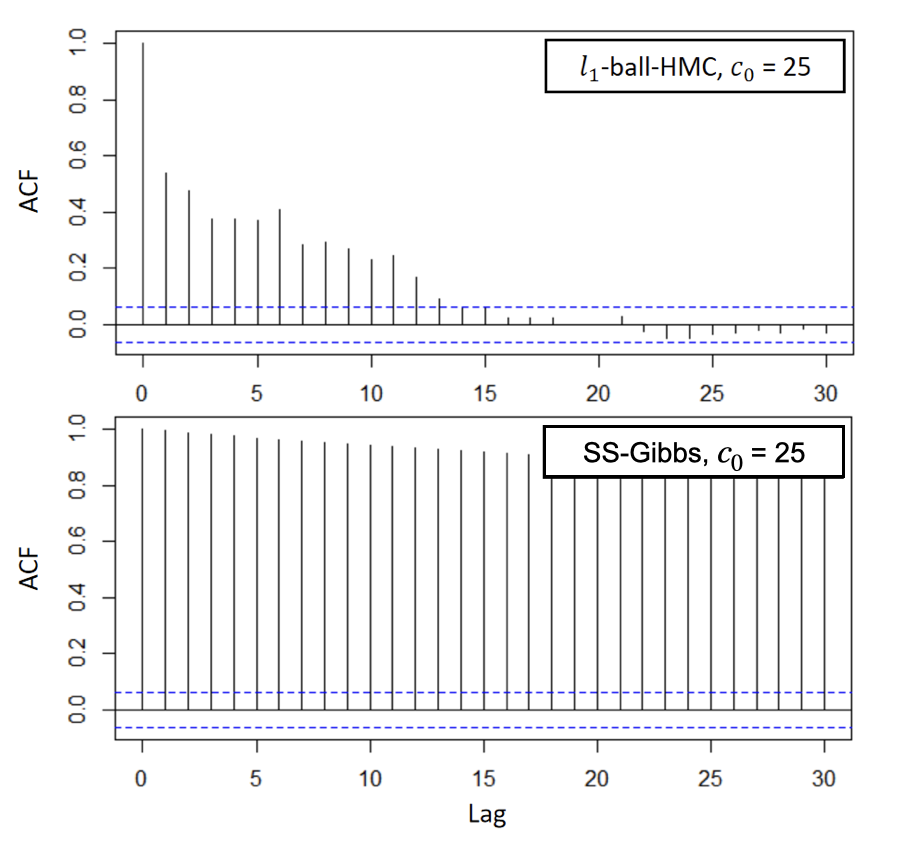}
		\caption{ACF plot of $|C|$ in the posterior when $c_0 = 50$.}
		\end{subfigure}		
		\caption{\label{fig: compare_sns}Comparing mixing performance of the $l_1$-ball-HMC (upper rows) with the SS-Gibbs (lower rows) in $n=200, p=500, c_0=25$ and $n=500,p=1000,c_0 = 50$. Under both settings, the $l_1$-ball-HMC enjoys faster mixing with rapid changes between different numbers of non-zeros $|C|$, while SS-Gibbs tends to be stuck at the same $|C|$ for a long time.}
	\end{figure}

On the mixing performance, Figure \ref{fig: compare_sns} shows that $l_1$-ball-HMC can quickly transition between states with different numbers of non-zeros $|C|$, whereas SS-Gibbs suffers from slow mixing with only a few changes in 1,000 iterations. This is reflected in \ref{table: compare_algo}(b) as the effective sample size from $l_1$-ball-HMC is almost an order higher than the one from SS-Gibbs.

\section{Benchmark in Linear Regression}

        As discussed in the theory section, in the linear regression, the recovery of $\theta^0$  requires some conditions on the cardinality of the true parameter $c_0$, and the sample size $n$. We now use numerical simulations to empirically estimate the sparsity detection limits and the required minimum sample size.
        
\begin{figure}[H]
                \centering
                \begin{subfigure}[t]{0.24\textwidth}
                        \centering \includegraphics[width=1.1\textwidth]{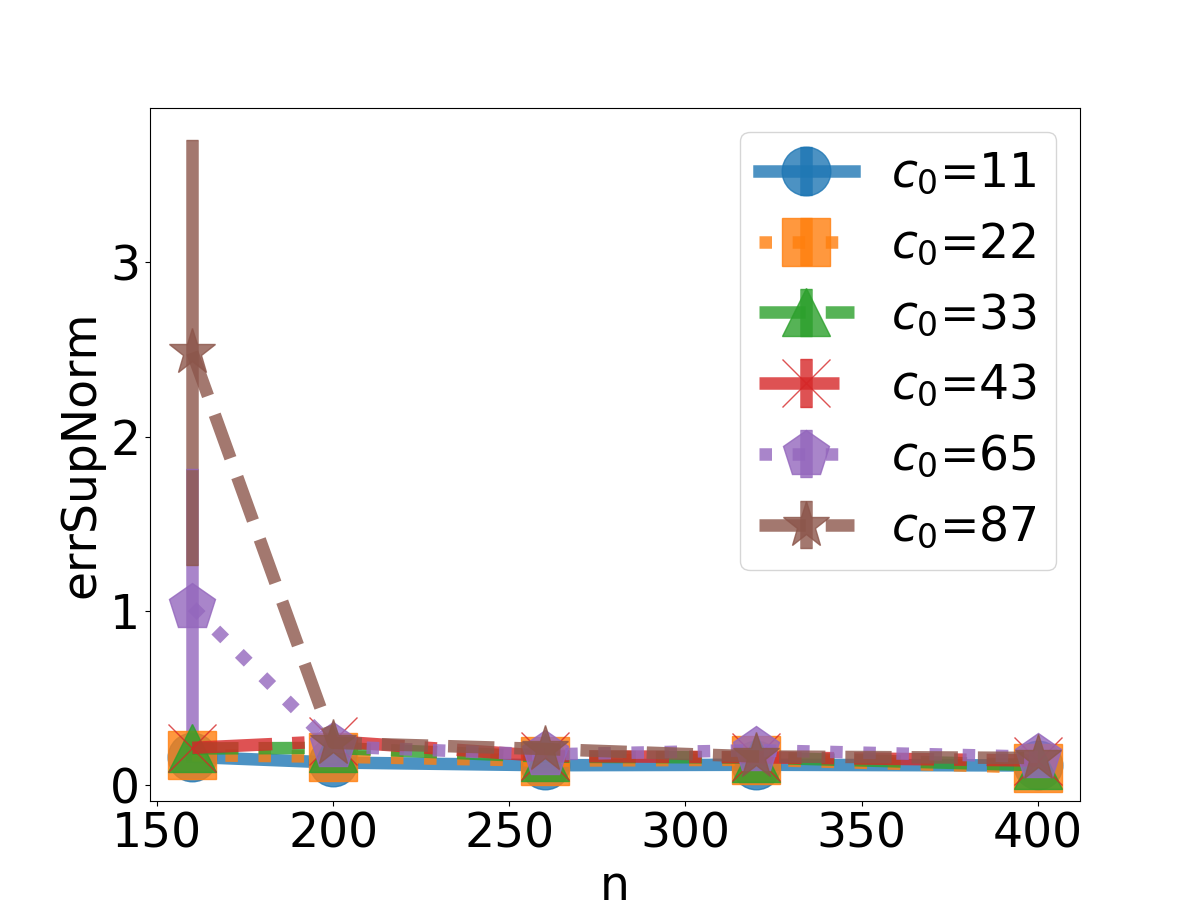}
                        \caption{$p=200$.}
                \end{subfigure}
                \begin{subfigure}[t]{0.24\textwidth}
                        \centering
                        \includegraphics[width=1.1\textwidth]{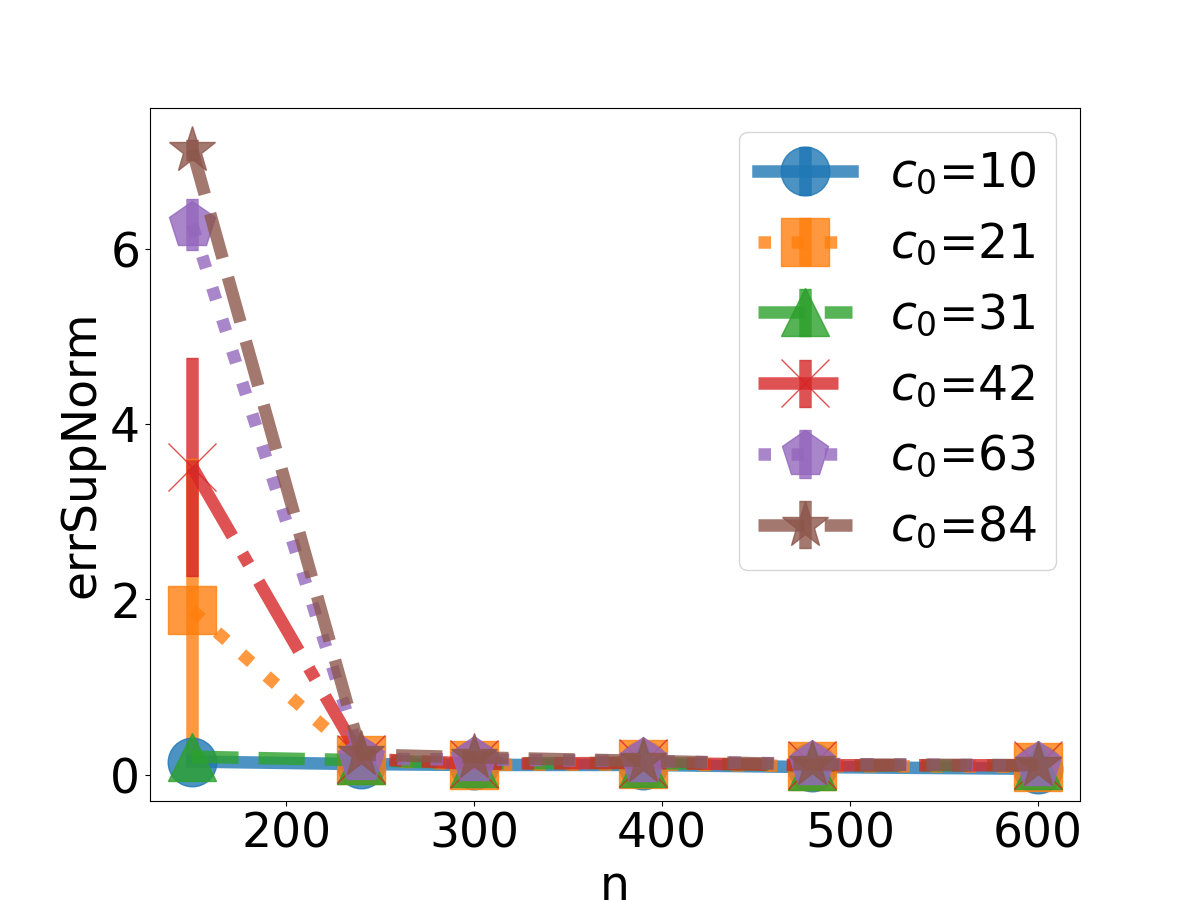}
                        \caption{$p=300$.}
                \end{subfigure}
                \begin{subfigure}[t]{0.24\textwidth}
                        \centering
                        \includegraphics[width=1.1\textwidth]{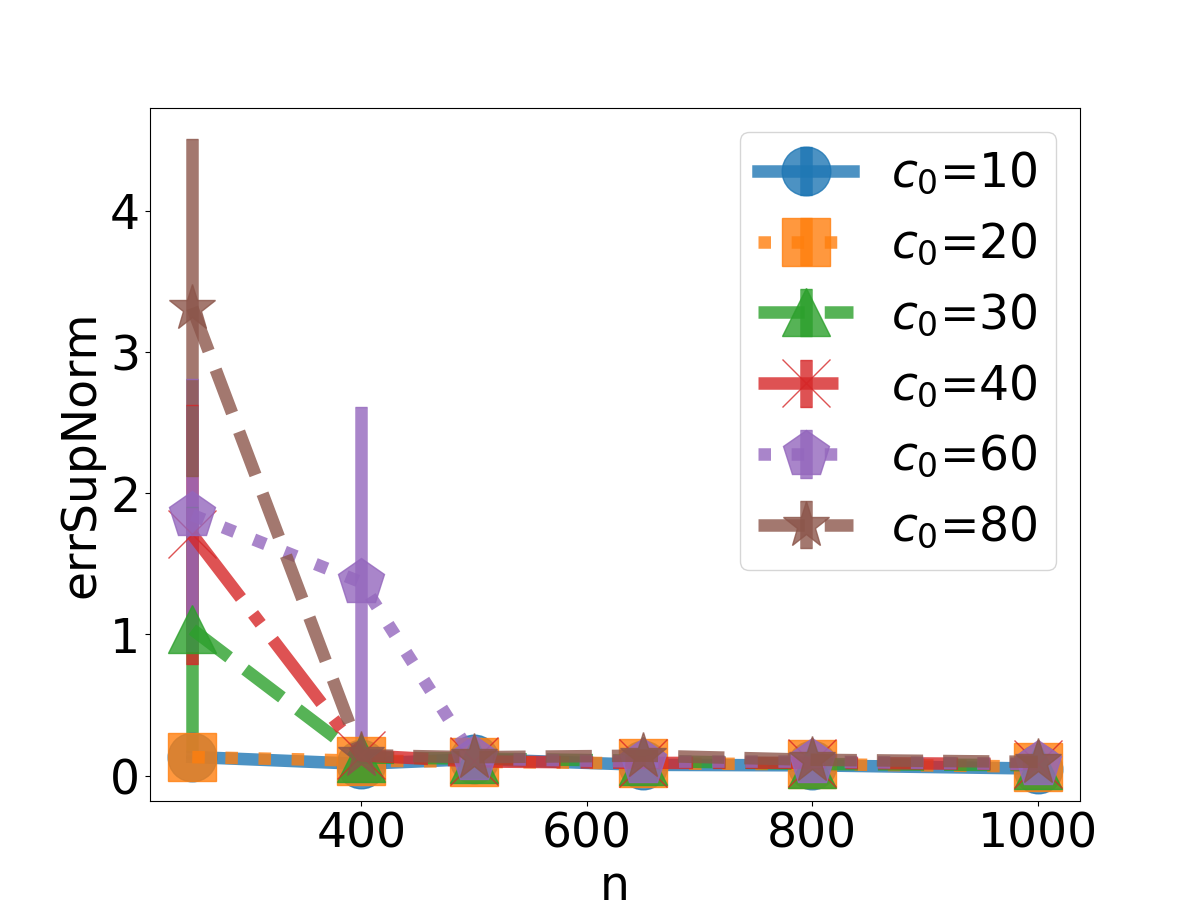}
                        \caption{$p=500$.}
                \end{subfigure}
                \begin{subfigure}[t]{0.24\textwidth}
                        \centering
                        \includegraphics[width=1.1\textwidth]{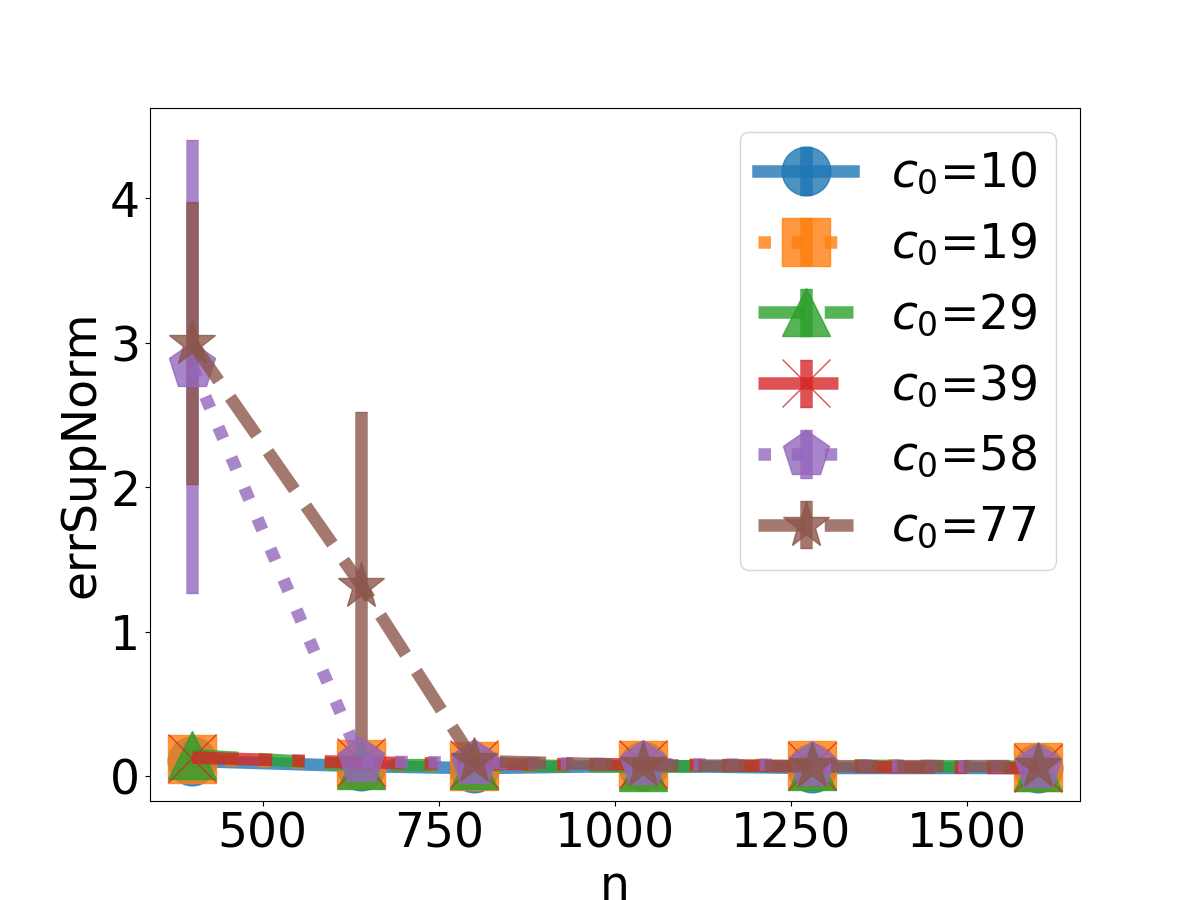}
                        \caption{$p=800$.}
                \end{subfigure}
                
                \caption{ The simulation in sparse regression shows the $l_1$-ball prior can correctly recover $\theta^0$ in  $l_\infty$ norm when $c_0 \lesssim 2 \sqrt{n/\log p}$.}
                \label{fig: errSupReg_p}
        \end{figure}

        For each experiment setting,  the $n$ rows of the design matrix $X\in\mathbb R^{n\times p}$ are independently drawn from $N(0, I_p)$. We generate the true $\theta^0$ according to the level of sparsity, where the non-zero entries are drawn from $N(5,1^2)$. We experiment with  $p=200,300,500$ and $800$, with $n$ being a multiple of $p$ and $c_0$ set to $25,50,75,100,150,200$ times $\sqrt{1/\log p}$, as corresponding to different degrees of sparsity. To be consistent with the theory result on regression, we benchmark the sup-norm $\sup_i\|\hat\theta_i-\theta^0_i\|$ between the posterior mean $\hat\theta$ and the oracle $\theta^0$. We plot the results in Figure~\ref{fig: errSupReg_p}, and make a few observations: (i) when $n \ge p$, all settings have low estimation errors close to zero; (ii) when $n<p$,  we have good result roughly when $c_0\le 2{\sqrt{n/\log p}}$. This range is coherent with our theoretic analysis.
        
        % Second, we experiment on change point detection. Note that in these settings we have square design matrix, i.e., $n=p$. We let $p$ vary from $100, 200, 300, 500$ and $800$ and the sparsity level $c_0$ be a multiple of ${\sqrt{p/\log p}}$. Figure~\ref{fig:errSupCGPL1} shows that the $l_1$-ball prior works  well in change point detection $c_0<2 \sqrt{p/\log p}$ as well.
        % \begin{figure}[H]
        %         \centering
        %         %        \begin{subfigure}[t]{.48\textwidth}
        %         \includegraphics[width = .5\textwidth]{simulationFig/errSupCGP}
        %         %                \caption{Change point detection.}
        %         %        \end{subfigure}
        %         %        \begin{subfigure}[t]{.48\textwidth}
        %         %                \includegraphics[width = .9\textwidth]{simulationFig/errSupL1}
        %         %                \caption{Linear trend filter.}
        %         %        \end{subfigure}
        %         \caption{       \label{fig:errSupCGPL1}The $l_1$-ball prior can recover $\theta^0$ in $l_\infty$   norm in change point detection. }%  and (b) linear trend filter $c_0 \lesssim 2\sqrt{p/\log p}$.}
        % \end{figure}
        
        Next, we compare the performance of the $l_1$-ball prior with the Bayesian lasso and horseshoe priors, over a range of different $p$, $n$ and $c_0$. We benchmark using the canonical normal means (hence $n=p$) problem $y_i = \theta_i+\epsilon_i$ for $i=1,\ldots,n$, and $\epsilon_i\stackrel{\text{iid}}\sim N(0,1^2)$, and compare with two Bayesian continuous priors: the horseshoe \citep{carvalho2010horseshoe} and the Bayesian lasso \citep{park2008bayesian}, implemented through \texttt{R} packages \texttt{horseshoe} and \texttt{monomvn}.  The true non-zero entries in $\theta$ are drawn from $N(10,1^2)$. We consider $n=200$ and $500$. For each $n$, we let the true cardinality be $c_0=5,10$ and $20$.  For the $l_1$-ball prior, we choose  $\beta_i \sim \text{DE}(0,\lambda_i \sigma)$ with scale $\lambda_i\stackrel{\text{iid}}\sim \text{Exp}(1)$, and $r\sim \text{Exp}(10)$. For the Bayesian Lasso, we choose prior $\theta_i \stackrel{\text{iid}}\sim \text{DE}(0, 0.5\sigma)$. For the horseshoe prior, we use the default scaled half-Cauchy $C^+(0,\sigma)$ prior for the global scale $\tau$. For all three methods, we use the Jeffreys prior for $\sigma^2$, $\pi_{\sigma^2}(\sigma^2)\propto \sigma^{-2}$. We run 10,000 MCMC steps for each model, and discard the first 5,000 steps as burn-in.

We use the posterior mean $\hat\theta$ to compute the mean squared error,  $||\hat\theta-\theta^0||_2^2$. We also compute the estimated cardinality $\hat c$ based on $\hat \theta$. Since the continuous shrinkage priors do not produce exact zero, we adopt the strategy in \cite{carvalho2010horseshoe}, the $i$-th entry  is viewed as non-zero if $|\hat\theta_i/y_i| > 0.5.$

\begin{figure}[H]
        \centering
        \begin{subfigure}[t]{0.35\textwidth}
                \centering
                \includegraphics[width=.85\textwidth]{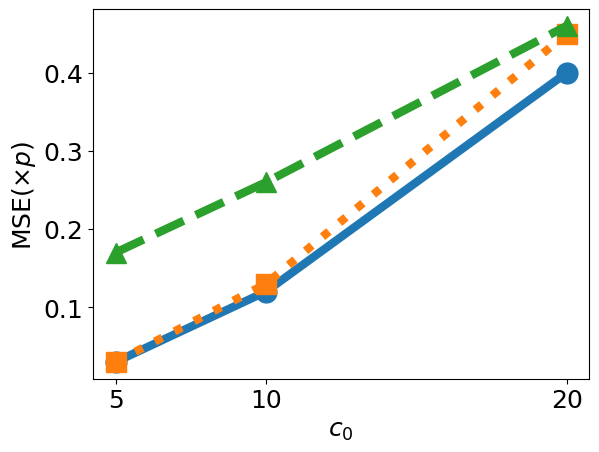}
                \caption{The mean squared error across $c_0=5,10,20$ when $n=200$.}
        \end{subfigure}
        \begin{subfigure}[t]{0.35\textwidth}
                \centering
                \includegraphics[width=.85\textwidth]{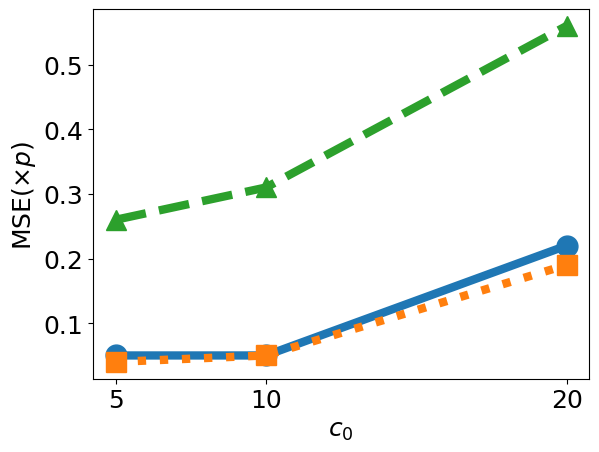}
                \caption{The mean squared error across $c_0=5,10,20$ when $n=500$.}
        \end{subfigure}
        \begin{subfigure}[t]{0.15\textwidth}
        \centering
        \includegraphics[width=.8\textwidth]{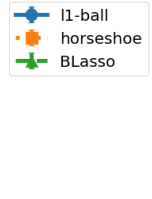}
\end{subfigure}

        \begin{subfigure}[t]{0.35\textwidth}
                \centering
                \includegraphics[width=.85\textwidth]{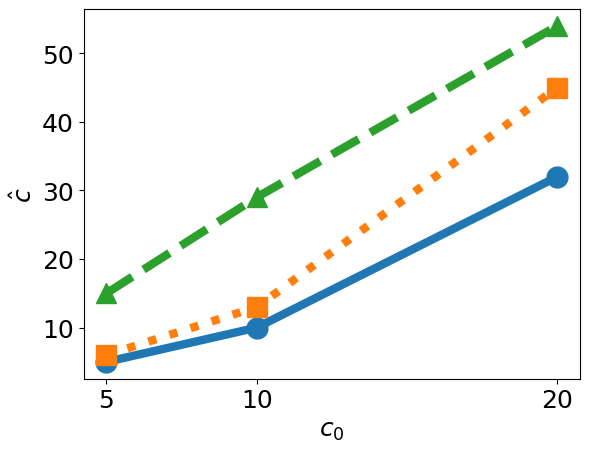}
                \caption{The estimated cardinality across $c_0=5,10,20$ when $n=200$.}
        \end{subfigure}
        \begin{subfigure}[t]{0.35\textwidth}
                \centering
                \includegraphics[width=.85\textwidth]{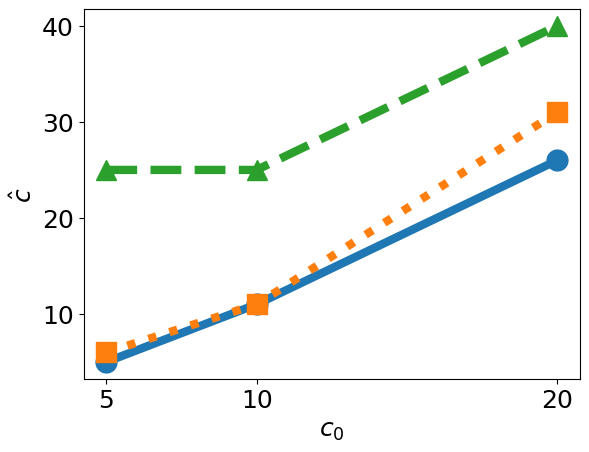}
                \caption{The estimated cardinality across $c_0=5,10,20$ when $n=500$.}
        \end{subfigure}
        \begin{subfigure}[t]{0.15\textwidth}
        \centering
        \includegraphics[width=.8\textwidth]{simulationFig/legend}
\end{subfigure}
        \caption{Comparing the $l_1$-ball, the horseshoe and the Bayesian lasso in different dimensionality and sparsity levels. The $l_1$-ball method gives a satisfactory estimation of the cardinality.}
        \label{fig: study1}
\end{figure}

As shown in Figure~\ref{fig: study1}, Panel (a) and (b), the $l_1$-ball  and the horseshoe methods are comparable in parameter estimation, with smaller errors than the Bayesian lasso. Panel (c) and (d) show that the $l_1$-ball method gives a satisfactory estimation of the cardinality, and the horseshoe produces similarly good results although with a slight over-estimation (which could be reduced with some further tuning on its global scale parameter).
 
\section{Comparison with Post-processing Methods}

\subsection{Methodological Comparison with Post-processing Methods}
There have been some works that post-process the posterior samples when modeling under continuous priors.  \cite{bondell2012consistent}  use a conjugate continuous prior for $\theta$ (without imposing any shrinkage apriori) to first estimate high posterior density region associated with $(1-\alpha)$ probability, within which they extract a sub-region by minimizing the $\|\theta\|_0$ ($l_0$-norm of $\theta$). Similarly, \cite{hahn2015decoupling} use a shrinkage prior for $\theta$ and estimate the posterior mean and variance of $\theta$, then find a summary point estimate $\hat \theta$ that minimizes a  loss function consisting of squared prediction error and $l_0$-norm. \cite{li2017variable} use a shrinkage prior for $\theta$, then use $2$-means to cluster each posterior sample into two groups, with the goal of finding a point estimate on the number of non-zero $\theta_j$'s as represented by the size of one cluster.

In the scope of variable selection in regression, those post-processing approaches produce a point estimate $\hat\theta$ \citep{hahn2015decoupling, li2017variable}, or a set of sparse $\theta$ associated with zero posterior probability \citep{bondell2012consistent}. In a diagram, those estimates are produced from two stages,
$$(i) \; \theta \sim \frac{\mathcal L(y;\theta)\pi_\theta(\theta)}{\int \mathcal  L(y;\theta)\pi_\theta(\theta) \textup{d}\theta}, \qquad (ii) \; \theta^*=T(\theta),$$
for some transformation $T$. While being well motivated for other purposes (such as ease of interpretation), the key issue is in the lack of a tractable probabilistic characterization on the transform $T$. As a result, these approaches cannot be used for uncertainty quantification, such as estimating the $(1-\alpha)$-credit interval on $\|\theta\|_0$ and $(1-\alpha)$-prediction interval for $x^{* \rm T}\theta$ for a new $x^*$. 

On the other hand, our proposed $l_1$-ball and generalized $l_1$-ball priors are fully Bayesian. The projection $\theta= P_{\mathbb{B}}(\beta)$ induces a proper combinatorial prior with positive probability in either $\theta$ or some transformation of $\theta$. With a likelihood function of $\theta$, one could estimate the canonical posterior distribution $\Pi(\theta \mid y)$ and conduct uncertainty quantification via standard Bayesian procedures. 
	The key idea is using $l_1$-ball projection $\theta= P_{\mathbb{B}}(\beta)$ as a many-to-one mapping, to {\em reparameterize} $\theta$. Similarly, in a diagram, our modeling framework is
	$$  \theta \sim \frac{\mathcal L(y;\theta)\pi_\theta(\theta)}{\int \mathcal L(y;\theta)\pi_\theta(\theta) \textup{d}\theta} , \qquad \text{where }\theta=P_{\mathbb B}(\beta).$$
In the above, optimization algorithms serve as means to compute such a reparameterization, and are only required when $P_{\mathbb B_{h,r}}(\beta)$ do not have closed-form solution (in the generalized $l_1$ cases).

\subsection{Numerical Comparison on Point Estimates}

Besides the difference in the capability of uncertainty quantification,
we further show performance differences in the accuracy of point estimates.

We extract zero/non-zero labels from Fr\'echet mean generated by the $l_1$-ball and point estimates generated by the joint set penalized credible regions method (PCR) \citep{bondell2012consistent} and the sequential 2-Means clustering (S2M) method \citep{li2017variable}. We then compare their false positive rates and false negative rates.

				The data are generated from $y \sim N(X\theta^0,\sigma^2I)$. We fix $\sigma=1$, and experiment with different settings of $n,p,c_0$ and signal strength in the non-zeros ($\theta_{C_0}$). We test with both independent design matrix $X_{i,j} \stackrel{\text{iid}}\sim N(0,1^2)$ and correlated design matrix with each row of $X$ from $N(0,I)$ or $N(0,\Sigma)$, $\Sigma_{j,k} = 0.5^{|j-k|}$. For the PCR, we obtain 5,000 MCMC samples from the conjugate Gaussian prior and use the joint credible set results. For the S2M,  we first obtain 5,000 MCMC samples under a horseshoe prior \citep{carvalho2010horseshoe}, then apply the sequential 2-means algorithm. The comparison results are listed in Table \ref{table: compare_post-processing} and \ref{table: compare_post-processing1}. The three methods are comparably good with large sample size, but the $l_1$-ball prior method outperforms the post-processing methods in small $(n,p)$ cases.

% \FloatBarrier
\begin{table}
	\parbox[t]{\linewidth}{\caption{\label{table: compare_post-processing}False positive rate and false negative rate under independent design matrix.}}\\
		\begin{tabular}{*{4}{l}}
			\toprule
			{$(n, p, c_0, \theta_{C_0})$} &{ $l_1$-ball} &  S2M & PCR \\ \midrule
			{(50, 300, 10, 5)}  & \textbf{(0.0, 0.0)}  &  (0.020, 0.60)   &  (0.021, 0.80)   \\ 
			{(50, 300, 10, 10)} &  \textbf{(0.0, 0.0)}   & (0.013, 0.30) & (0.020, 0.60)      \\
			{(50, 300, 30, 10)}  & (0.085,0.77)  & \textbf{(0.081, 0.73)} & (0.104, 0.93) \\
			{(200, 1000, 10, 5)}  & \textbf{(0.0, 0.0)}  &  \textbf{(0.0, 0.0)}  & ({0.01}, 0.10)     \\ 
			{(200, 1000, 10, 10)} &  \textbf{(0.0, 0.0)}   &  \textbf{(0.0, 0.0)} & (0.003, 0.30)      \\
			{(200, 1000, 50, 10)}  & \textbf{(0.017, 0.34)}  &(0.025, 0.48) &  (0.043, 0.82) \\
			\bottomrule
		\end{tabular}
		\end{table}
		\vfill
\begin{table}
	\parbox[t]{\linewidth}{
		\caption{\label{table: compare_post-processing1}False positive rate and false negative rate under correlated design matrix.}}\\
		\begin{tabular}{*{4}{l}}
			\toprule
				{$(n, p, c_0, \theta_{C_0})$} &{ $l_1$-ball} &  S2M & PCR \\ \midrule
			{(50, 300, 10, 5)}  &\textbf{(0.0, 0.0)}  &\textbf{(0.0, 0.0)} & (0.010, 0.30)    \\ 
			{(50, 300, 10, 10)} &\textbf{(0.0,0.0)}  & \textbf{(0.0, 0.0)}   &  (0,006, 0.20)  \\
			{(50, 300, 30, 10)}  &\textbf{(0.037, 0.33)}  & (0.044, 0.40) & (0,070, 0.63)\\
			{(200, 1000, 10, 5)}  &\textbf{(0.0, 0.0)}  &  \textbf{(0.0, 0.0)}  &   \textbf{(0.0, 0.0)} \\ 
			{(200, 1000, 10, 10)} &\textbf{(0.0, 0.0)}   & \textbf{(0.0, 0.0)}  &  \textbf{(0.0, 0.0)} \\
			{(200, 1000, 50, 10)}  &\textbf{(0.001, 0.02)}  &\textbf{(0.001, 0.02)}   &  (0.013, 0.26) \\
			\bottomrule
		\end{tabular}
\end{table}
%\FloatBarrier
\section{Change Point Detection}
 
To show that the exact zeros are essential for the success of the $l_1$-tricks, we first experiment with  a change point detection model and compare the results with the  continuous shrinkage prior.

  We use the simulated data with $y_t \mid \mu_t \sim N(\mu_t,\sigma^2)$ for $t=1,\ldots,100$, where  
  $\mu_t$ is piecewise constant from $\{30, 10, 40,20\}$ with three change points at $t\in\{20,40,80\}$. In order to compare with the continuous shrinakge prior, we re-parameterize this as a linear regression problem using $\theta_t = \mu_t-\mu_{t-1}$:
    \be
        y_t = \sum_{i=1}^t\theta_i +  \epsilon_t, \quad\epsilon_t\sim N(0,\sigma^2),\quad t=1,\ldots,100,
        \ee
        where we use $\sigma^2=10$ during the data generation. This enables us to impose sparsity  on $\theta_t$, as the curve is a flat line in $[t,t+d]$ if $\theta_t=\theta_{t+1}=\ldots=\theta_{t+d}=0$, and nonzero values only occur at the sudden changes. We use the $l_1$-ball prior on $\theta_i$; to compare, we also test the model with a horseshoe prior on $\theta_i$ \citep{carvalho2010horseshoe}. In both cases, we use the Jeffreys prior $\pi_{\sigma^2}(\sigma^2)\propto 1/\sigma^2$.

\begin{figure}[H]
                \centering
                \begin{subfigure}[t]{0.95\textwidth}
                        \centering
                        \includegraphics[width=0.4\textwidth, height = .18\textwidth]{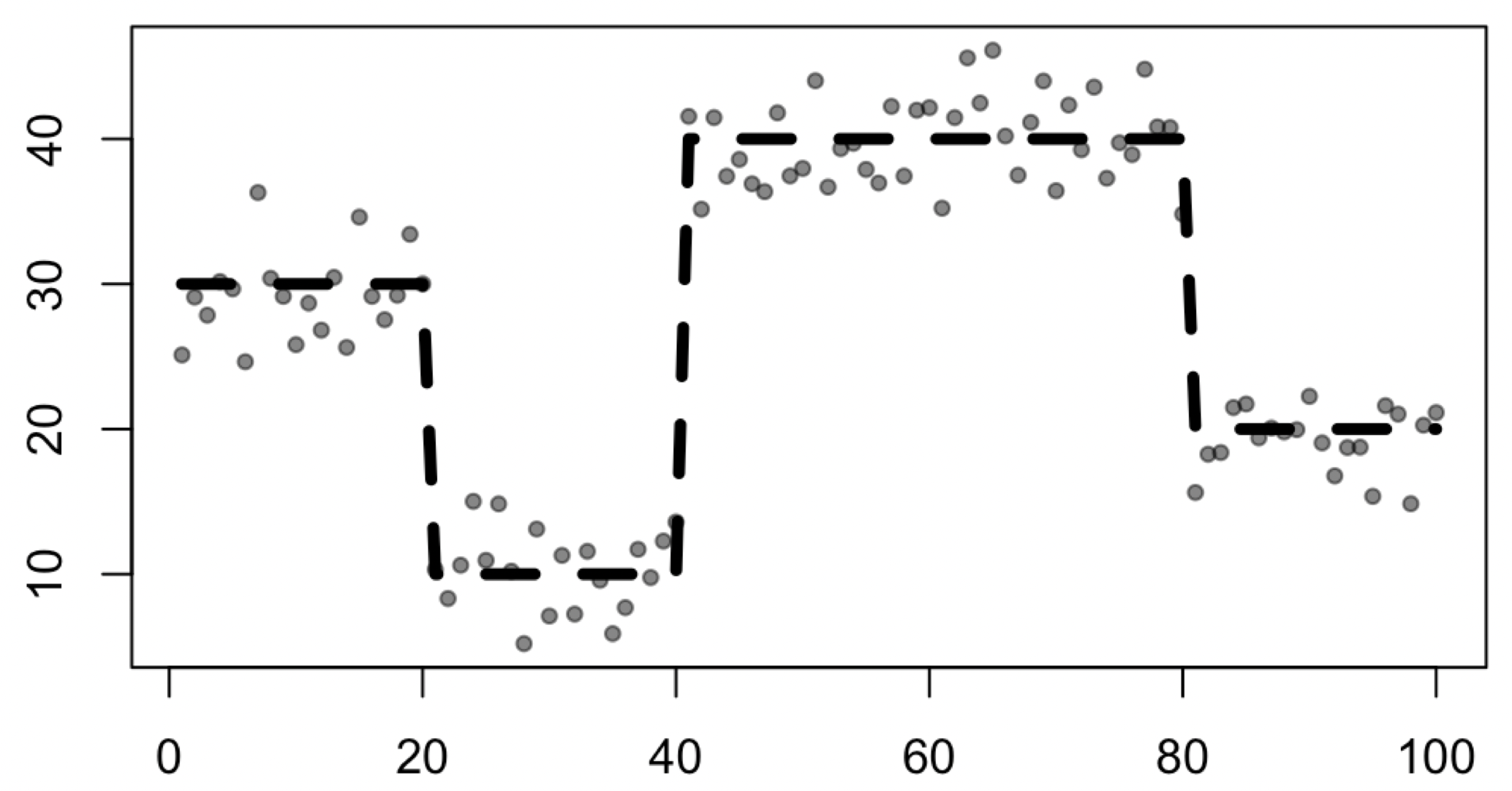}
                        \caption{The simulated data, with dashed line representing $\mu_t$, dots representing data $y_t$.}
                \end{subfigure}
                \begin{subfigure}[t]{0.48\textwidth}
                        \centering
                        \includegraphics[width=0.8\textwidth, height = .35\textwidth]{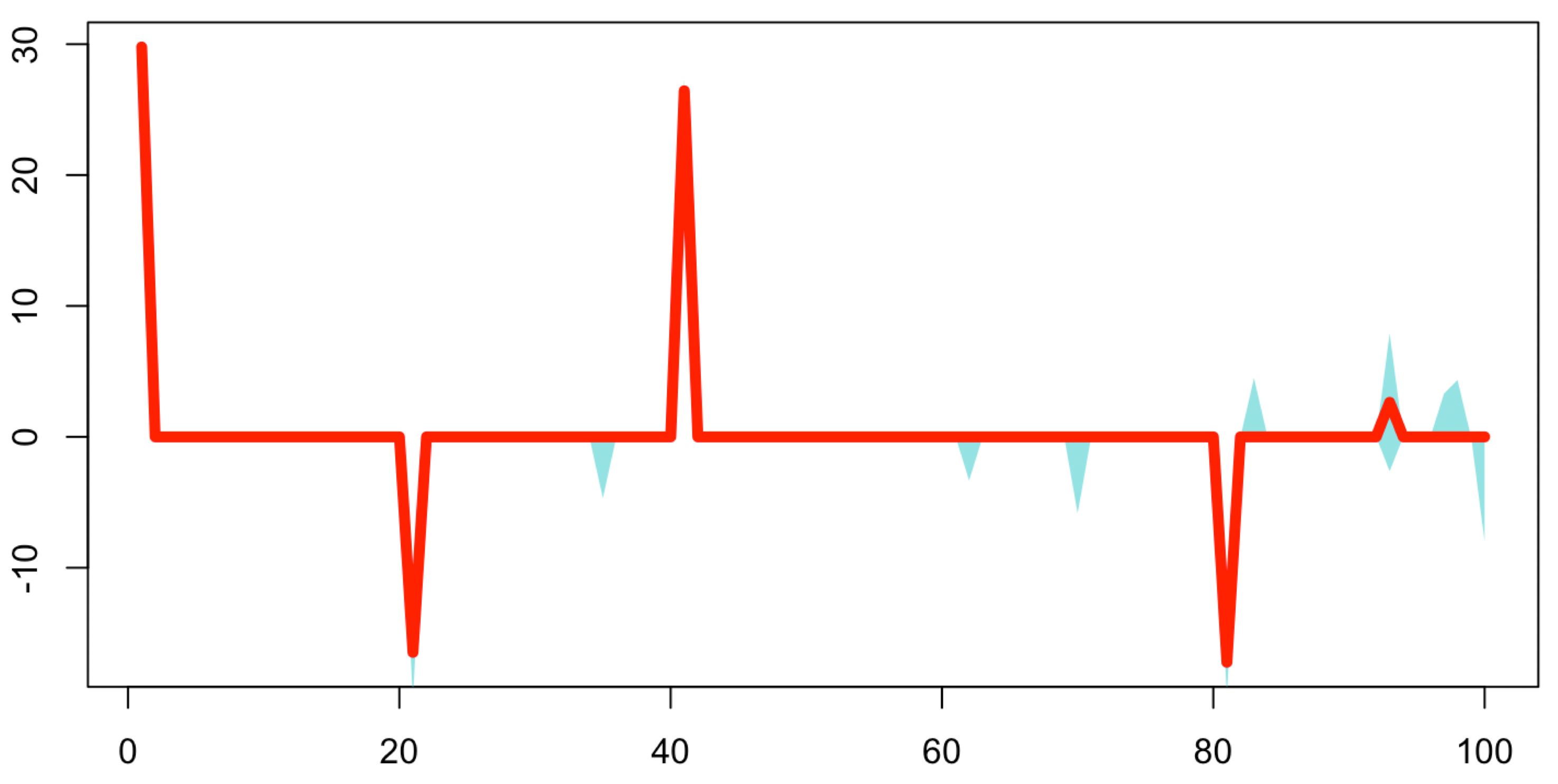}
                        \caption{The $l_1$-ball Fr\'echet mean of $\theta$ (red solid line) with 95$\%$ credible band (blue shadow).}
                \end{subfigure}
                \begin{subfigure}[t]{0.48\textwidth}
                        \centering
                        \includegraphics[width=0.8\textwidth, height = .35\textwidth]{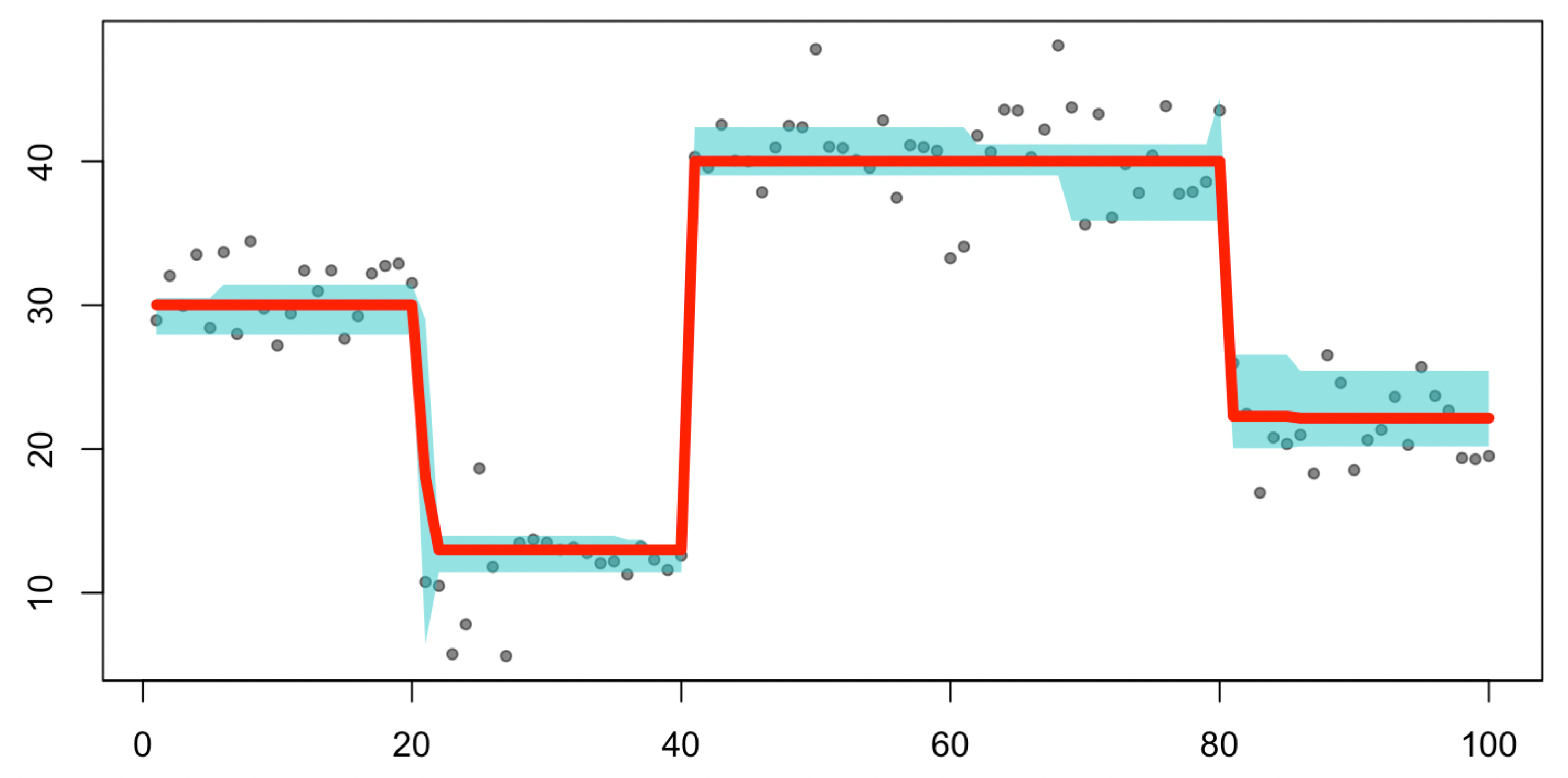}
                        \caption{The Fr\'echet mean curve of the $l_1$-ball prior (red solid line), with 95$\%$ credible band (blue shadow), using the highest posterior kernel region.}
                \end{subfigure}
                \begin{subfigure}[t]{0.48\textwidth}
                        \centering
                        \includegraphics[width=0.8\textwidth, height = .35\textwidth]{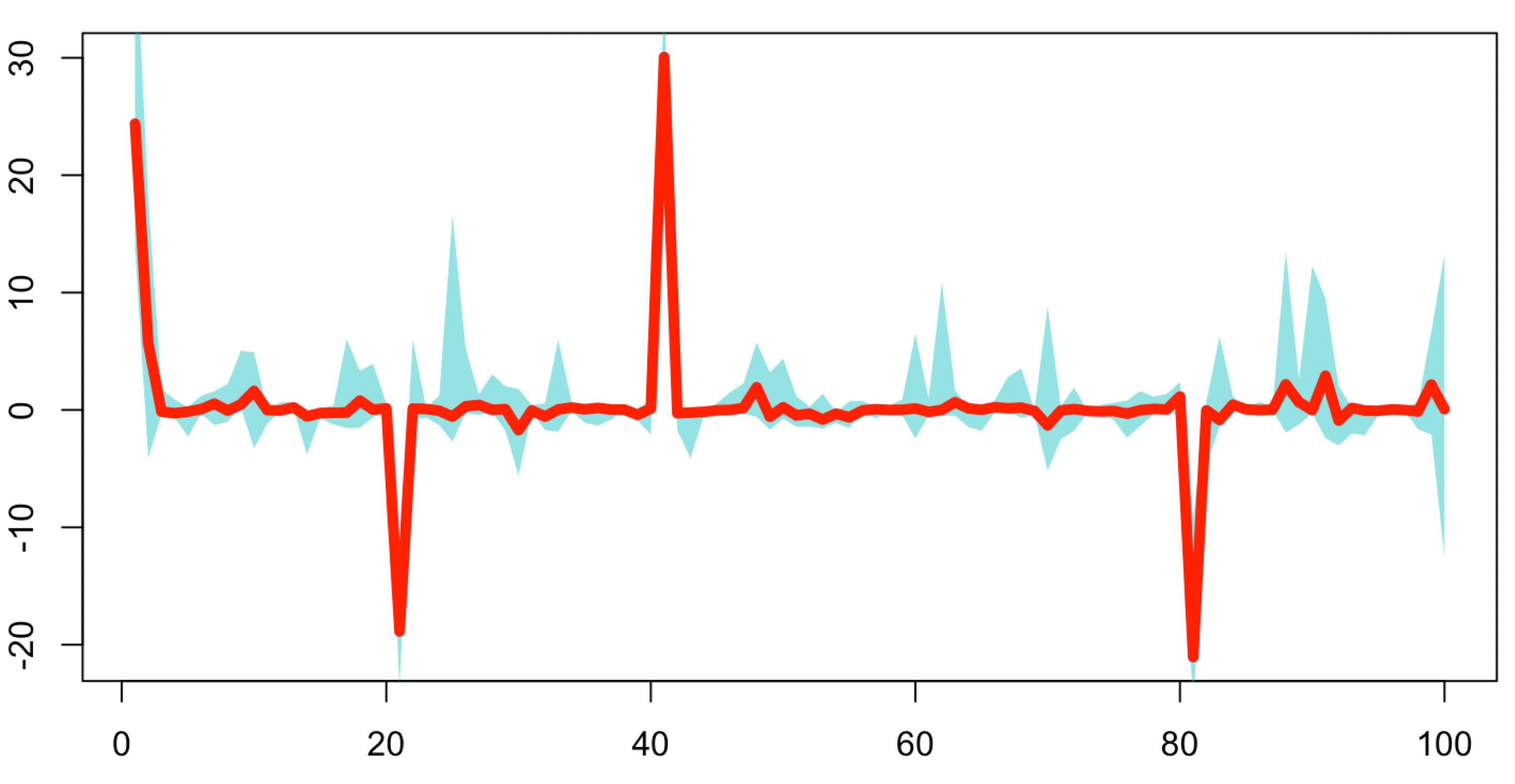}
                        \caption{The horseshoe Fr\'echet mean of $\theta$ (red solid line) with 95$\%$ credible band (blue shadow).}
                \end{subfigure}
                \begin{subfigure}[t]{0.48\textwidth}
                        \centering
                        \includegraphics[width=0.8\textwidth, height = .35\textwidth]{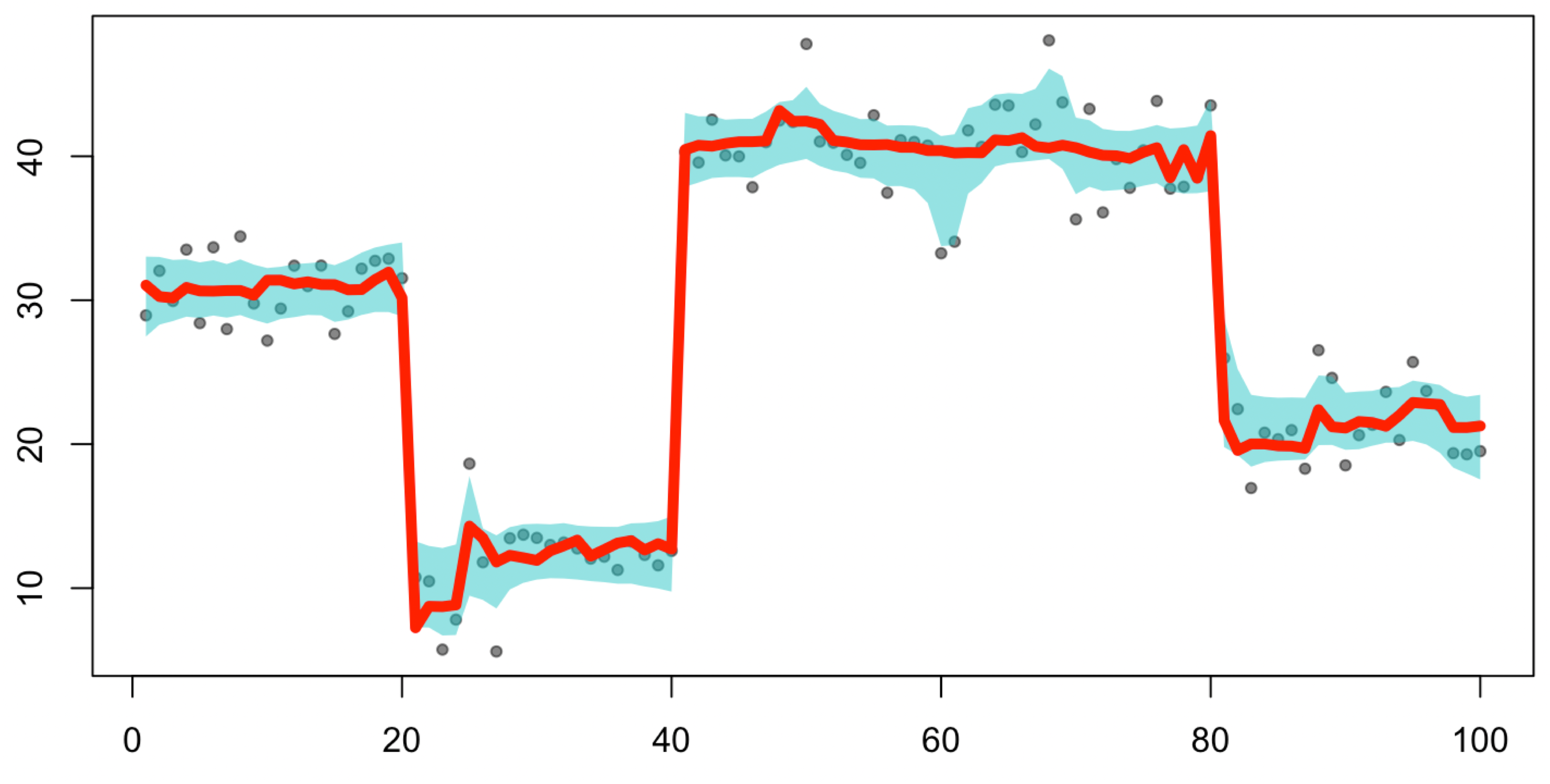}
                        \caption{The Fr\'echet mean curve of the horseshoe prior  (red solid line), with 95$\%$ credible band (blue shadow), using the highest posterior kernel region.}
                \end{subfigure}
                        
                \caption{Comparing the performances of applying $l_1$-ball prior and continuous shrinkage prior in the change point detection model: the $l_1$-ball produces a step function with a few steps corresponding to the major changes, whereas the continuous shrinkage prior produces small increments/decrements that accumulate over time to a non-trivial departure from a step function.}
                \label{fig: EG_CGP}
        \end{figure}

        As shown in Figure~\ref{fig: EG_CGP} (c), under the $l_1$-ball prior, we obtain the posterior curves in step functions, as desired in this model. On the other hand, the horseshoe prior could not produce a step function, due to the small increments/decrements accumulating over time (e), leading to a clear departure from a step function curve. 

        To be fair, this is an expected result as the continuous shrinkage prior is not designed for handling such a problem. In fact, comparing Panels b and d, the horseshoe prior here has a good performance in the uncertainty quantification on each of $\theta_i$. However, a key difference is in the {\em joint} probability of all $\theta_i$'s --- in this case, the horseshoe prior does not have a large probability for the neighboring $\theta_i$'s to have $\sum_{i=t}^{t+d}|\theta_i|=0$; whereas the $l_1$-ball does have this property, since all these small $\beta_i$'s with $|\beta_i|\le \tilde\mu$ are now reduced to exactly zero.
        
\section{Additional Experiments}

\subsection{Structured Sparsity: Inducing Dependency among Zeros}

We want to show how the $l_1$-ball prior can easily incorporate structured sparsity assumption \citep{hoff2017lasso, griffin2019structured}, where those zeros may have an inherent dependency structure.

To give more specifics, we present an application of improving network estimation on human brain functional connectivity, using prior information from the structural connectivity. The raw data of the former are an affinity  matrix with scores $A_{i,j}\in[-1,1]$ between 1,000 voxels collected from a functional magnetic resonance imaging (fMRI) on tracking their blood oxygen levels, and the latter is from a diffusion tensor imaging (DTI) that measures the white matter tractography in terms of observed probability $S_{i,j}\in [0,1]$ that two voxels are anatomically connected (with $S_{i,i}=0$ on the diagonal) \citep{cole2021surface}.

        \begin{figure}[H]
                \centering
                \begin{subfigure}[t]{0.2\textwidth}
                        \centering
                        \includegraphics[width=1\textwidth, trim={0 0 2.5cm 0},clip]{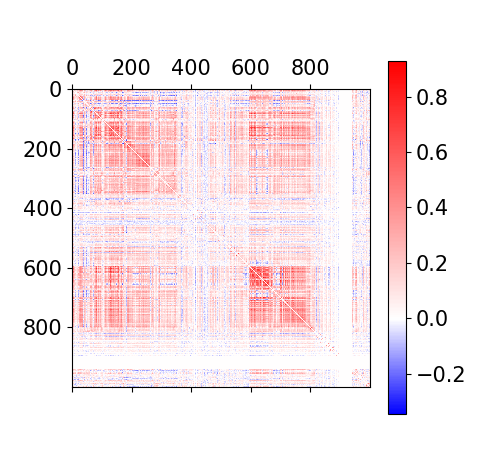}
                        \caption{The raw affinity matrix $A$ produced from fMRI.}
                \end{subfigure}\;
                \begin{subfigure}[t]{0.2\textwidth}
                        \centering
                        \includegraphics[width=1\textwidth, trim={0 0 2.5cm 0},clip]{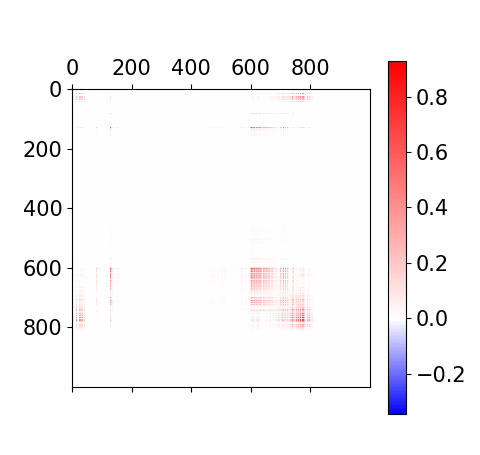}
                        \caption{The observed structural connectivity matrix $S$ from DTI.}
                \end{subfigure}\;
                \begin{subfigure}[t]{0.2\textwidth}
                        \centering  \includegraphics[width=1\textwidth, trim={0 0 2.5cm 0},clip]{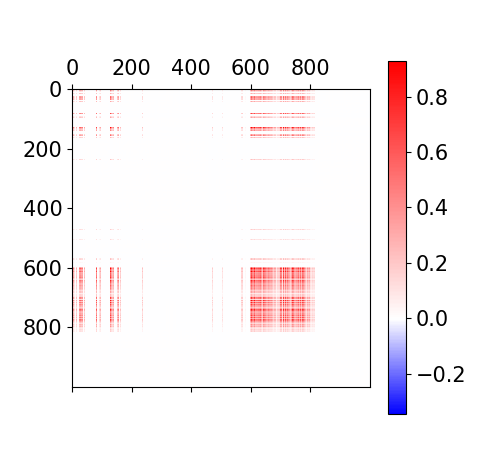}
                        \caption{Estimated functional connectivity using  $l_1$-ball prior with correlated $\beta$.}
                \end{subfigure}\;
                \begin{subfigure}[t]{0.24\textwidth}
                        \centering
                        \includegraphics[width=1.05\textwidth]{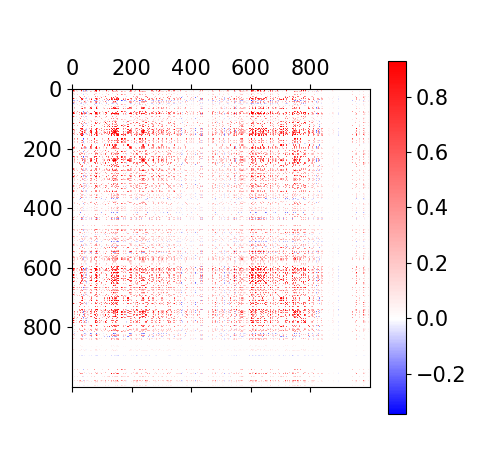}
                        \caption{Estimated functional connectivity using  uncorrelated shrinkage.}
                \end{subfigure}
                \caption{\label{fig: correlated_sc_fc} Improving low-rank functional connectivity estimates [from (a) to (c)] by inducing dependency among the zeros under an $l_1$-ball prior. The dependency comes from a correlation structure on $\beta$ according to the structural connectivity (b), and project to $\theta$. In comparison, uncorrelated shrinkage (d) does not enjoy such a structured sparsity, as it tends to pick up the large values from (a). Fr\'echet means are shown in (c) and (d).} 
        \end{figure} 

Due to that the affinity scores in $A$ are calculated based on some heuristic post-processing of the multivariate time series data from fMRI, there are often a large number of spurious associations. Therefore, it is useful to borrow information from the structural connectivity to model functional connectivity  \citep{honey2009predicting,bassett2018nature,zhu2014fusing}. That is, when the structural connectivity is small $S_{i,j}\approx 0$, then the chance of finding a functional connectivity should be small (whereas an $S_{i,j}\approx 1$ does not necessarily mean a high functional connnectivty). Following common low dimensional modeling strategy \citep{hoff2002latent} for a network, we use
\bel
\label{equ: correlated_brain}
A &= \sum_{k=1}^d \lambda_k\theta_k\theta_k^T + \mathcal E,\quad\mathcal E_{i,j}\stackrel{\text{iid}}{\sim} N(0,\sigma^2), \; \mathcal E_{j,i} = \mathcal E_{i,j} \t{ for } i<j,\\
\theta_k&=P_{\mathbb{B}_r}(\beta_k), \quad \beta_k\sim N(0, J-S + \kappa I ),\\
\lambda&=P_{\mathbb{B}_{\tilde r}}(\gamma), \quad  \gamma_k\stackrel{\text{iid}}\sim \t{Exp}(1) \quad \t{ for } k = 1,\ldots,d.
\eel
where $J$ is a matrix of ones, and $\kappa\ge 0$ is a constant to make the covariance positive definite; for our $S$, we use $\kappa=0$. We use $d=10$ and an $l_1$-ball prior to induce some $\lambda_k=0$ [in the Fr\'echet mean, we have effectively $2$ non-zero $\lambda_k$'s]. In the above model, when $S_{i,j}\approx 0$, we would have $\beta_{k,i}$ and $\beta_{k,j}$ strongly correlated for $k=1,\ldots, d$, hence $\theta_{k,i}$ and $\theta_{k,j}$ have a large chance to be simultaneously zero a priori; on the other hand, when $S_{i,j} \approx 1$, $\theta_{k,i}$ and $\theta_{k,j}$ are less correlated, hence $S_{i,j}$ has less influece on the estimate.

Figure \ref{fig: correlated_sc_fc} shows how this model borrows information from structural connectivity (panel b) to make a sparse estimate on the functional connectivity (panel c). Compared to the raw affinity matrix (panel a), the majority of the functional connectivity among the first 400 voxels are shrunk to zero, since there is little structural connectivity. To compare, we apply the same model except with $\beta_k \sim N(0,I)$ and plot the estimate (in panel d); without imposing dependency among those zeros, the estimated matrix is close to picking up the large affinity scores from $A$.

\vspace{-0.5cm}

\subsection{Additional Results of Discontinuous Gaussian Process Regression}

\vspace{-0.5cm}

      \begin{figure}[H] 
      \begin{minipage}{.9\textwidth}
      	                      \begin{subfigure}[t]{.3\textwidth}
                        \includegraphics[width=1\linewidth, height =4.0cm]{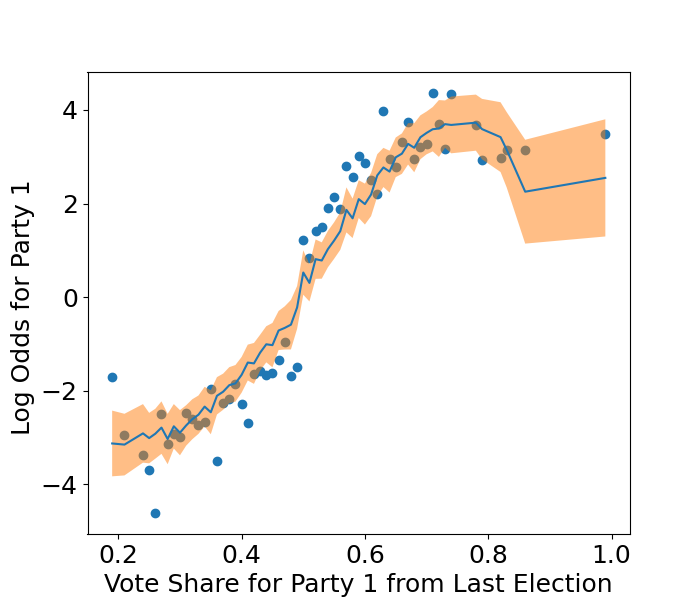}
                \end{subfigure}
                \;
                      \begin{subfigure}[t]{.3\textwidth}
                        \includegraphics[width=1\linewidth, height =4.0cm]{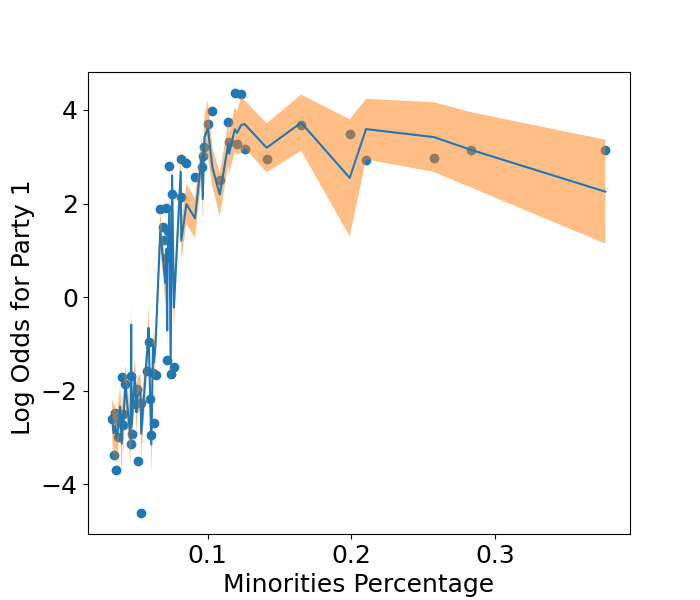}
                \end{subfigure}
                \;
                     \begin{subfigure}[t]{.3\textwidth}
                        \includegraphics[width=1\linewidth, height =4.0cm]{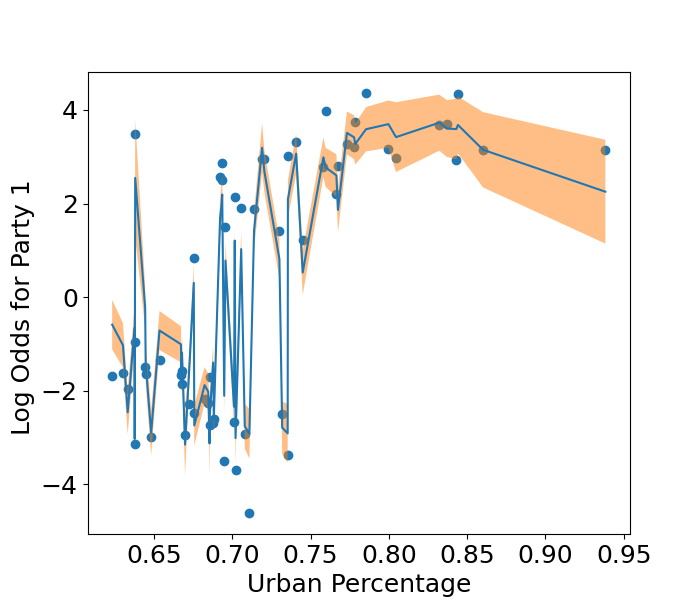}
                \end{subfigure}
                                        \caption*{Fitting a continuous Gaussian process with squared exponential covariance.}
      \end{minipage}
            \begin{minipage}{.9\textwidth}
                \begin{subfigure}[t]{.3\textwidth}
                        \includegraphics[width=1\linewidth, height =4.0cm]{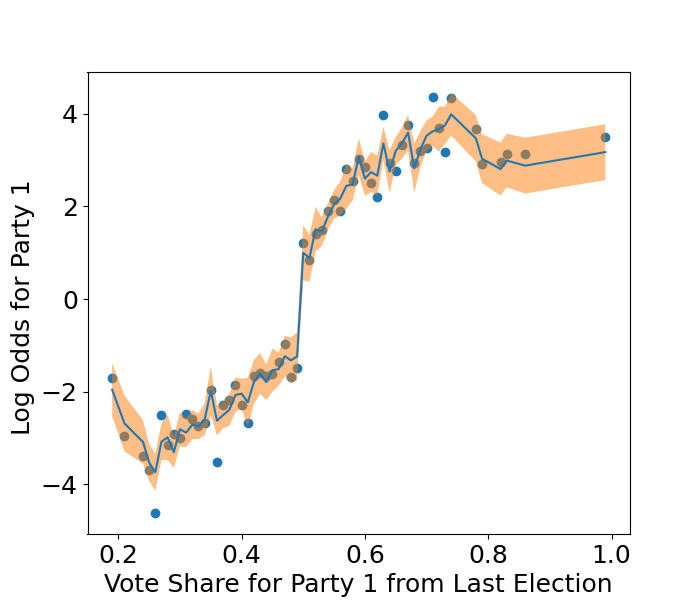}
                \end{subfigure}
                                \;
                     \begin{subfigure}[t]{.3\textwidth}
                        \includegraphics[width=1\linewidth, height =4.0cm]{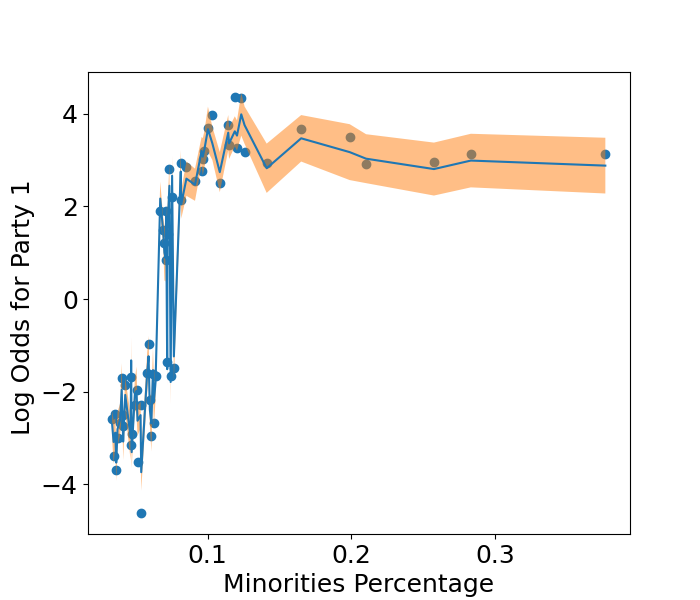}
                \end{subfigure}\;
                     \begin{subfigure}[t]{.3\textwidth}
                        \includegraphics[width=1\linewidth, height =4.0cm]{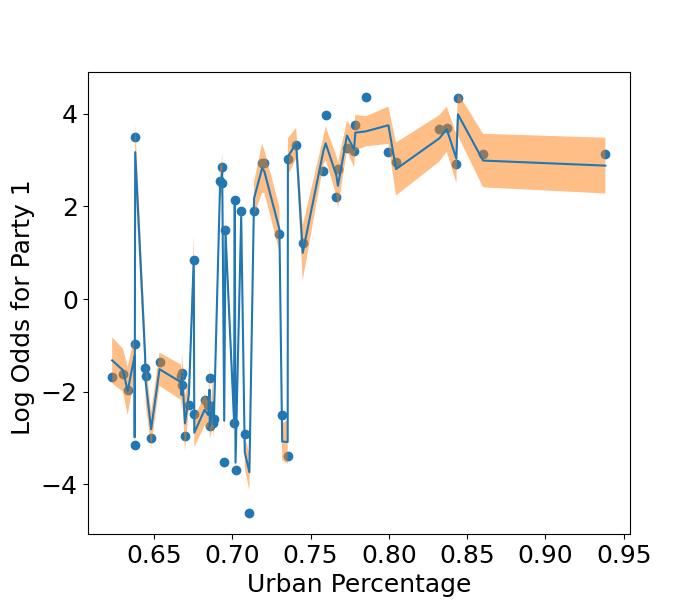}
                \end{subfigure}\\\centering
                                     \begin{subfigure}[t]{.3\textwidth}
                        \includegraphics[width=1\linewidth]{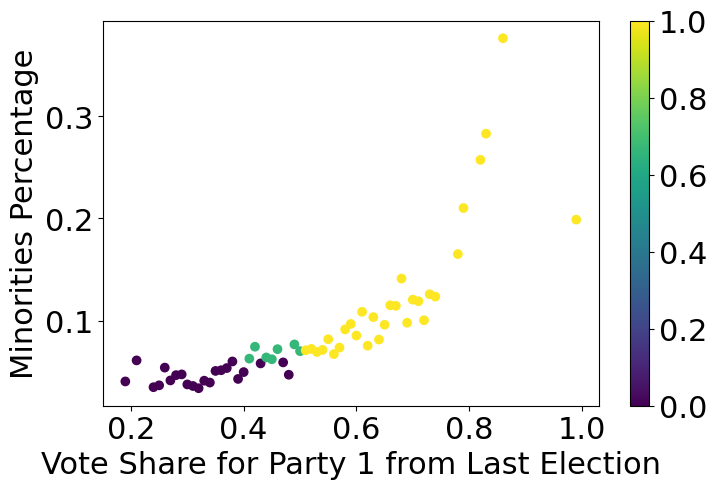}
                \end{subfigure}
                        \caption*{Fitting a discontinuous Gaussian process with latent jittering coordinates (with the values of $\eta_i$ shown in the last panel).}
                      \end{minipage}
                                \caption{Discontinuous Gaussian process regression on the election data. With a latent jittering coordinate $\eta_i$ regularized by a generalized $l_1$-ball prior, we can change a Gaussian process with squared exponential covariance function (first row) to have  discontinuities, giving an improved fit to the data (second row). The fitted curves and 95\% point-wise credible bands are shown.\label{fig: GPR_full}}
        \end{figure}

Figure \ref{fig: GPR_full} compares the fitting continuous and discontinuous Gaussian process regression models to the election data, using all three predictors. The root-mean-square deviation (RMSD) is 0.315 for the discontinuous model, and is 0.427 for the continuous one. The discontinuous Gaussian process  finds three distinct values in $\eta_i$'s.

% To compare with sparse estimation using only the dMRI data ($F$ matrix), we also apply model [\eqref{equ: correlated_brain}] with isotropic prior on $\beta$ (letting $\Sigma_\beta = I$). We fix $d$ = 2 in both models. Figure \ref{fig: correlated_sc_fc} compares the results of using and not using SC as the covariance matrix in $\pi_\beta$. By introducing dependence in $\beta$ with prior knowledge of SC, we successfully construct sparse eigenvectors that eliminate redundant and spurious association (Panel c). On the other hand, an isotropic $\pi_\beta$ (Panel d) just selects those large signals in the raw correlation matrix, but might include misdetected association and lacks interpretation in structural level. 
        
\vspace{-0.8cm}

\subsection{Additional Results of the Sparse Change Detection Data Application}

\vspace{-0.3cm}

\begin{figure}[H]
\centering
        \begin{subfigure}[b]{.3\textwidth}
\centering
        \includegraphics[height=3.3cm, width=4cm]{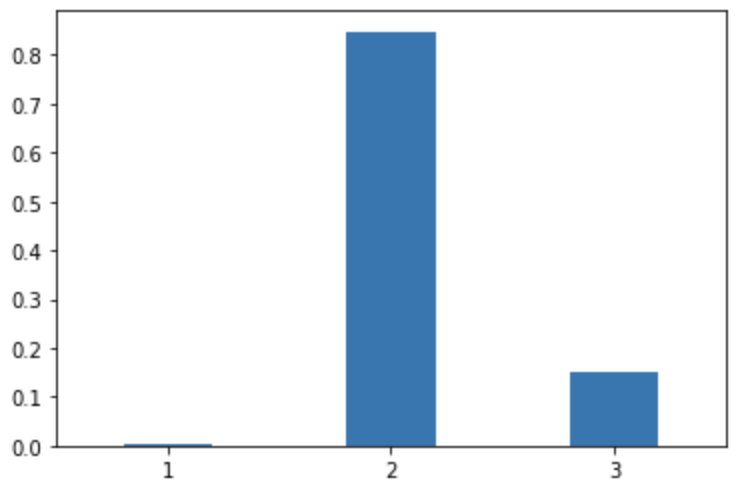}
        \caption{The posterior distribution of the rank $d$.}
\end{subfigure}
% \begin{subfigure}[htbp]{.24\textwidth}
% \centering
%         \includegraphics[height=3.5cm, width=4.2cm]{figure/loading_1}
%         \caption{The image of the first loading.}
% \end{subfigure}
\begin{subfigure}[b]{.3\textwidth}
\centering
        \includegraphics[height=3.3cm, width=4.2cm]{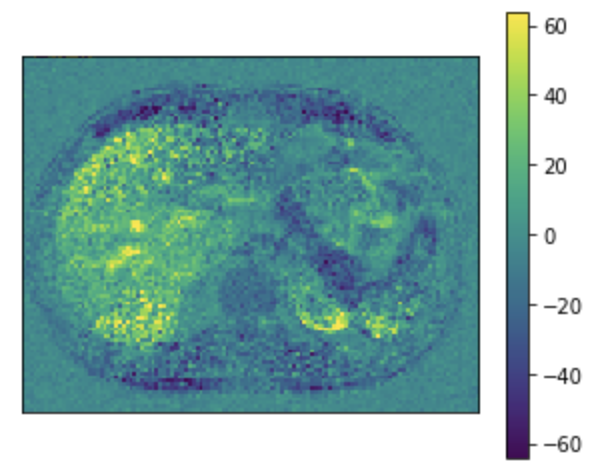}
        \caption{The image of the latent component $C_1$.}
\end{subfigure}
\quad
\begin{subfigure}[b]{.3\textwidth}
\centering
        \includegraphics[height=3.3cm, width=4.2cm]{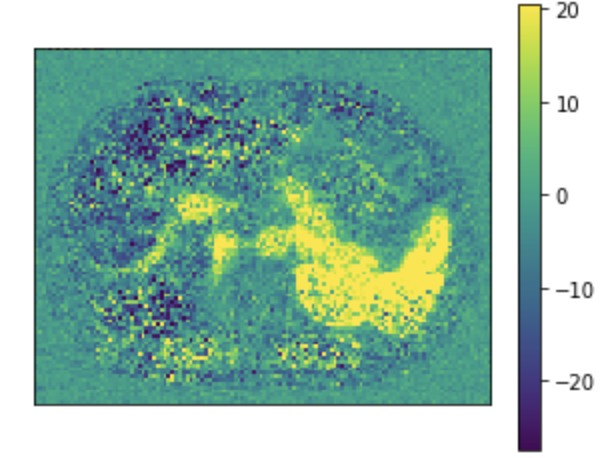}
        \caption{The image of the latent component $C_2$}
\end{subfigure}
\quad
        \begin{subfigure}[b]{.3\textwidth}
\centering
        \includegraphics[height=3.3cm, width=4.2cm]{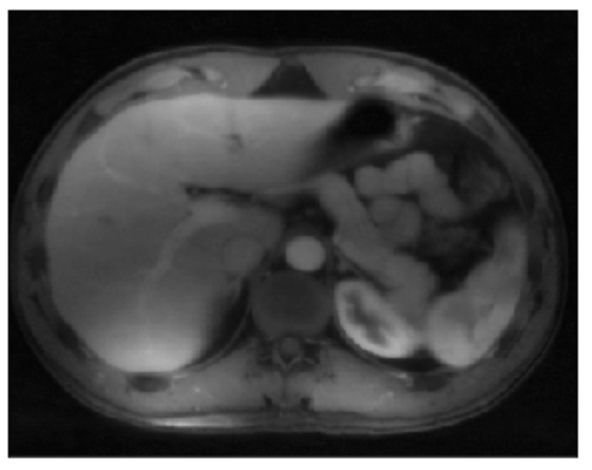}
        \caption{Estimated background at $t=15$.}
\end{subfigure}
\quad
        \begin{subfigure}[b]{.3\textwidth}
\centering
        \includegraphics[height=3.3cm, width=4.2cm]{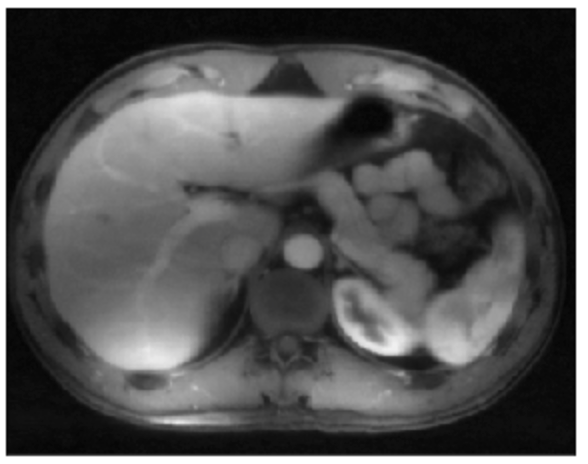}
        \caption{Estimated background at $t=35$.}
\end{subfigure}
\quad
        \begin{subfigure}[b]{.3\textwidth}
\centering
        \includegraphics[height=3.3cm, width=4.2cm]{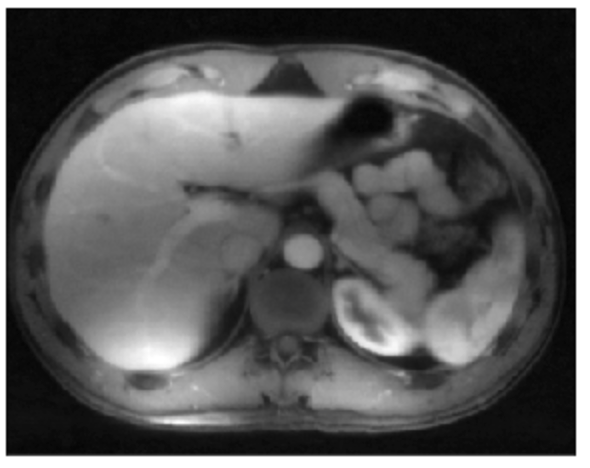}
        \caption{Estimated background at $t=75$.}
\end{subfigure}
\caption{\label{fig: video_L} The low-rank modeling of the video background $L$,
 where $L$ is a flattened matrix containing all the background images over $75$ time points. The posterior distribution is concentrated at a low rank $d=2$ for the matrix $L$ (panel a). This captures the subtle changes that happen in the background, such as 
the brightness between (d) and (f).
}
\end{figure}

\begin{figure}[H]
     \begin{subfigure}[b]{.3\textwidth}
\centering
        \includegraphics[height=3.3cm, width=4.2cm]{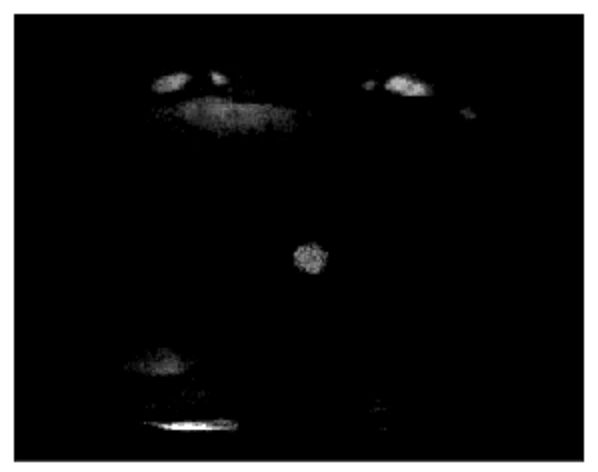}
        \caption{The sparse change at $t=15$.}
\end{subfigure}
\quad
        \begin{subfigure}[b]{.3\textwidth}
\centering
        \includegraphics[height=3.3cm, width=4.2cm]{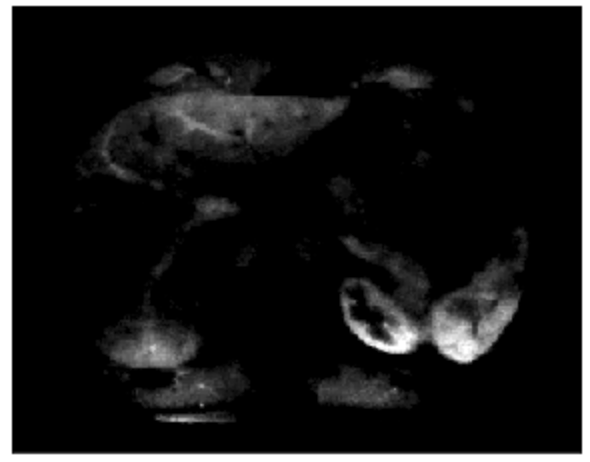}
        \caption{The sparse change at $t=35$.}
\end{subfigure}
\quad
        \begin{subfigure}[b]{.3\textwidth}
\centering
        \includegraphics[height=3.3cm, width=4.2cm]{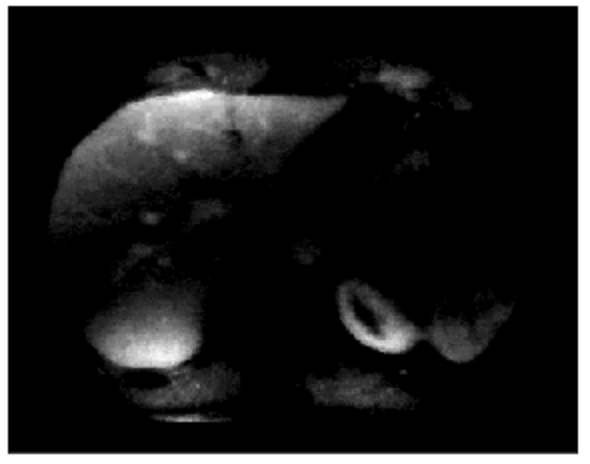}
        \caption{The sparse change at $t=75$.}
\end{subfigure}
\caption{\label{fig:video_simple_S}The sparse change  $S_t$ with a time-invariant background. Compared to Figure 5 in the article, these sparse estimates fail to reveal details under the slowly changing background: (a) does not highlight the aorta enhancement; (b) and (c) involve too much light change, so that the detailed liver and portal vein enhancement are blurred. }
\end{figure}

\subsection{Simulation on Rank Estimation}
\begin{figure}[H]
	\centering
		\begin{subfigure}[htbp]{.44\linewidth}
			\centering
			\includegraphics[width=.99\linewidth]{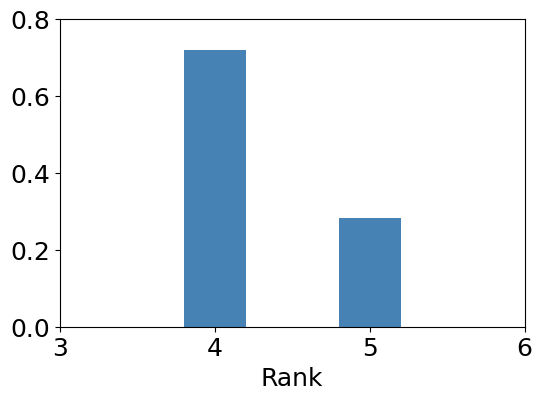}
			\caption{Posterior distribution of rank$(L)$, where the truth is 5.}
		\end{subfigure}		\quad\quad
		\begin{subfigure}[htbp]{.44\linewidth}
			\centering
			\includegraphics[width=.99\linewidth]{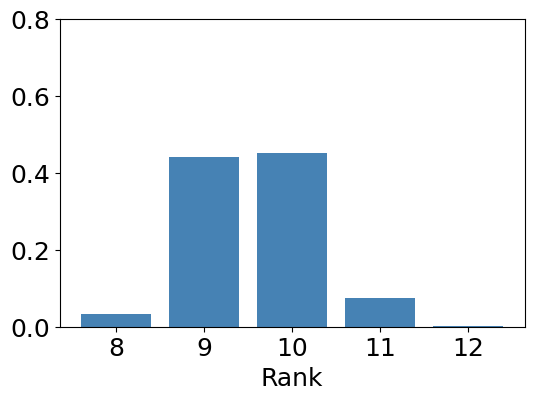}
			\caption{Posterior distribution of rank$(L)$, where the truth is  10.}
		\end{subfigure}	  
		\caption{\label{fig: rank_estimation}Simulation for the rank estimation using the nuclear-norm $l_1$-ball prior.}
	\end{figure}

{We now use a simulation to empirically illustrate the performance of rank estimation with the generalized $l_1$-ball prior using the nuclear norm. We first generate a set of $p$-element vectors $\phi_1,\ldots,\phi_d$ with $\phi_{k,i}\stackrel{\text{iid}}\sim N(0,5^2)$, and  a set of $T$-element vectors $\tilde \alpha_1,\ldots,\tilde \alpha_d$ with $\tilde \alpha_{k,i}\stackrel{\text{iid}}\sim N(0,1^2)$. Then we obtain a $p\times T$ matrix
\(
M = \sum_{k=1}^d  \phi_k (\tilde\alpha_{k})^{\rm T} + \mathcal E,
\)
where each entry in the noise matrix $\mathcal E$ is generated from iid $N(0,1^2)$. We model the simulated data by

\[
\begin{aligned}
&	M =  L + \mathcal E, \\
& L = \argmin_{Z\in\mathbb R^{p\times T},\|Z\|_*\le r}\|Z-\beta\|_F^2,\quad \beta_{ij}\stackrel{\text{iid}}{\sim}N(0,\sigma_\beta^2).
\end{aligned}
\]
where we set $\sigma^2_\beta = 5^2$, $r\sim \text{Exp}(10)$. 
Figure \ref{fig: rank_estimation} shows the posterior distribution of the rank in the $(d, T) = (5, 25)$ and $(10,50)$ settings, with $p=25$. In both cases, the nuclear-norm based $l_1$-ball prior successfully recovers the true rank.}

% \appendix

% \section*{Appendix}

% \subfile{sections/supp/proof}
% \subfile{sections/supp/linear_regression}
% \subfile{sections/supp/sparse_change}
% \subfile{sections/supp/change-point}
\bibliographystyle{chicago}
\bibliography{reference2.bib} 
\end{document}